\documentclass{aa}

\pdfoutput=1
\usepackage[varg]{txfonts}
\usepackage{graphicx}
\usepackage{array}
\usepackage{multirow}
\usepackage{microtype}
\usepackage[english]{babel}
\usepackage{relsize}
\usepackage{multicol}
\usepackage{pifont}

\usepackage{xcolor}
\usepackage[normalem]{ulem}
\usepackage{enumitem}
\usepackage{colortbl}
\usepackage{arydshln}

\usepackage{natbib}
\bibpunct{(}{)}{;}{a}{}{,} % to follow the A&A style

\newcommand*{\norm}[1]{\left\lVert#1\right\rVert}
\DeclareMathOperator*{\RMSE}{RMSE}
\DeclareMathOperator*{\SAM}{SAM}
\newcommand{\mycomment}[1]{}

\newcommand{\ocfigure}[4][tbp]{
    \begin{figure}[#1]
        \resizebox{\hsize}{!}{\includegraphics{#2}}
        \caption{#3}
        \label{#4}
    \end{figure}
}

\newcommand{\tcfigure}[4][tbp]{
    \begin{figure*}[#1]
        \centering
        \includegraphics[width=17cm]{#2}
        \caption{#3}
        \label{#4}
    \end{figure*}
}

\newcommand{\wfigure}[4][tbp]{
    \begin{figure*}[#1]
        \sidecaption
        \includegraphics[width=12cm]{#2}
        \caption{#3}
        \label{#4}
    \end{figure*}
}

\renewcommand{\b}[1]{{#1}}

\newcommand{\num}[1]{{#1}}

\newcommand{\cmark}{\ding{51}}

\usepackage{amstext}

\begin{document}

\title{
    Neural network for determining an asteroid mineral composition from reflectance spectra
    }

\author{
    David Korda\inst{1}
    \and
    Antti Penttil{\" a}\inst{2}
    \and
    Arto Klami\inst{3}
    \and
    Tom{\' a}{\v s} Kohout\inst{1, 4}
}

\offprints{David Korda, \\ \email{david.korda@helsinki.fi}}

\institute{
    Department of Geosciences and Geography, P.O.~Box~64, FI-00014, University of Helsinki, Finland
    \and
    Department of Physics, P.O.~Box~64, FI-00014, University of Helsinki, Finland
    \and
    Department of Computer Science, P.O.~Box~68, FI-00014, University of Helsinki, Finland
    \and
    Institute of Geology, Czech Academy of Sciences, Rozvojov{\' a}~269, CZ-16500, Prague, Czech Republic
} 

\abstract
{Chemical and mineral compositions of asteroids reflect the formation and history of our Solar System. This knowledge is also important for planetary defence and in-space resource utilisation. In the next years, space missions will generate extensive spectral datasets from asteroids or planets with spectra that will need to be processed in real time.}
{We aim to develop a fast and robust neural-network-based method for deriving the mineral modal and chemical compositions of silicate materials from their visible and near-infrared spectra. The method should be able to process raw spectra without significant pre-processing.}
{We designed a convolutional neural network with two hidden layers for the analysis of the spectra, and trained it using labelled reflectance spectra. For the training, we used a dataset that consisted of reflectance spectra of real silicate samples stored in the RELAB and C-Tape databases, namely olivine, orthopyroxene, clinopyroxene, their mixtures, and olivine-pyroxene-rich meteorites.}
{We used the model on two datasets. First, we evaluated the model reliability on a test dataset where we compared the model classification with known compositional reference values. The individual classification results are mostly within 10 \b{percentage-point} intervals around the correct values. Second, we classified the reflectance spectra of S-complex (Q-type and V-type, also including A-type) asteroids with known Bus--DeMeo taxonomy classes. The predicted mineral chemical composition of S-type and Q-type asteroids agree with the chemical composition of ordinary chondrites. The modal abundances of V-type and A-type asteroids show a dominant contribution of orthopyroxene and olivine, respectively. Additionally, our predictions of the mineral modal composition of S-type and Q-type asteroids show an apparent depletion of olivine related to the attenuation of its diagnostic absorptions with space weathering. This trend is consistent with previous results of the slower pyroxene response to space weathering relative to olivine.}
{The neural network trained with real silicate samples and their mixtures is applicable for a quantitative mineral \b{evaluation of spectra of asteroids that are rich in dry silicates}. The modal abundances and mineral chemistry of common silicates (olivine and pyroxene) can be derived with an accuracy better than 10 \b{percentage points}. The classification is fast and has a relatively small computer-memory footprint. Therefore, our code is suitable for processing large spectral datasets in real time.}


\authorrunning{D. Korda et al.}
\maketitle

%%%%%%%%%%%%%%%%%%%%%%%%%%%%%%%%%%%%%%%%%%%%%%%%%%

\section{Introduction}

Ordinary chondrites are the most common type of meteorites. They originate from the S-complex asteroids and are one of the oldest materials in our Solar System. Their dominant constituents are olivine, orthopyroxene, FeNi metal, and troilite. The Fe content of the silicate minerals encodes information about redox conditions in the early Solar System. The mineralogical and chemical compositions of asteroids are also important in in-space resource utilisation and planetary defence.

Olivine and pyroxene have strong absorption bands in their near-infrared reflectance spectrum. The absorption bands are due to Fe$^{2+}$ cations in crystalline structure. The three overlapping absorption bands of olivine are located approximately at the 1~\textmu{}m wavelength. The two strong absorption bands of pyroxene are located approximately at the 1 and 2~\textmu{}m wavelengths, and one weak band of iron-rich pyroxene is approximately at the 1.25~\textmu{}m wavelength. The properties of the individual bands change with the chemical composition of olivine and pyroxene, while the overall spectral shape depends on the \b{ratio of the olivine to pyroxene modal abundances} or on the \b{grain sizes of the minerals}. For this reason, previous authors have been focusing on establishing relations between the properties of the bands and the corresponding composition. \citet{Adams_1974} pointed out a correlation between the chemical composition of pyroxene and the position of the minima of the two strongest pyroxene bands. \citet{Cloutis_1986} found a relation between the area of the olivine-orthopyroxene mixture absorption bands and the modal composition of the mixture. Because the central wavelength and the area of the absorption bands are sensitive to the modal and chemical compositions of the minerals, it is important to model the bands properly. A model-fitting tool of the reflectance spectra was derived by \cite{MGM}, who introduced modified Gaussians to model the absorption bands in logarithmic reflectance. These methods have been intensively used and tested \citep[e.g.][]{Gaffey_1984, McSween_1991, Fornasier_2003, Kanner_2007, Dunn_2010b, Popescu_2012, Han_2020}. Therefore, the chemical composition of olivine and pyroxene, and their abundance in mixtures, can in principle be obtained from the reflectance spectra through spectral deconvolution.

\begin{table*}
    \caption{Basic information about our dataset.}
    \label{tab:data}
    \centering
    \begin{tabular}{l c c c c c c c}
        \hline\hline
        & OL & OPX & CPX & OL + OPX & OL + CPX$^{*}$ & OPX + CPX & OL + OPX + CPX$^{**}$\\
        \hline
        Number & 100 & 102 & 108 & 62 & 3 & 53 & 82\\
        \hdashline
        OL frac. [vol\%] & 100 & N/A & N/A & \phantom{0}4.9--90.2 & 14.5--91.4 & N/A & 10.7--94.7\\
        OPX frac. [vol\%] & N/A & 100 & N/A & \phantom{0}9.8--95.1 & N/A & \phantom{0}9.6--89.6 & \phantom{0}2.6--86.1\\
        CPX frac. [vol\%] & N/A & N/A & 100 & N/A & \phantom{0}8.6--85.5 & 10.4--90.4 & \phantom{0}0.2--41.5\\
        \hdashline
        Fa & \phantom{0}3.1--100\phantom{.} & N/A & N/A & \phantom{0}9.6--40.0 & 31.2--69.0 & N/A & \phantom{0}7.7--31.2\\
        Fs (OPX) & N/A & \phantom{0}7.1--100\phantom{.} & N/A & \phantom{0}7.1--42.7 & N/A & \phantom{0}9.5--63.5 & 12.4--28.9\\
        Fs (CPX) & N/A & N/A & \phantom{0}5.0--93.0 & N/A & 13.6--24.0 & \phantom{0}8.0--33.0 & \phantom{0}5.2--13.8\\
        En (CPX) & N/A & N/A & \phantom{0}0.0--72.0 & N/A & 37.0--47.3 & 29.4--53.7 & 41.3--57.3\\
        Wo (CPX) & N/A & N/A & \phantom{0}7.0--54.0 & N/A & 39.0--39.1 & 36.6--49.0 & 32.3--48.0\\
        \hline
    \end{tabular}
    
    \vspace{1ex}
    {\raggedright N/A stands for `not applicable'. $^{*}$Data include two brachinites and one SNC achondrite. $^{**}$Data include 45 ordinary chondrites, seven HED meteorites, five lodranites, and three SNC achondrites. \par}
\end{table*}

The deconvolution methods usually consist of two steps: (1)~separation of a spectrum into continuum and absorption bands by fitting, and (2)~application of empirical relations between the band parameters of the fit and the chemical composition. However, for each mineral mixture, we need to know a priori which minerals we expect to be present. This results in a different number of fitting parameters for fitting the spectrum in each case. Another limiting factor is \b{that the band depth depends on the grain size}, which results in a varying relative contribution of individual bands in the mineral-mixture spectrum. Without a precise fit of the whole spectrum and a correct separation of the bands and the continuum, no precise chemical composition can be derived.

In the past 50 years, several asteroid classification schemes (taxonomies) were introduced \citep[e.g.][]{Tholen_1984, Bus_1999, DeMeo_2009}. The taxonomies are based on observed asteroid spectra of (typically) a few hundred asteroids in a given number of colours (filters) or in a continuous spectral range (with most of the observations \b{taken from visible up to} 1~\textmu{}m) and are based on spectral features rather than on the features in the asteroid composition space. Therefore, taxonomies can tell us whether the spectrum of the observed asteroid contains absorption bands of specific minerals, but they contain little information about the quantitative mineral abundances or mineral chemical compositions of the asteroid.

For this reason, spectrum deconvolution methods have been applied on asteroid spectra to gain the quasi-quantitative metrics of their mineral and chemical composition. In the next years, spacecraft missions to asteroids or planets (e.g. Hera, BepiColombo, or Psyche) will provide us with high-resolution disk-resolved reflectance spectra that will increase the need for precise spectrum-processing methods even further to determine subtle variations in the surface compositions with little a priori knowledge.

Machine learning is based on flexible models that are trained using observed data. Deep learning \citep{Goodfellow_2016} considers specific machine-learning models, neural networks, that consist of multiple layers of simple computational steps. Neural networks have been successfully used in astronomy \citep[e.g.][]{Angel_1990, Odewahn_1992, Snider_2001, Pearson_2018, Hon_2018, Ni_2021}, geophysics and planetary science \citep[e.g.][]{Rigol_2003, Ross_2018, Civilini_2021}, or asteroid studies \citep[e.g.][]{Howell_1994, Misra_2008, Lieu_2019, Wallace_2021, Penttila_2021, Penttila_2022} because they combine a high degree of flexibility with efficient processing of new observations. Therefore, they are ideal candidates for the processing of reflectance spectra on the ground and in flight using devices embedded on-board of spacecraft. 

The aim of this work is to create a neural-network model to estimate the modal abundances of olivine, orthopyroxene, and clinopyroxene and their chemical compositions based on the recorded spectrum. We evaluate the reliability of error metrics on a subset of test spectra. Subsequently, we apply the method to derive the composition of S$^*$-complex asteroids from their reflectance spectra. Within this work, we use the term S$^*$-complex asteroids to refer to asteroids that are rich in dry silicates such as olivine and pyroxenes. This includes the S, Q, R, O, and V types and their subtypes as defined by \citet{DeMeo_2009}. Additionally, we include A and Sa types because they are assumed to have an olivine-rich composition.

The Python scripts \b{and} datasets \b{including all sample spectra and composition information} we used in this study \b{together with} metadata can be downloaded from GitHub repository\footnote{\url{https://github.com/Sirrah91/Asteroid-spectra}}.

%%%%%%%%%%%%%%%%%%%%%%%%%%%%%%%%%%%%%%%%%%%%%%%%%%

\section{Data}

\tcfigure{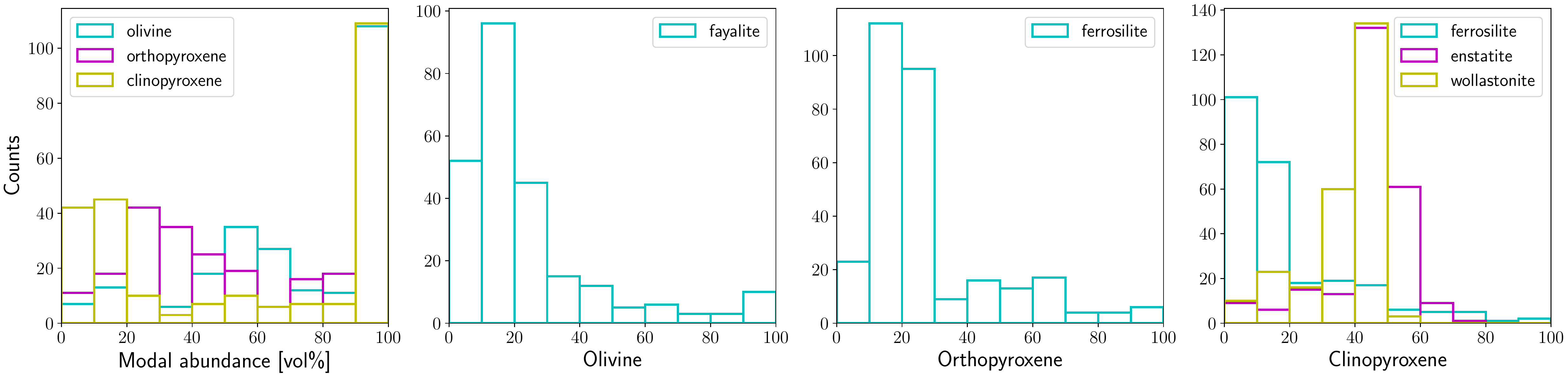}
{Mineralogical distributions of the data.}
{fig:spectra_min_hist}

Representative reflectance spectra of olivine~(OL; 100 samples), orthopyroxene~(OPX; 102 samples), clinopyroxene~(CPX; 108 samples), laboratory silicate mixtures~(137 samples), and meteorites~(45 ordinary chondrites, seven HED, five lodranites, four SNC achondrites, and two brachinites; 63 samples) were recovered from the RELAB\footnote{\url{http://www.planetary.brown.edu/relabdocs/relab.htm}} and C-Tape\footnote{\url{https://uwinnipeg.ca/c-tape/sample-database.html}} databases.

The spectra were selected based on the following criteria: (1)~the wavelength range \b{must cover} 450--2450~nm \b{interval with} a step size of 15~nm or smaller to match available asteroid spectra in \citet{DeMeo_2009} and \citet{Binzel_2019}, (2)~both mineral modal and elemental compositions should be documented, for instance by X-ray diffraction~(XRD), full electron microprobe analysis~(EMPA), or at least cation ratios. Alternatively, a reasonable confidence to published composition information should exist, for example through information of the sample locality, (3)~the fayalite number of olivine and the ferrosilite number of pyroxene is higher than 3 and 5, respectively, as we found that samples with lower content lack significant diagnostic features, (4)~the sample does not contain more than 35~\b{vol}\% \b{of} additional phases that contain strong absorption bands in the visible to near-infrared range (e.g. plagioclase), as we found that a higher content of these phases can have a noticeable effect on the overall spectral shape; spectrally neutral or opaque phases such as metals or sulphides were ignored as they do not alter silicate absorptions, and (5)~visual inspection of the spectra shows no signs of terrestrial weathering, contamination, or other artefacts.

The mineral composition must be documented with high precision to match the reflectance spectrum to a specific material composition. For a modal mineral composition, we used data from X-ray diffraction with an uncertainty of about 3 percentage points\ (pp; a percentage point is the unit for the arithmetic difference of two percentages). When analytic results were available in weight percent, we converted them into volumetric modal abundances. For the conversion, we determined the mineral density of a given composition from the linear interpolation of  the corresponding end-member densities. The densities of end-members and molar masses we used are listed in Table~\ref{tab:constants}. The elemental mineral composition was either gathered from the RELAB database (when available) or from the literature and is typically measured using EMPA with a typical uncertainty of a few pp.

The aim of our model is to detect the composition of olivine-pyroxene-rich mixtures independently of their grain size or spectrally neutral components. This we accomplished during training by including samples with varying grain sizes and forms (e.g. powders or slabs) and normalising modal abundances to olivine-orthopyroxene-clinopyroxene alone, ignoring other spectrally neutral phases. This approach resulted in a high variation of the continuum shape in the training dataset and subsequently resulted in locking our model on diagnostic band shapes and their relative proportions rather than on absolute values (e.g. band depths) or continuum shapes. The general overview of our sample set is presented in Table~\ref{tab:data}, and the full sample list together with analytical results and the physical state is given in Appendix~\ref{sect:supplement} and additionally as an Excel spreadsheet in \b{the GitHub repository}.

The mineralogical distributions of the data are plotted in Fig.~\ref{fig:spectra_min_hist}. Although our data cover the full range of individual modal abundances, most of our samples are pure minerals or mixtures with compositions similar to those of ordinary chondrites. Compositions of olivine and orthopyroxene have a peak in relatively low-iron contents, but all chemistry combinations are contained in our dataset. In the chemistry of clinopyroxene, individual components also cover most of the solid solution of clinopyroxene, but samples peak at the high calcium and low iron part of the compositional space. This bias is caused by the natural availability of samples in the RELAB database.

The peaks in the mineral chemical compositions can cause biases in the predictions of our model when spectra of samples with compositions very different from the composition of the training samples are evaluated. For these samples, the model may predict a composition that is artificially closer to the peaks. On the other hand, our model was made to be used on S-complex asteroids, whose composition is comparable to that of ordinary chondrites, and all the peaks in the compositions correspond to ordinary chondrites. For this reason, we do not expect significantly biased results for S-complex asteroids.

The reflectance spectra of the samples were interpolated to the wavelength grid from 450~nm to 2450~nm with a resolution of 5~nm. This choice enabled us to use the model directly on asteroid spectra without superfluous interpolation, and it keeps the information of high-resolution laboratory measurements. To reduce the noise level of the re-sampled spectra, we performed a convolution filtering on the spectra. The convolution kernel was a Gaussian with $\sigma = 7$~nm. To highlight relative differences among the spectra, the denoised spectra were normalised at 550~nm. \b{The processed spectra} are plotted in Fig.~\ref{fig:spectra}.

\wfigure{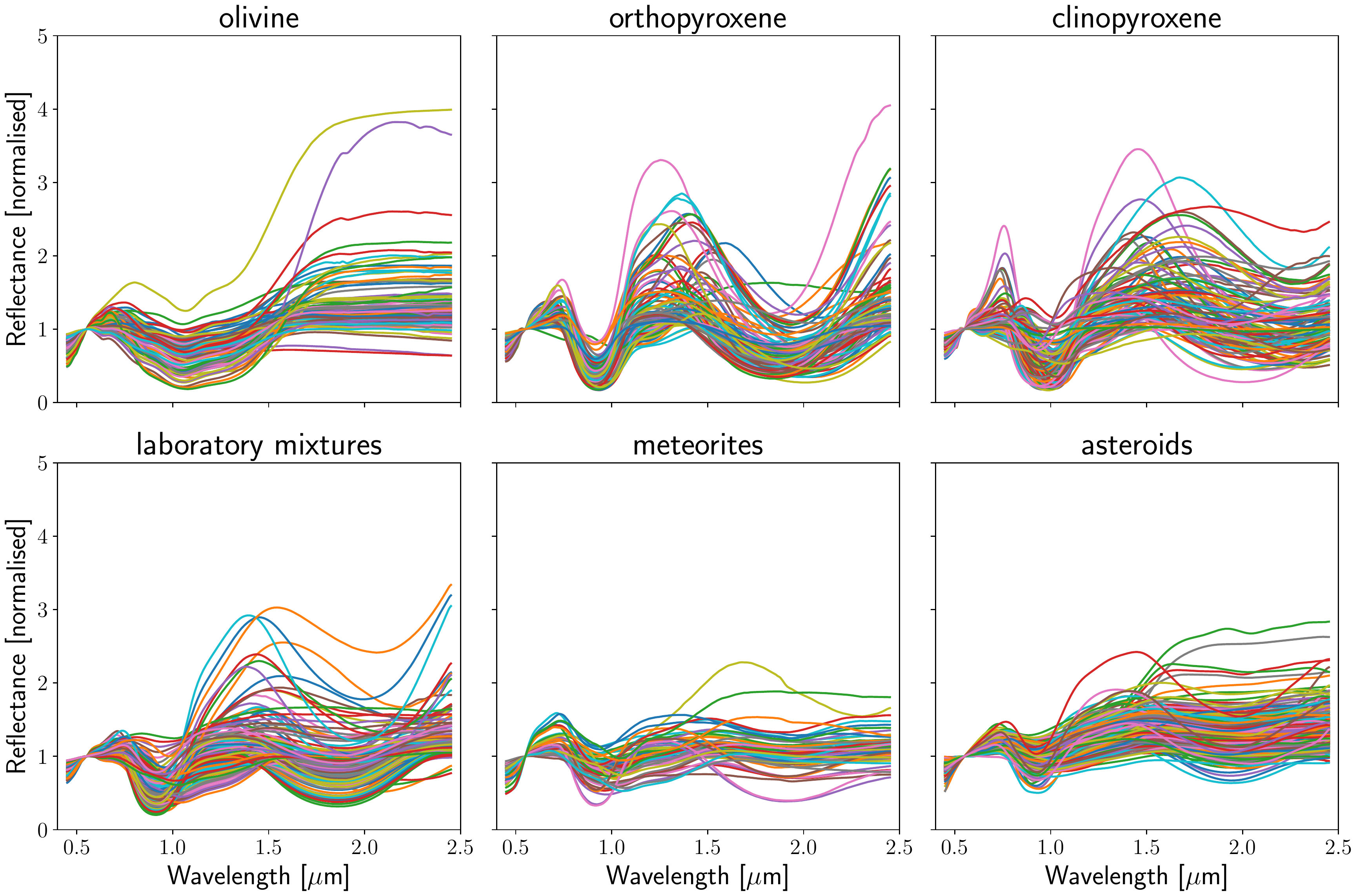}
{Denoised and normalised spectra of silicates, their mixtures, and asteroids utilised in this study.}
{fig:spectra}

%%%%%%%%%%%%%%%%%%%%%%%%%%%%%%%%%%%%%%%%%%%%%%%%%%

\section{Neural network}

We modelled the unknown mapping between the spectra and the composition using a feed-forward convolutional neural network. The network takes the spectrum as input and propagates it through layers of non-linear transformations consisting of individual computational units called neurons, where the parameters of the transformation are learnt from data.

Each layer performs a simple non-linear transformation on its inputs $\vec{h}$ in the form $\vec{o} = \mathrm{f}\!\left( \tens{W} \vec{h} + \vec{b} \right)$, where $\vec{o}$ is the layer output, $\tens{W}$ is the weights, $\vec{b}$ is the bias, and $\mathrm{f}$ is the non-linear activation function. The trainable parameters of the network are the weights and biases of all layers, whereas the choice of the activation functions and the layer architecture are considered design choices. The first hidden layer takes the spectrum as the input, and the last layer provides the predicted abundances as its output. We built the neural network using the Keras library \citep{keras} in Python.

%%%%%%%%%%%%%%%%%%%%%%%%%

\subsection{Model architecture}

The neural network takes as inputs the reflectance spectra at given wavelengths. The wavelengths range from 450~nm to 2450~nm with a step of 5~nm, resulting in input vectors of 401~values.

The model architecture was chosen using a validation and random search, following the procedure described in Sect.~\ref{sect:hp}. The final network consisted of two convolutional hidden layers and an output layer. A convolutional layer performs a convolution on its input. The layer uses the given number of convolution kernels (filters) with the given size. Each kernel is adapted to a different pattern in data. Both convolutional layers use kernels of width five and consist of 24 and 8 kernels, and the output of the second convolutional layer is flatted as input for the fully connected output \b{layer}. Throughout the network, we used the rectified linear unit~(ReLU) $\mathrm{f}\!\left( x \right) = \max \left( 0, x \right)$ as the activation function, except for the output layer, which used the sigmoid activation $\mathrm{f}\!\left( x \right) = \tfrac{1}{1 + \exp \left( -x \right)}$ with further normalisation described next.

The outputs of our model are modal abundances of olivine, orthopyroxene, and clinopyroxene in volume percent together with their chemistry (fayalite~[Fa] and forsterite~[Fo] numbers for olivine; ferrosilite~[Fs] and enstatite [En] numbers for orthopyroxene; and ferrosilite, enstatite, and wollastonite~[Wo] numbers for clinopyroxene). We normalised the output of the sigmoid function for each group of parameters separately (i.e. separately for the modal abundances and separately for the end-members of each mineral). This guarantees that the outputs are directly related to the modal and chemical compositions of the minerals. Additionally, the mineral modal composition has a meaning of a volume fraction normalised to studied minerals, for example the olivine volume fraction in percent is defined as $\tfrac{100 \times \mathrm{OL}}{\mathrm{OL} + \mathrm{OPX} + \mathrm{CPX}}$. Similarly, the mineral chemical composition is normalised to the given end-members, for example Fs of orthopyroxene is defined as $\tfrac{\mathrm{100 \times Fs_{OPX}}}{\mathrm{Fs_{OPX}} + \mathrm{En_{OPX}}}$ , while Fs of clinopyroxene is $\tfrac{\mathrm{100 \times Fs_{CPX}}}{\mathrm{Fs_{CPX}} + \mathrm{En_{CPX}} + \mathrm{Wo_{CPX}}}$.

%%%%%%%%%%%%%%%%%%%%%%%%%

\subsection{Loss function and accuracy metrics}

We assumed $K$ different minerals with $C_k$ chemical end-members. Each sample $i$ has a modal volume fraction $w^{ik}$ of the $k$th mineral, and the mole fraction $z^{ikc}$ of the $c$th end-member. Hence, 
\begin{equation}
    \forall i, k, c \colon w^{ik} \geq 0; z^{ikc} \geq 0; \quad \sum \limits_k w^{ik} = \sum \limits_c z^{ikc} = 1,
\end{equation}
which are also the basic properties of our output activation function. The loss function $\mathcal{L}$ should reflect the variable modal composition of various minerals and must be small when the predicted modal $\vec{w}_{\rm{pred}}$ and chemical $\vec{z}_{\rm{pred}}$ compositions are both close to the actual values $\vec{w}_{\rm{true}}$ and $\vec{z}_{\rm{true}}$. We defined the loss function in the form of
\begin{align}
    \mathcal{L} &= \mathlarger{\sum}\limits_i \norm{\vec{w}^{i:}_{\rm{true}} - \vec{w}^{i:}_{\rm{pred}}}^2 + \alpha \mathlarger{\sum}\limits_k \mathlarger{\sum}\limits_i w^{ik}_{\rm{true}} \norm{\vec{z}^{ik:}_{\rm{true}} - \vec{z}^{ik:}_{\rm{pred}}}^2 + \nonumber\\
    &+ g\!\left(\vec{w}_{\rm{true}}, \vec{z}_{\rm{pred}}\right),
\end{align}
where $\text{the colon}$ stands for all possible indices at the given position. The first term controls the predictions of the modal compositions, and the second term fits the chemical compositions of individual minerals. $\alpha$ is the trade-off parameter between the fit precision of the modal and chemical compositions. The function $g$ adds an additional large penalisation if the predicted solution is in a region with non-existent mineral solid solution (e.g. when predicted $\mathrm{Wo} > 60$). If the abundance of the $k$th mineral $w^{ik}_{\rm{true}}$ is small, its influence on the spectrum is small as well. This is reflected in the second term, where a small modal abundance automatically leads to a small influence of the loss-function value. On the other hand, the predictions of the chemical compositions are potentially very inaccurate in these cases.

Our task is ultimately that of non-linear regression. We therefore used three standard measures of regression accuracy to evaluate the accuracy of the predictions: the \b{root-mean-square error}~($\RMSE$), the \b{coefficient of determination}~($R^2$), and the spectral angle mapper~($\SAM$). The vector accuracy metrics computed from the test data give us information about the model generalisation (i.e. how successful the model is when \b{analysing} new observations). Namely, $\RMSE$ gives us an estimation of the errors in the predicted values for each of the value separately, $R^2$ provides the meaning of which part of the variations in the data is explained by the model, and $\SAM$ quantifies the similarity of the actual and predicted values in terms of the angle between these two vectors. We note that the metrics do not account for various modal abundances, therefore, a mismatch in a low-abundance mineral can significantly contribute to the final mineral mixture metrics.

%%%%%%%%%%%%%%%%%%%%%%%%%

\subsection{Training and regularisation}
\label{sect:training}

Neural networks are trained using iterative gradient descent algorithms, where the gradients of the loss with respect to the weights \b{and biases} are obtained using the backpropagation algorithm, a special case of automatic differentiation. We followed standard practices and applied the Keras library to solve the optimisation problem and the Adam optimiser \citep{adam} with an initial learning rate of 0.0005, which was selected together with the model architecture using the validation data.

For the training and evaluation of the model, we split the available data randomly into three parts: training (60\%, 306 spectra), validation (20\%, 102 spectra), and test (20\%, 102 spectra). This was done in a stratified manner to retain the relative numbers of each type of data (columns in Table~\ref{tab:data}). Only the training data were used to update the weights \b{and biases}, whereas the validation data were only used to select the model architecture (see Sect.~\ref{sect:hp}) and to determine when to terminate the optimisation. For the final evaluation, we chose the model at the point of the iterative training that performed best on the validation data. The test samples were only used at the very end to evaluate the performance of this final model.

We used two types of regularisation to minimise overfitting for the training data: $L_1$ (absolute value) regularisation for the weights and biases to encourage small weights, and dropout regularisation \citep{Srivastava_2014}, which randomly sets selected connections between the neurons as zeros during training. The parameters for these regularising elements were selected together with the model architecture. In addition, we used the commonly employed batch normalisation between the layers. 

In addition to the olivine-orthopyroxene-clinopyroxene model, we also tried different mineral configurations. These models included an olivine-orthopyroxene-only model or a more complex olivine-orthopyroxene-clinopyroxene-plagioclase model. Compared with the presented baseline model, the additional models resulted in reliability matrices that were similar to or worse than that of the baseline model. \b{These models} are described separately in Appendix~\ref{sect:app_models}.

%%%%%%%%%%%%%%%%%%%%%%%%%

\subsection{Hyperparameters}
\label{sect:hp}

The exact neural network architecture (including e.g. kernel widths and the number of convolutional channels), the hyperparameters controlling the optimisation routine, and the regularisation techniques and their parameters were selected based on the performance on the validation data using a random search over possible architectures and parameter choices. For each hyperparameter, we chose a range of values and trained over 2,500 models with unique randomly chosen hyperparameters, selecting the architecture with the lowest validation $\RMSE$ as the final \b{choice}. The final parameters and the ranges of possible values are summarised in Table~\ref{tab:hp}. Because we considered fairly small neural networks that are appropriate for the limited number of training samples, the random search was sufficient to discover a good architecture. More advanced architecture search techniques, such as Bayesian optimisation \citep{Kandasamy_2018}, could be used to reduce the computational cost of the search process.

\begin{table}
    \caption{Hyperparameters and other properties of the neural network.}
    \label{tab:hp}
    \centering
    \begin{tabular}{l l l}
        \hline\hline
        \multicolumn{1}{c}{Property} & 
        \multicolumn{1}{c}{Value} & 
        \multicolumn{1}{c}{Range}\\
        \hline
        Network type & conv. & conv., dense\\
        Number of hidden layers & 2 & 1--3\\
        Nodes in the input l. & 401 & fixed\\
        Nodes/filters in hid. l. & 24 and 8 & 4--32\\
        Nodes in the output l. & 10 & fixed\\
        Conv. kernel size & 5 & 3--5\\
        Hid. l. activation & ReLU & ReLU, ELU\\
        Out. l. activation & sigmoid & softmax, sigmoid\\
        Dropout rate (in.-hid.) & 0.0 & 0.0--0.2\\
        Dropout rate (hid.-hid.) & 0.3 & 0.0--0.3\\
        Dropout rate (hid.-out.) & 0.4 & 0.2--0.4\\
        Training algorithm & Adam & Adam, SGD\\
        Max. number of epochs & 5000 & fixed\\
        Batch size & 8 & 8--1024\\
        Learning rate & 0.0005 & 0.01--0.0001\\
        $L_1$ trade-off parameter & 0.005 & 0--0.001\\
        $\alpha$ trade-off parameter & 0.1 & 0.01--1\\
        \hline
    \end{tabular}
\end{table}

%%%%%%%%%%%%%%%%%%%%%%%%%%%%%%%%%%%%%%%%%%%%%%%%%%

\section{Results}

We trained the neural-network model using a training set of 306 spectra of natural silicate samples and their physical mixtures, including meteorites. \b{The hyperparameters of the model were set based on a model performance on a validation set of 102 spectra.}

%%%%%%%%%%%%%%%%%%%%%%%%%

\subsection{Evaluation on the test data}

\wfigure{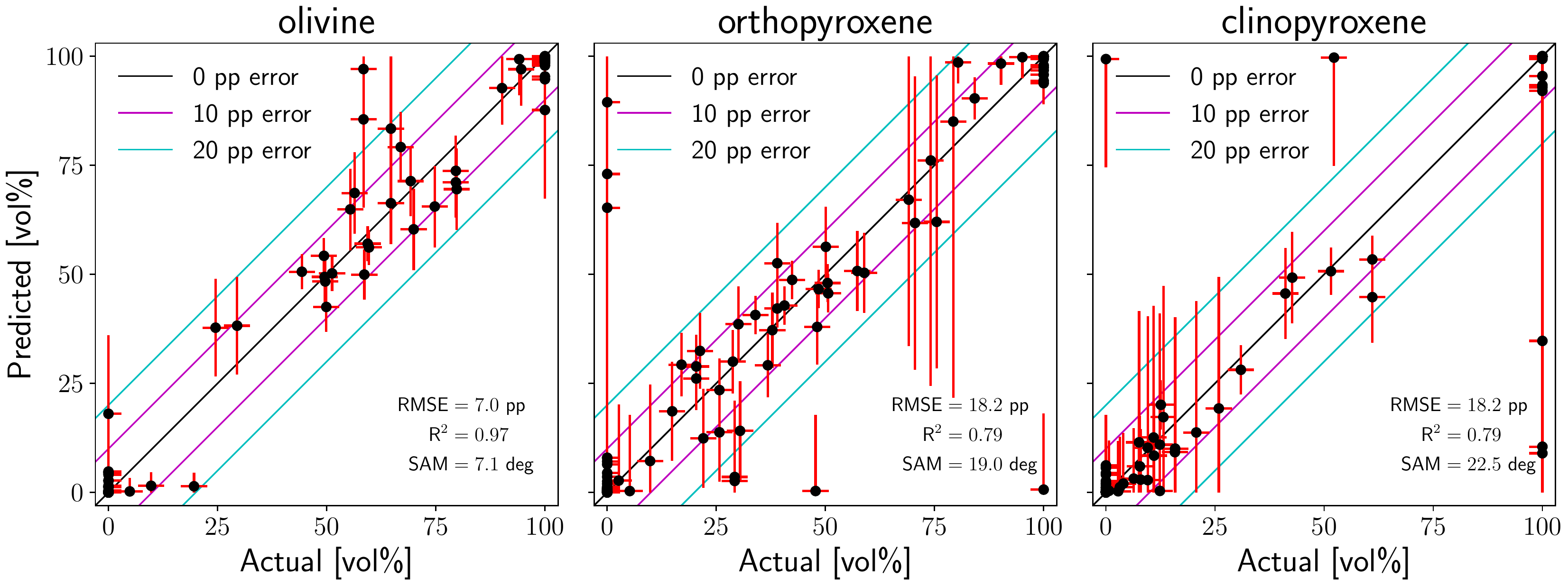}
{Scatter plots of the true and predicted modal compositions of the test data. The diagonal lines delimit the accuracy of the predictions. Error bars in actual values (horizontal) are considered as a conservative analytical error of 3~pp, and in predicted values (vertical), they correspond to the RMSE errors given in Table~\ref{tab:pred_error}.}
{fig:modal}

\ocfigure{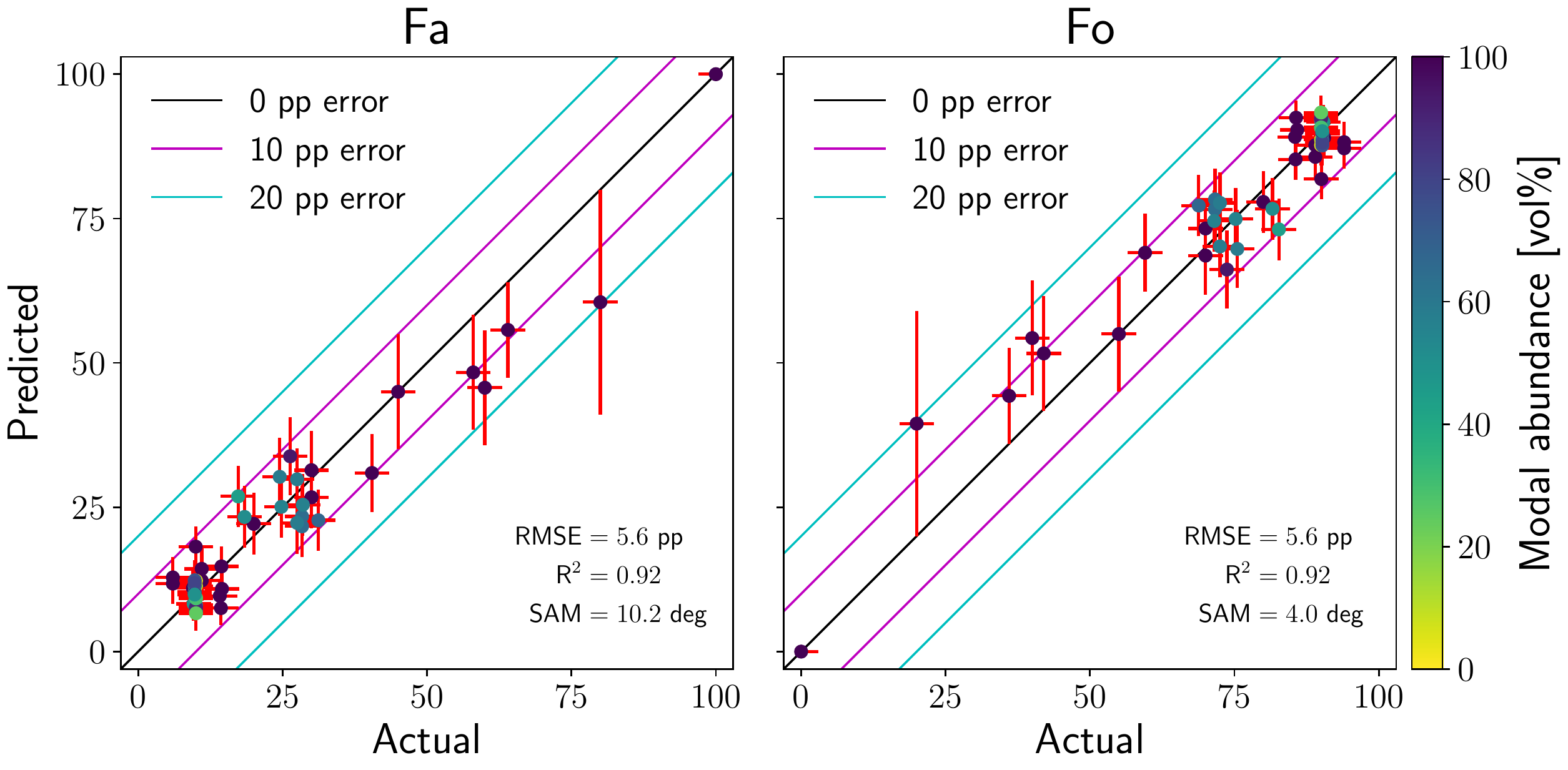}
{Scatter plots of the true and the predicted chemical compositions of olivine in the test data. Left: Iron content. Right: Magnesium content. The colours of the points correspond to the actual modal abundance of olivine in the samples.}
{fig:OL}

\ocfigure{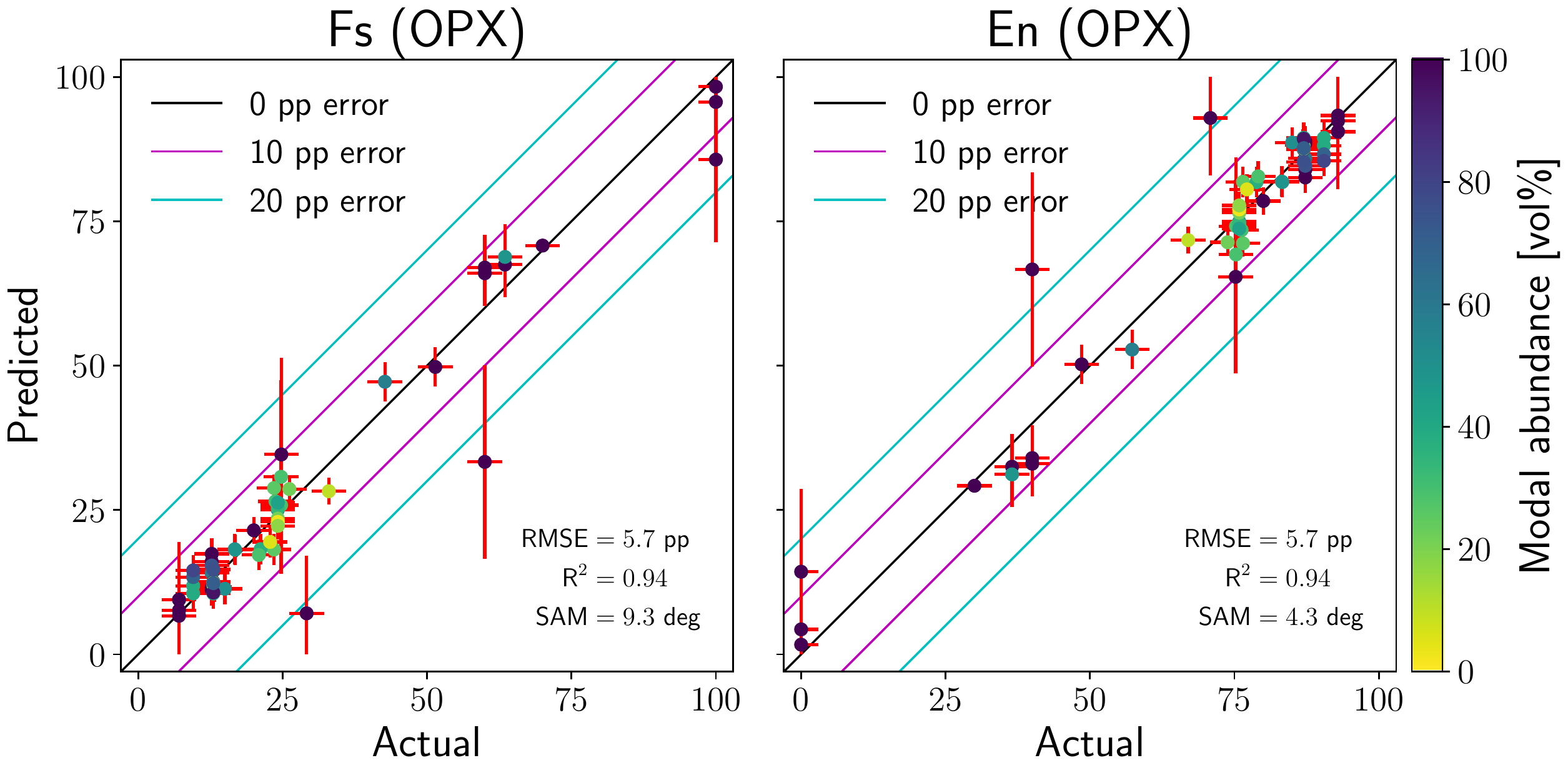}
{Scatter plots of the true and predicted chemical compositions of orthopyroxene in the test data. Left: Iron content. Right: Magnesium content. The colours of the points correspond to the actual modal abundance of orthopyroxene in the samples.}
{fig:OPX}

\wfigure{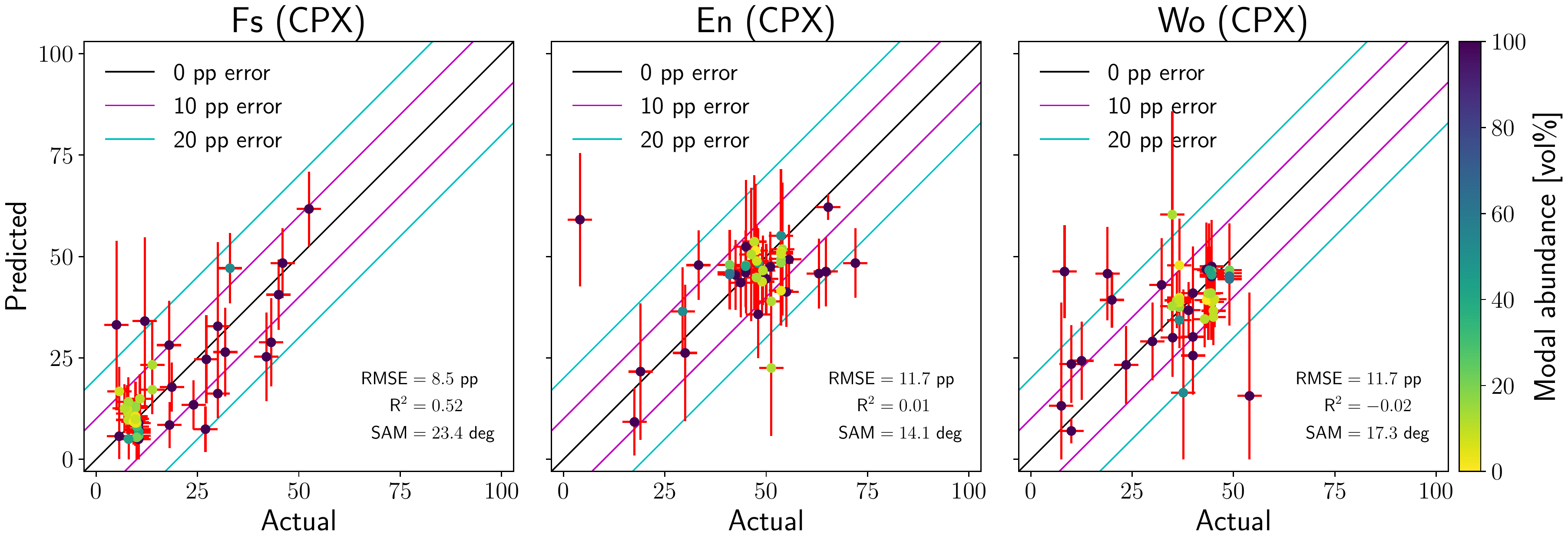}
{Scatter plots of the true and predicted chemical compositions of clinopyroxene in the test data. Left: Iron content. Middle: Magnesium content. Right: Calcium content. The colours of the points correspond to the actual modal abundance of clinopyroxene in the samples.}
{fig:CPX}

\ocfigure{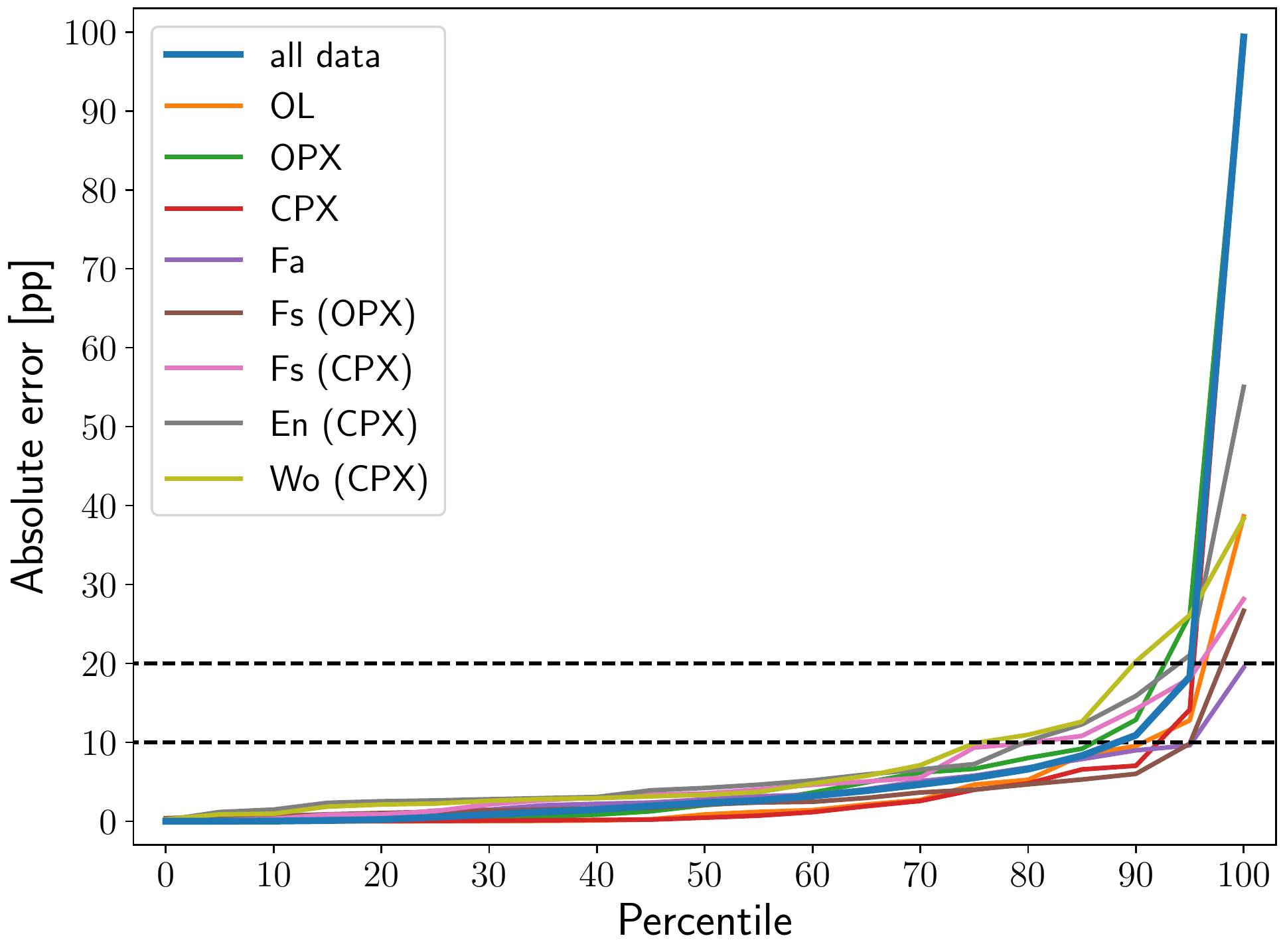}
{Quantiles of the absolute errors between the actual and predicted values in pp. The dashed black lines indicate the 10~pp and 20~pp error.}
{fig:quantile}

\begin{table}
    \caption{Accuracy metric values of the model on the test data.}
    \label{tab:metrics}
    \centering
    \begin{tabular}{l c c c}
        \hline\hline
        \multicolumn{1}{c}{} & 
        \multicolumn{1}{c}{OL} &
        \multicolumn{1}{c}{OPX} &
        \multicolumn{1}{c}{CPX}\\
        \hline
        \multicolumn{4}{c}{$\RMSE$ [pp]}\\
        \hline
        modal & 7.0 & 18.2 (6.5)${^*}$ & 18.2 (3.5)$^{*}$\\
        \hdashline
        Fa & 5.6 & N/A & N/A\\
        Fo & 5.6 & N/A & N/A \\
        \hdashline
        Fs & N/A & \phantom{0}5.7 \phantom{(6.5)${^*}$} & \phantom{0}8.5 \phantom{(6.5)${^*}$}\\
        En & N/A & \phantom{0}5.7 \phantom{(6.5)${^*}$} & 11.7 \phantom{(6.5)${^*}$}\\
        Wo & N/A & N/A & 11.6 \phantom{(6.5)${^*}$}\\
        \hline
        %%%%%
        \multicolumn{4}{c}{$R^2$}\\
        \hline
        modal & 0.97 & 0.79 (0.97)$^{*}$ & \phantom{-}0.79 (0.99)$^{*}$\\
        \hdashline
        Fa & 0.92 & N/A & N/A\\
        Fo & 0.92 & N/A & N/A\\
        \hdashline
        Fs & N/A & 0.94 \phantom{(0.97)$^{*}$} & \phantom{-}0.52 \phantom{(0.97)$^{*}$}\\
        En & N/A & 0.94 \phantom{(0.97)$^{*}$} & \phantom{-}0.01 \phantom{(0.97)$^{*}$}\\
        Wo & N/A & N/A & -0.02 \phantom{(0.97)$^{*}$}\\
        \hline
        %%%%%
        \multicolumn{4}{c}{$\SAM$ [deg]}\\
        \hline
        modal & \phantom{0}7.1 & 19.0 (6.8)$^{*}$ & 22.5 (4.4)$^{*}$\\
        \hdashline
        Fa & 10.2 & N/A & N/A\\
        Fo & \phantom{0}4.0 & N/A & N/A\\
        \hdashline
        Fs & N/A & \phantom{0}9.3 \phantom{(6.5)${^*}$} & 23.4 \phantom{(6.5)${^*}$}\\
        En & N/A & \phantom{0}4.3 \phantom{(6.5)${^*}$} & 14.1 \phantom{(6.5)${^*}$}\\
        Wo & N/A & N/A & 17.3 \phantom{(6.5)${^*}$}\\
        \hline
    \end{tabular}
            
    \vspace{1ex}
    {\raggedright $^{*}$Without the orthopyroxene-clinopyroxene outliers. \par}
\end{table}

\begin{table*}
    \caption{Error estimates computed from predicted values at the given interval. The values are the RMSE in percentage points. In brackets we quote the number of points from which the error was estimated. Typical values for ordinary chondrites (S- and Q-type asteroids) are highlighted.}
    \label{tab:pred_error}
    \centering
    \begin{tabular}{l l l l l l l l l l l}
        \hline\hline
        \multicolumn{1}{c}{Value} & 
        \multicolumn{1}{c}{0--10} &
        \multicolumn{1}{c}{10--20} &
        \multicolumn{1}{c}{20--30} &
        \multicolumn{1}{c}{30--40} &
        \multicolumn{1}{c}{40--50} &
        \multicolumn{1}{c}{50--60} &
        \multicolumn{1}{c}{60--70} &
        \multicolumn{1}{c}{70--80} &
        \multicolumn{1}{c}{80--90} &
        \multicolumn{1}{c}{90--100}\\
        \hline
        OL [vol\%] & \phantom{0}3.1 (54) & 18.0 (1) & N/A (0) & 11.2 (2) & \cellcolor[gray]{0.8}\phantom{0}5.7 (4) & \cellcolor[gray]{0.8}\phantom{0}4.1 (5) & \phantom{0}9.3 (6) & \phantom{0}8.1 (4) & 20.3 (3) & \phantom{0}8.3 (23)\\
        OPX [vol\%] & 17.5 (45) & 11.4 (4) & \cellcolor[gray]{0.8}\phantom{0}7.3 (6) & \cellcolor[gray]{0.8}\phantom{0}8.7 (4) & \phantom{0}4.4 (7) & \phantom{0}9.2 (4) & 33.6 (4) & 51.6 (2) & 63.4 (2) & \phantom{0}4.9 (24)\\
        CPX [vol\%] & \cellcolor[gray]{0.8}11.6 (65) & \cellcolor[gray]{0.8}30.1 (9) & \phantom{0}5.7 (2) & 65.3 (1) & 10.5 (3) & \phantom{0}5.4 (2) & N/A (0) & N/A (0) & N/A (0) & 24.8 (20)\\
        \hdashline
        Fa & \phantom{0}3.0 (13) & \cellcolor[gray]{0.8}\phantom{0}3.5 (16) & \cellcolor[gray]{0.8}\phantom{0}5.3 (11) & \phantom{0}6.8 (4) & 10.0 (3) & 8.3 (1) & 19.5 (1) & N/A (0) & N/A (0) & \phantom{0}0.0 (1)\\
        Fo & \phantom{0}0.0 (1) & N/A (0) & N/A (0) & 19.5 (1) & 8.3 (1) & 10.0 (3) & \phantom{0}6.8 (4) & \cellcolor[gray]{0.8}\phantom{0}5.3 (11) & \cellcolor[gray]{0.8}\phantom{0}3.5 (16) & \phantom{0}3.0 (13)\\
        \hdashline
        Fs (OPX) & 10.0 (5) & \cellcolor[gray]{0.8}\phantom{0}2.7 (27) & \cellcolor[gray]{0.8}\phantom{0}2.3 (16) & 16.8 (3) & \phantom{0}3.4 (2) & N/A (0) & \phantom{0}5.7 (4) & \phantom{0}0.8 (1) & 14.3 (1) & \phantom{0}3.3 (2)\\
        En (OPX) & \phantom{0}3.3 (2) & 14.3 (1) & \phantom{0}0.8 (1) & \phantom{0}5.7 (4) & N/A (0) & \phantom{0}3.4 (2) & 16.8 (3) & \cellcolor[gray]{0.8}\phantom{0}2.3 (16) & \cellcolor[gray]{0.8}\phantom{0}2.7 (27) & 10.0 (5)\\
        \hdashline
        Fs (CPX) & \cellcolor[gray]{0.8}\phantom{0}5.6 (19) & \cellcolor[gray]{0.8}\phantom{0}6.0 (16) & 10.9 (6) & 20.7 (3) & \phantom{0}8.6 (3) & N/A (0) & \phantom{0}9.2 (1) & N/A (0) & N/A (0) & N/A (0)\\
        En (CPX) & \phantom{0}8.3 (1) & N/A (0) & 16.8 (3) & 10.8 (3) & \cellcolor[gray]{0.8}\phantom{0}8.6 (28) & 16.4 (12) & \phantom{0}3.1 (1) & N/A (0) & N/A (0) & N/A (0)\\
        Wo (CPX) & \phantom{0}3.0 (1) & 25.5 (3) & \phantom{0}9.6 (6) & \cellcolor[gray]{0.8}\phantom{0}6.9 (18) & \cellcolor[gray]{0.8}11.5 (19) & N/A (0) & 25.4 (1) & N/A (0) & N/A (0) & N/A (0)\\
        \hline
    \end{tabular}
\end{table*}

\begin{table}
    \caption{Part of predictions that is within the given error interval.}
    \label{tab:quantile}
    \centering
    \begin{tabular}{l c c c c}
        \hline\hline
        & \multicolumn{4}{c}{absolute error}\\
        \hline
        \multicolumn{1}{c}{Value} & 
        \multicolumn{1}{c}{5~pp} &
        \multicolumn{1}{c}{10~pp} &
        \multicolumn{1}{c}{15~pp} &
        \multicolumn{1}{c}{20~pp}\\
        \hline
        all data & \phantom{0}72\% & \phantom{0}89\% & \phantom{0}93\% & \phantom{0}95\% \\
        \hdashline
        OL [vol\%] & \phantom{0}78\% & \phantom{0}91\% & \phantom{0}95\% & \phantom{0}98\% \\
        OPX [vol\%] & \phantom{0}65\% & \phantom{0}86\% & \phantom{0}92\% & \phantom{0}93\% \\
        CPX [vol\%] & \phantom{0}80\% & \phantom{0}94\% & \phantom{0}95\% & \phantom{0}95\% \\
        \hdashline
        Fa & \phantom{0}69\% & \phantom{0}95\% & \phantom{0}97\% & 100\% \\
        \hdashline
        Fs (OPX) & \phantom{0}83\% & \phantom{0}95\% & \phantom{0}96\% & \phantom{0}98\% \\
        \hdashline
        Fs (CPX) & \phantom{0}63\% & \phantom{0}80\% & \phantom{0}91\% & \phantom{0}95\% \\
        En (CPX) & \phantom{0}57\% & \phantom{0}78\% & \phantom{0}89\% & \phantom{0}93\% \\
        Wo (CPX) & \phantom{0}61\% & \phantom{0}76\% & \phantom{0}87\% & \phantom{0}89\% \\
        \hline
    \end{tabular}
        
    \vspace{1ex}
    {\raggedright Fo and En~(OPX) have the same absolute errors as Fa and Fs~(OPX), respectively. \par}
\end{table}

The reliability tests were performed on the \b{test set of 102 spectra} and are plotted in Figs.~\ref{fig:modal}--\ref{fig:CPX}. In all the plots, the horizontal axis corresponds to the true value and the vertical axis to the predicted value. The diagonal lines delimit the prediction accuracy intervals. Error bars in actual values (horizontal) are considered as the conservative analytical error of \num{3}~pp, and in the predicted values (vertical), they correspond to the RMSE errors given in Table~\ref{tab:pred_error}. Additionally, in Figs.~\ref{fig:OL}--\ref{fig:CPX}, the colour of the points indicates the modal abundance of the mineral. The colour map is the same in all figures. The $\RMSE$, $R^2$, and $\SAM$ values for all predicted quantities can be found in the corresponding subplots and in Table~\ref{tab:metrics}.

The predicted modal abundances (see Fig.~\ref{fig:modal}) are mostly within the 10~pp error interval. Six distinct outliers significantly worsen the accuracy metrics. Five of them are due to the mismatch between orthopyroxene and clinopyroxene (see the 0, 50, and 100 actual abundances of OPX and CPX in Fig.~\ref{fig:modal}), and one olivine-orthopyroxene-clinopyroxene mixture was predicted to be almost pure olivine because the 2~\textmu{}m pyroxene band is very shallow. When we computed the $\RMSE$ without the orthopyroxene-clinopyroxene outliers, the estimated errors of the silicate modal abundances were \num{6.5}\b{~pp} and \num{3.5}\b{~pp} for orthopyroxene and clinopyroxene, respectively.

The error estimates (RMSE) for different predicted intervals are listed in Table~\ref{tab:pred_error}. These errors show us that within the typical composition of ordinary chondrites, our model predictions are mostly better than 10~pp. High errors in some bins are usually due to a single outlier and a small number of points in the bins.

The $\RMSE$ metric is very sensitive to outliers and overestimates the real prediction error of orthopyroxene and clinopyroxene abundances. To better present the errors between the true and the predicted values, we therefore computed the quantiles of the absolute error. They are plotted in Fig.~\ref{fig:quantile} and summarised in Table~\ref{tab:quantile}, and they are shown in the density plots in Fig.~\ref{fig:density}. \b{Table~\ref{tab:quantile}} shows that more than \num{75}\% of the predictions in each composition separately are within the 10~pp error interval, and more than \num{85}\% of the predictions are within the 15~pp error interval. The predicted chemical compositions are very accurate for olivine (Fig.~\ref{fig:OL}) and orthopyroxene (Fig.~\ref{fig:OPX}), where \num{95}\% of the predictions are within the 10~pp error interval for both. The predictions of the clinopyroxene composition (Fig.~\ref{fig:CPX}) are less accurate; only \num{80}\% of the predicted Fs, \num{78}\% of the predicted En, and \num{76}\% of the predicted Wo are within the 10~pp error interval. The precision of the model is limited mostly by the number of samples with intermediate abundances of clinopyroxene, by the lack of olivine-clinopyroxene mixtures, and also by the number of samples in general. When we assume a normal distribution for the absolute errors, the 1-$\sigma$ errors are \num{2.5, 5.7, and 2.1}~\b{pp} for the modal compositions of olivine, orthopyroxene, and clinopyroxene, respectively, and \num{4.9, 3.3, 5.5, 6.4, and 5.9}~\b{pp} for fayalite, OPX ferrosilite, CPX ferrosilite, CPX enstatite, and CPX wollastonite, respectively.

%%%%%%%%%%%%%%%%%%%%%%%%%

\subsection{Classification of asteroid spectra}

The asteroid dataset consists of 387 reflectance spectra of S$^*$-complex asteroids \citep{DeMeo_2009, Binzel_2019} with assigned Bus--DeMeo taxonomy classes \citep{DeMeo_2009}. The data include spectra of 147~S-, three~Sa-, 41~Sq-, 17~Sqw-, 31~Sr-, eight~Srw-, three~Sv-, two~Svw-, 57~Sw-, 42~Q-, one~Qw-, 26~V-, two~Vw-, and seven~A-type asteroids. These spectra are in the range from 450~nm to 2450~nm and have a resolution of 10~nm. As in the case of mineral spectra, we resampled the spectra to a resolution of 5~nm\b{, denoised them with the Gaussian kernel, and finally normalised them at 550~nm} (see the bottom right panel in Fig.~\ref{fig:spectra}).

The S-type asteroids are the most common asteroids in the near-Earth region and are very common in the inner and middle Main Belt \citep{Binzel_2019}. Their reflectance spectra are dominated by the 1 and 2\textmu{}m bands of olivine and pyroxene. The Q-type asteroids are very similar in composition to the S-type asteroids. The difference is in the slope of the spectrum. The spectra of Q-type asteroids have a neutral slope, while the spectra of the S-type asteroids have a red slope. Reflectance spectra of the V-type asteroids show a narrow 1\textmu{}m band and a wide 2\textmu{}m band. These suggest a high abundance of pyroxene. The A-type asteroids are dominated by the 1\textmu{}m band of olivine and a high red slope of the spectrum. We refer to Fig.~15 of \citet{DeMeo_2009} for the shape of reflectance spectra of individual taxonomy classes. 

The predictions of the neural-network model for the asteroid spectra show that the S-type asteroids are depleted in olivine compared to Q-type asteroids, but are of the same mineral chemical composition, and both are comparable with the chemical composition of ordinary chondrites. The V-type asteroids are mainly made of orthopyroxene (Fs about \num{31}) and clinopyroxene, and the A-type asteroids are made of almost pure olivine (Fa about \num{26}). Standard deviations indicate that the mineral chemical compositions are mostly homogeneous within the asteroid types, while modal abundances are potentially more various. The statistical information about the predictions is listed in Tables~\ref{tab:ast_types} and \ref{tab:ast_types_S}.

\begin{table}
    \small
    \caption{Mean values and standard deviations of the model predictions for the asteroid spectra.}
    \label{tab:ast_types}
    \centering
    \begin{tabular}{l c c c c}
        \hline\hline
        \multicolumn{1}{c}{} & 
        \multicolumn{1}{c}{S} &
        \multicolumn{1}{c}{Q} &
        \multicolumn{1}{c}{V} &
        \multicolumn{1}{c}{A}\\
        \hline
        OL [vol\%] & 34 $\pm$ 21 & 65 $\pm$ 19 & \phantom{0}8 $\pm$ 15 & 99.1 $\pm$ \phantom{0}1.8 \\
        OPX [vol\%] & 26 $\pm$ 14 & 19 $\pm$ 14 & 70 $\pm$ 22 & \phantom{0}0.2 $\pm$ \phantom{0}0.2 \\
        CPX [vol\%] & 40 $\pm$ 18 & 16 $\pm$ 12 & 22 $\pm$ 20 & \phantom{0}0.7 $\pm$ \phantom{0}1.6 \\
        \hdashline
        Fa & 21.3 $\pm$ \phantom{0}5.3 & 26.5 $\pm$ \phantom{0}5.5 & 14.1 $\pm$ \phantom{0}7.0 & 26 $\pm$ 11 \\
        Fo & 78.7 $\pm$ \phantom{0}5.3 & 73.5 $\pm$ \phantom{0}5.5 & 85.9 $\pm$ \phantom{0}7.0 & 74 $\pm$ 11 \\
        \hdashline
        Fs (OPX) & 17.9 $\pm$ \phantom{0}5.0 & 23.8 $\pm$ \phantom{0}5.2 & 30.7 $\pm$ \phantom{0}7.2 & 15.5 $\pm$ \phantom{0}1.5 \\
        En (OPX) & 82.1 $\pm$ \phantom{0}5.0 & 76.2 $\pm$ \phantom{0}5.2 & 69.3 $\pm$ \phantom{0}7.2 & 84.5 $\pm$ \phantom{0}1.5 \\
        \hdashline
        Fs (CPX) & 11.3 $\pm$ \phantom{0}4.3 & 16.1 $\pm$ \phantom{0}9.4 & 11.5 $\pm$ \phantom{0}5.3 & \phantom{0}9.4 $\pm$ \phantom{0}1.5 \\
        En (CPX) & 46.3 $\pm$ \phantom{0}2.6 & 44.9 $\pm$ \phantom{0}7.8 & 49.8 $\pm$ \phantom{0}4.9 & 45.4 $\pm$ \phantom{0}0.7 \\
        Wo (CPX) & 42.4 $\pm$ \phantom{0}3.5 & 39.0 $\pm$ \phantom{0}5.5 & 38.6 $\pm$ \phantom{0}9.2 & 45.2 $\pm$ \phantom{0}0.8 \\

        \hline
    \end{tabular}
\end{table}

\begin{table*}
    \small
    \caption{Mean values and standard deviations of the model predictions for the S-complex asteroid spectra.}
    \label{tab:ast_types_S}
    \centering
    \begin{tabular}{l c c c c c c c c c}
        \hline\hline
        \multicolumn{1}{c}{} & 
        \multicolumn{1}{c}{S} &
        \multicolumn{1}{c}{Sw} &
        \multicolumn{1}{c}{Sq} &
        \multicolumn{1}{c}{Sqw} &
        \multicolumn{1}{c}{Sr} &
        \multicolumn{1}{c}{Srw} &
        \multicolumn{1}{c}{Sv} &
        \multicolumn{1}{c}{Svw} &
        \multicolumn{1}{c}{Sa}\\
        \hline
        Number & 147 & 57 & 41 & 17 & 31 & 8 & 3 & 2 & 3 \\
        \hdashline
        OL [vol\%] & 30 $\pm$ 20 & 36 $\pm$ 17 & 48 $\pm$ 20 & 50 $\pm$ 24 & 24 $\pm$ 17 & 20 $\pm$ 18 & 40.5 $\pm$ \phantom{0}5.5 & 15 $\pm$ 13 & 84 $\pm$ 25 \\
        OPX [vol\%] & 28 $\pm$ 13 & 19 $\pm$ 11 & 19 $\pm$ 10 & 11.7 $\pm$ \phantom{0}7.1 & 38.2 $\pm$ \phantom{0}8.9 & 34 $\pm$ 14 & 34.5 $\pm$ \phantom{0}5.5 & 36.0 $\pm$ \phantom{0}7.6 & \phantom{0}1.2 $\pm$ \phantom{0}0.4 \\
        CPX [vol\%] & 42 $\pm$ 18 & 45 $\pm$ 16 & 33 $\pm$ 18 & 38 $\pm$ 23 & 38 $\pm$ 17 & 46 $\pm$ 12 & 25.0 $\pm$ \phantom{0}4.8 & 49.0 $\pm$ \phantom{0}5.6 & 15 $\pm$ 25 \\
        \hdashline
        Fa & 20.9 $\pm$ \phantom{0}4.6 & 21.6 $\pm$ \phantom{0}4.3 & 23.3 $\pm$ \phantom{0}4.6 & 24.1 $\pm$ \phantom{0}3.9 & 18.5 $\pm$ \phantom{0}5.8 & 23 $\pm$ 12 & 13.7 $\pm$ \phantom{0}4.9 & 16.9 $\pm$ \phantom{0}4.1 & 30.4 $\pm$ \phantom{0}8.5 \\
        Fo & 79.1 $\pm$ \phantom{0}4.6 & 78.4 $\pm$ \phantom{0}4.3 & 76.7 $\pm$ \phantom{0}4.6 & 75.9 $\pm$ \phantom{0}3.9 & 81.5 $\pm$ \phantom{0}5.8 & 77 $\pm$ 12 & 86.3 $\pm$ \phantom{0}4.9 & 83.1 $\pm$ \phantom{0}4.1 & 69.6 $\pm$ \phantom{0}8.5 \\
        \hdashline
        Fs (OPX) & 18.3 $\pm$ \phantom{0}4.9 & 14.7 $\pm$ \phantom{0}4.3 & 21.0 $\pm$ \phantom{0}4.1 & 16.5 $\pm$ \phantom{0}3.6 & 19.2 $\pm$ \phantom{0}5.4 & 16.3 $\pm$ \phantom{0}3.3 & 17.3 $\pm$ \phantom{0}2.1 & \phantom{0}7.7 $\pm$ \phantom{0}2.2 & 18.4 $\pm$ \phantom{0}1.8 \\
        En (OPX) & 81.7 $\pm$ \phantom{0}4.9 & 85.3 $\pm$ \phantom{0}4.3 & 79.0 $\pm$ \phantom{0}4.1 & 83.5 $\pm$ \phantom{0}3.6 & 80.8 $\pm$ \phantom{0}5.4 & 83.7 $\pm$ \phantom{0}3.3 & 82.7 $\pm$ \phantom{0}2.1 & 92.3 $\pm$ \phantom{0}2.2 & 81.6 $\pm$ \phantom{0}1.8 \\
        \hdashline
        Fs (CPX) & 11.3 $\pm$ \phantom{0}4.0 & 10.3 $\pm$ \phantom{0}3.1 & 12.4 $\pm$ \phantom{0}4.8 & 11.5 $\pm$ \phantom{0}2.9 & 11.4 $\pm$ \phantom{0}4.0 & 10.9 $\pm$ \phantom{0}5.2 & \phantom{0}5.6 $\pm$ \phantom{0}1.3 & \phantom{0}6.6 $\pm$ \phantom{0}2.9 & 11.1 $\pm$ \phantom{0}2.4 \\
        En (CPX) & 46.3 $\pm$ \phantom{0}2.6 & 45.6 $\pm$ \phantom{0}1.3 & 47.1 $\pm$ \phantom{0}3.0 & 45.7 $\pm$ \phantom{0}1.4 & 47.1 $\pm$ \phantom{0}2.2 & 45.3 $\pm$ \phantom{0}2.4 & 47.6 $\pm$ \phantom{0}0.6 & 46.9 $\pm$ \phantom{0}1.2 & 45.4 $\pm$ \phantom{0}1.1 \\
        Wo (CPX) & 42.4 $\pm$ \phantom{0}3.2 & 44.1 $\pm$ \phantom{0}2.0 & 40.4 $\pm$ \phantom{0}4.1 & 42.8 $\pm$ \phantom{0}2.6 & 41.5 $\pm$ \phantom{0}4.4 & 43.8 $\pm$ \phantom{0}2.8 & 46.8 $\pm$ \phantom{0}0.7 & 46.5 $\pm$ \phantom{0}1.7 & 43.6 $\pm$ \phantom{0}1.4 \\
        \hline
    \end{tabular}
\end{table*}

%%%%%%%%%%%%%%%%%%%%%%%%%%%%%%%%%%%%%%%%%%%%%%%%%%

\section{Discussion}

\subsection{Biases}

\tcfigure{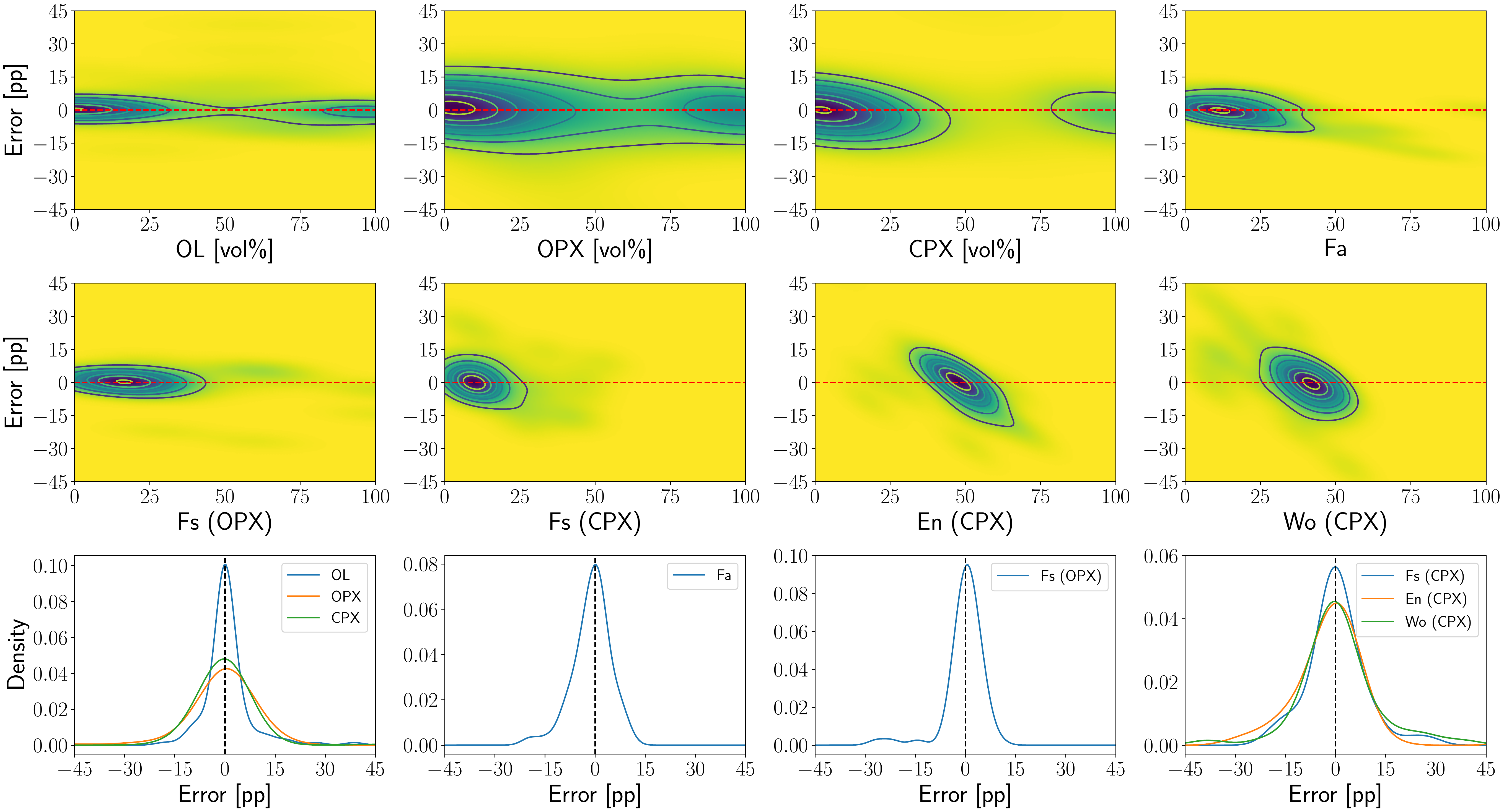}
{Density plots based on error predictions (predicted minus actual) and specific compositions. The dashed lines delimit zero error. Top and middle rows: Concentrations of the test data as a function of composition and error. Bottom row: Error distributions. Biases are present when the error is dependent on the composition, the \b{mean value of error} is significantly non-zero, or when the error distributions are asymmetrical.}
{fig:density}

The density plots in Fig.~\ref{fig:density} visualise the approximate distributions of our test data and \b{dependence between} the error and composition. In this way, the density plots show biases in the predictions for the test data. The top and middle rows of the density plot show the approximate distributions of the points in error-composition space. A possible trend in the dependence of the error on composition\b{, }non-zero mean value of error, or a significantly asymmetric error distribution \b{indicate} biased predictions. The bottom row in Fig.~\ref{fig:density} show that the error distributions peaked at 0\b{~pp} error and do not show significant asymmetries. No error-composition dependences are observed in the modal abundances. For olivine, only about 10\% of the data have $\mathrm{Fa} > 50,$ which might cause the slight depletion in the prediction of iron in the high-iron $\left( 50 < \mathrm{Fa} < 80 \right)$ samples. This observation was made on four points and thus cannot be proved. The prediction of the chemical composition of orthopyroxene does not contain any biases. For clinopyroxene, we observe trends in the enstatite and wollastonite numbers. They are probably caused by the distribution of our training data and by the coupling between enstatite and wollastonite via the normalisation of the chemical composition. Even other metrics indicate that the predictions of the clinopyroxene chemical composition are the least accurate. For these reasons, the predictions of the clinopyroxene composition of our model are reliable only in a limited region of approximately $\left( 40 < \mathrm{En, Wo} < 60 \right)$. This region is valid for ordinary chondrites or achondrites, for instance, and thus also for S-complex asteroids.

%%%%%%%%%%%%%%%%%%%%%%%%%

\subsection{Asteroid spectra and the effects of space weathering}

Our model output indicates that S-type asteroids are relatively depleted in olivine abundance compared to Q-type asteroids, while the mineral chemistry remains comparable. At first glance, this is a surprising finding. However, space weathering provides an explanation for this observation. \b{Based on their spectra,} it is generally thought that Q-type asteroids represent fresh material \b{that is} similar to \b{the material} of ordinary chondrites \citep{Binzel_2010}. The S-type asteroid spectra deviate from the Q type in the way in which they show more signs of space weathering, that is, the silicate absorption bands are more attenuated and the overall spectral slope is redder; see Fig.~1 in \citet{Binzel_2010}. It is also known that olivine undergoes space weathering changes on a shorter timescale than pyroxene \citep{Sasaki_2002, Marchi_2005, Quadery_2015, Chrbolkova_2021}. Moreover, our model is trained on non-weathered data, hence the model interprets the weaker olivine bands as a lower olivine abundance. Therefore, we explain our model results so that our model does not actually indicate the physical absence of olivine, but rather that accelerated space weathering masks its diagnostic spectral absorptions relative to pyroxene. This is further supported by the fact that despite the relatively lower observed olivine modal abundance, its chemistry is not significantly altered. Fig.~\ref{fig:PCA} supports this as well: The apparent depletion of the olivine abundance correlates with the space-weathering trend (red arrow) defined by \citet{Binzel_2019} in the PCA-based asteroid spectrum classification.

To further test the space-weathering effects on our neural-network output, we used the Chelyabinsk dataset (olivine-pyroxene mixtures) from \citet{Kohout_2020}, which includes laboratory space-weathered, impact-melted, and shock-darkened mixture series, and also a pure olivine and pure pyroxene dataset from \citet{Chrbolkova_2021}, which includes a space-weathered series.

The OL, OPX, and CPX-normalised compositions of the Chelyabinsk meteorite are based on \citet{Reddy_2014}. The XRD is consistent with our model output for a fresh Chelyabinsk spectrum (\b{see} Table~\ref{tab:chelyabinsk}). The model predictions for the \citet{Kohout_2020} dataset are listed in the bottom part of Table~\ref{tab:chelyabinsk}. We note that the predictions of the CPX chemical composition can be ignored for the fresh sample because the predicted CPX abundance is only \num{3.9}~\b{vol}\%. With increasing laboratory-induced space weathering, the relative modal abundance of olivine in Chelyabinsk as indicated by our model systematically decreases from \num{66.6 to 2.8}~vol\%, similarly to the trend observed in S-type asteroids, while the olivine and orthopyroxene chemistries show weaker systematic decreases that are within its 1-$\RMSE$ error. In contrast, for the Chelyabinsk mixtures with spectrally neutral impact-melted or shock-darkened materials (IM and SD series), the modal abundances as well as the silicate chemistry remain similar (i.e. \num{62.8--68.2~vol\% and 58.0--67.9~vol\%} for the olivine fraction in IM and SD, respectively) and mostly within 1~$\RMSE$ to the fresh Chelyabinsk XRD-derived values (see Appendix~\ref{sect:chelyabinsk}). This demonstrates that especially the predictions of the modal abundances of our model are indeed sensitive to the relative changes in individual mineral absorption bands and are robust against global spectral attenuation (e.g. caused by shock or melting). This is consistent with the finding of \citet{Kohout_2020}, who showed that the relative depth and area of the 2\textmu{}m and 1\textmu{}m bands do not depend on the amount of shock-darkened material, but that space weathering emphasises the 2\textmu{}m band (see their Fig.~4).

\citet{Chrbolkova_2021} irradiated olivine and pyroxene samples with H$^+$, He$^+$, Ar$^+$, and with a femtosecond pulse laser; see their Fig.~2 for the spectra. Especially ion irradiation is widely used to simulate space weathering \citep[e.g. ][]{Loeffler_2009, Lantz_2017}. We made predictions for these spectra and found that regardless of method or intensity, the predicted modal abundances were very close to those of a pure sample. The minimum olivine fraction in the pure olivine samples was \num{98.5}~vol\% (laser irradiated, 375.0~J\,cm$^{-2}$). Comparable results were obtained for pure pyroxene, where \num{78}\% of the samples were predicted with a pyroxene fraction of 99~vol\% and more. The predictions for pyroxene also showed a possible mismatch between pyroxene and olivine in cases of extreme space weathering. The samples with very high laser irradiation (450 and 1800~J\,cm$^{-2}$) had broadened the 1\textmu{}m band, which was interpreted as \num{11 and 34}~vol\% of olivine. For all experiments, predictions for the iron contents in the samples were weakly dependent on the degree of weathering. Namely, there was a positive trend for olivine (which is the opposite trend from what we observed for Chelyabinsk) and a negative trend for pyroxene. Moreover, the prediction of the pyroxene composition was offset to significantly lower values. These effects are demonstrated in Tables~\ref{tab:kachr_ol} and \ref{tab:kachr_px}, which list the predictions for the laser-irradiated spectra. The results from the other experiments are summarised in Appendix~\ref{sect:kachr}.

\begin{table}
    \caption{Modal abundances and mineral chemical composition of the Chelyabinsk meteorite.}
    \label{tab:chelyabinsk}
    \centering
    \begin{tabular}{l c c c c c c c}
        \hline\hline
        \multicolumn{1}{c}{} &
        \multicolumn{3}{c}{modal [vol\%]} & 
        \multicolumn{1}{c}{} &
        \multicolumn{1}{c}{OPX} &
        \multicolumn{2}{c}{CPX}\\
        & OL & OPX & CPX & Fa & Fs & Fs & Wo\\
        \hline
        XRD & \textcolor{red}{66.2} & 33.8 & \phantom{0}0.0 & 28.6 & 23.9 & N/A & N/A\\
        \hdashline
        fresh & \textcolor{red}{66.6} & 29.5 & \phantom{0}3.9 & 26.5 & 20.4 & 10.1 & 38.2 \\
        SW 400 & \textcolor{red}{54.0} & 32.0 & 14.0 & 23.4 & 14.1 & 13.5 & 40.2 \\
        SW 500 & \textcolor{red}{11.1} & 42.8 & 46.1 & 20.6 & 11.0 & 16.0 & 39.9 \\
        SW 600 & \textcolor{red}{\phantom{0}2.8} & 50.0 & 47.3 & 16.3 & 10.6 & 14.7 & 42.8 \\
        % SW 700 & \textcolor{red}{\phantom{0}3.1} & 53.8 & 43.2 & 18.0 & 11.5 & 16.4 & 41.6 \\
        \hline
    \end{tabular}
        
    \vspace{1ex}
    {\raggedright XRD: \citet{Reddy_2014}. Bottom part: Model predictions. Space-weathering designation from \citet{Kohout_2020}. Error bars as derived from the model testing are \num{7.0, 6.5, 5.6, and 5.7}\b{~pp} for OL, OPX, Fa, and Fs (OPX), respectively.\par}
\end{table}

\begin{table}
    \caption{Modal abundances and mineral chemical composition of the space-weathered olivine samples.}
    \label{tab:kachr_ol}
    \centering
    \begin{tabular}{l c c c c}
        \hline\hline
        \multicolumn{1}{c}{} &
        \multicolumn{3}{c}{modal [vol\%]} & 
        \multicolumn{1}{c}{}\\
        & OL & OPX & CPX & Fa\\
        \hline
        actual & 100.0 & \phantom{00}0.0 & \phantom{00}0.0 & \phantom{00}9.9\\
        \hdashline
        fresh & \phantom{0}98.9 & \phantom{0}\phantom{0}1.0 & \phantom{0}\phantom{0}0.1 & \phantom{0}10.3 \\
        \phantom{00}1.7 J\,cm$^{-2}$ & \phantom{0}99.1 & \phantom{0}\phantom{0}0.8 & \phantom{0}\phantom{0}0.1 & \phantom{0}11.1 \\
        \phantom{00}2.4 J\,cm$^{-2}$ & \phantom{0}99.4 & \phantom{0}\phantom{0}0.6 & \phantom{0}\phantom{0}0.0 & \phantom{0}12.0 \\
        \phantom{00}3.8 J\,cm$^{-2}$ & \phantom{0}99.5 & \phantom{0}\phantom{0}0.5 & \phantom{0}\phantom{0}0.0 & \phantom{0}16.2 \\
        \phantom{00}4.6 J\,cm$^{-2}$ & \phantom{0}99.5 & \phantom{0}\phantom{0}0.5 & \phantom{0}\phantom{0}0.0 & \phantom{0}15.6 \\
        \phantom{00}6.7 J\,cm$^{-2}$ & \phantom{0}99.6 & \phantom{0}\phantom{0}0.4 & \phantom{0}\phantom{0}0.0 & \phantom{0}17.2 \\
        \phantom{0}10.4 J\,cm$^{-2}$ & \phantom{0}99.6 & \phantom{0}\phantom{0}0.4 & \phantom{0}\phantom{0}0.0 & \phantom{0}18.2 \\
        \phantom{0}15.0 J\,cm$^{-2}$ & \phantom{0}99.6 & \phantom{0}\phantom{0}0.4 & \phantom{0}\phantom{0}0.0 & \phantom{0}20.9 \\
        \phantom{0}23.4 J\,cm$^{-2}$ & \phantom{0}99.6 & \phantom{0}\phantom{0}0.4 & \phantom{0}\phantom{0}0.0 & \phantom{0}21.7 \\
        \phantom{0}30.6 J\,cm$^{-2}$ & \phantom{0}99.6 & \phantom{0}\phantom{0}0.4 & \phantom{0}\phantom{0}0.0 & \phantom{0}22.3 \\
        \phantom{0}60.0 J\,cm$^{-2}$ & \phantom{0}99.6 & \phantom{0}\phantom{0}0.3 & \phantom{0}\phantom{0}0.1 & \phantom{0}20.2 \\
        \phantom{0}93.8 J\,cm$^{-2}$ & \phantom{0}99.4 & \phantom{0}\phantom{0}0.4 & \phantom{0}\phantom{0}0.1 & \phantom{0}19.1 \\
        375.0 J\,cm$^{-2}$ & \phantom{0}98.5 & \phantom{0}\phantom{0}1.3 & \phantom{0}\phantom{0}0.2 & \phantom{0}14.3 \\
        \hline
    \end{tabular}
        
    \vspace{1ex}
    {\raggedright Space-weathering designation from \citet{Chrbolkova_2021}.\par}
\end{table}

\begin{table}
    \caption{Modal abundances and mineral chemical composition of the space-weathered pyroxene samples.}
    \label{tab:kachr_px}
    \centering
    \begin{tabular}{l c c c c}
        \hline\hline
        \multicolumn{1}{c}{} &
        \multicolumn{3}{c}{modal [vol\%]} & 
        \multicolumn{1}{c}{}\\
        & OL & OPX & CPX & Fs (OPX)\\
        \hline
        actual & \phantom{00}0.0 & \phantom{0}94.4 & \phantom{00}5.6 & \phantom{0}32.9\\
        \hdashline
        fresh & \phantom{0}\phantom{0}0.0 & \phantom{0}87.4 & \phantom{0}12.6 & \phantom{0}15.1 \\
        \phantom{000}4.5 J\,cm$^{-2}$ & \phantom{0}\phantom{0}0.2 & \phantom{0}92.9 & \phantom{0}\phantom{0}6.9 & \phantom{0}16.5 \\
        \phantom{000}5.6 J\,cm$^{-2}$ & \phantom{0}\phantom{0}0.2 & \phantom{0}95.9 & \phantom{0}\phantom{0}3.9 & \phantom{0}16.9 \\
        \phantom{00}12.5 J\,cm$^{-2}$ & \phantom{0}\phantom{0}0.6 & \phantom{0}95.8 & \phantom{0}\phantom{0}3.6 & \phantom{0}19.4 \\
        \phantom{00}18.0 J\,cm$^{-2}$ & \phantom{0}\phantom{0}0.4 & \phantom{0}97.7 & \phantom{0}\phantom{0}1.8 & \phantom{0}17.6 \\
        \phantom{00}28.1 J\,cm$^{-2}$ & \phantom{0}\phantom{0}0.8 & \phantom{0}96.3 & \phantom{0}\phantom{0}2.8 & \phantom{0}16.3 \\
        \phantom{00}36.7 J\,cm$^{-2}$ & \phantom{0}\phantom{0}1.0 & \phantom{0}95.4 & \phantom{0}\phantom{0}3.6 & \phantom{0}17.3 \\
        \phantom{00}50.0 J\,cm$^{-2}$ & \phantom{0}\phantom{0}1.4 & \phantom{0}93.5 & \phantom{0}\phantom{0}5.0 & \phantom{0}14.4 \\
        \phantom{00}72.0 J\,cm$^{-2}$ & \phantom{0}\phantom{0}1.4 & \phantom{0}93.3 & \phantom{0}\phantom{0}5.4 & \phantom{0}14.7 \\
        \phantom{0}112.5 J\,cm$^{-2}$ & \phantom{0}\phantom{0}1.3 & \phantom{0}91.2 & \phantom{0}\phantom{0}7.5 & \phantom{0}13.9 \\
        \phantom{0}200.0 J\,cm$^{-2}$ & \phantom{0}\phantom{0}3.2 & \phantom{0}88.7 & \phantom{0}\phantom{0}8.1 & \phantom{0}14.1 \\
        \phantom{0}450.0 J\,cm$^{-2}$ & \phantom{0}11.3 & \phantom{0}83.0 & \phantom{0}\phantom{0}5.7 & \phantom{0}15.4 \\
        1800.0 J\,cm$^{-2}$ & \phantom{0}34.1 & \phantom{0}61.5 & \phantom{0}\phantom{0}4.4 & \phantom{0}14.1 \\
        \hline
    \end{tabular}
        
    \vspace{1ex}
    {\raggedright Space-weathering designation from \citet{Chrbolkova_2021}. Actual modal abundances are in wt\%.\par}
\end{table}

The mineral chemical compositions of the mean S- and mean Q-type asteroids are comparable within standard deviations and also with ordinary chondrites. This is shown in Fig.~\ref{fig:Fa_vs_Fs}, where the boxes delimit individual types of ordinary chondrites. However, \b{the Fa and Fs values of the mean S-type asteroid} are slightly lower than those of \b{the mean Q-type asteroid} (see Tables~\ref{tab:ast_types} and \ref{tab:ast_types_S}). We hypothesise that this might be related to the observation that Q-type asteroids are mostly present in the near-Earth asteroid population; see Fig.~6 in \citet{Binzel_2019}. Moreover, LL types dominate the ordinary chondrite meteorite population \citep{Grady_2000}. Therefore, Q types might be slightly more oxidised (LL-like) than the mean \b{S type}.

Another finding is that the mineral chemical distributions of various types of S-complex asteroids are comparable (see Table~\ref{tab:ast_types_S}). The individual types are not related to specific compositions, but rather to the appearance of spectra. The spectra of S-complex asteroids differ mostly in the shapes of the continua or the relative depths of the absorption bands. This can be caused by additional featureless minerals or space weathering. The similarity in mineral chemical composition can also be seen in the S column of Table~\ref{tab:ast_types}, which is indicated by the low values of the standard deviations of the chemical compositions. As we showed for the Chelyabinsk meteorite, the predictions of our model are only slightly sensitive to global attenuation. When we compare the mean predicted values for a type and its weathered version (e.g. S and Sw), we observed similar trends as for the samples of \citet{Chrbolkova_2021}. The prediction for fayalite numbers is slightly higher for weathered types, while orthopyroxene forsterite numbers are always much smaller. This may also indicate that the method used by \citet{Chrbolkova_2021} corresponds more closely to the physical reality of asteroid space weathering.

V-type asteroids are classified to be predominantly composed of pyroxenes, which correlates with high PC2' values (Fig.~\ref{fig:PCA}). In contrast, A-type asteroids are predicted to be composed exclusively of olivine, consistent with their low PC2' values in the DeMeo classification.

These tests show that the model predicts a reasonable chemical composition of olivine even for weathered samples. Predictions for the chemical composition of pyroxene are more sensitive to space weathering and can be offset to lower values. Furthermore, the predicted modal abundances of low-weathered asteroids (e.g. Q or Sq) or asteroids with a mono-mineral composition (e.g. containing one dominant phase, such as pyroxene in V, or olivine in A or Sa) are not significantly influenced by space weathering. On the other hand, the predicted modal abundances of S-type asteroids (i.e. weathered mixtures of olivine and pyroxene) lead to an artificial distortion of the modal ratio of pyroxene to olivine in favour of pyroxene because olivine is more strongly weathered and hence its diagnostic absorption bands are more strongly compressed than those of pyroxene.

\ocfigure{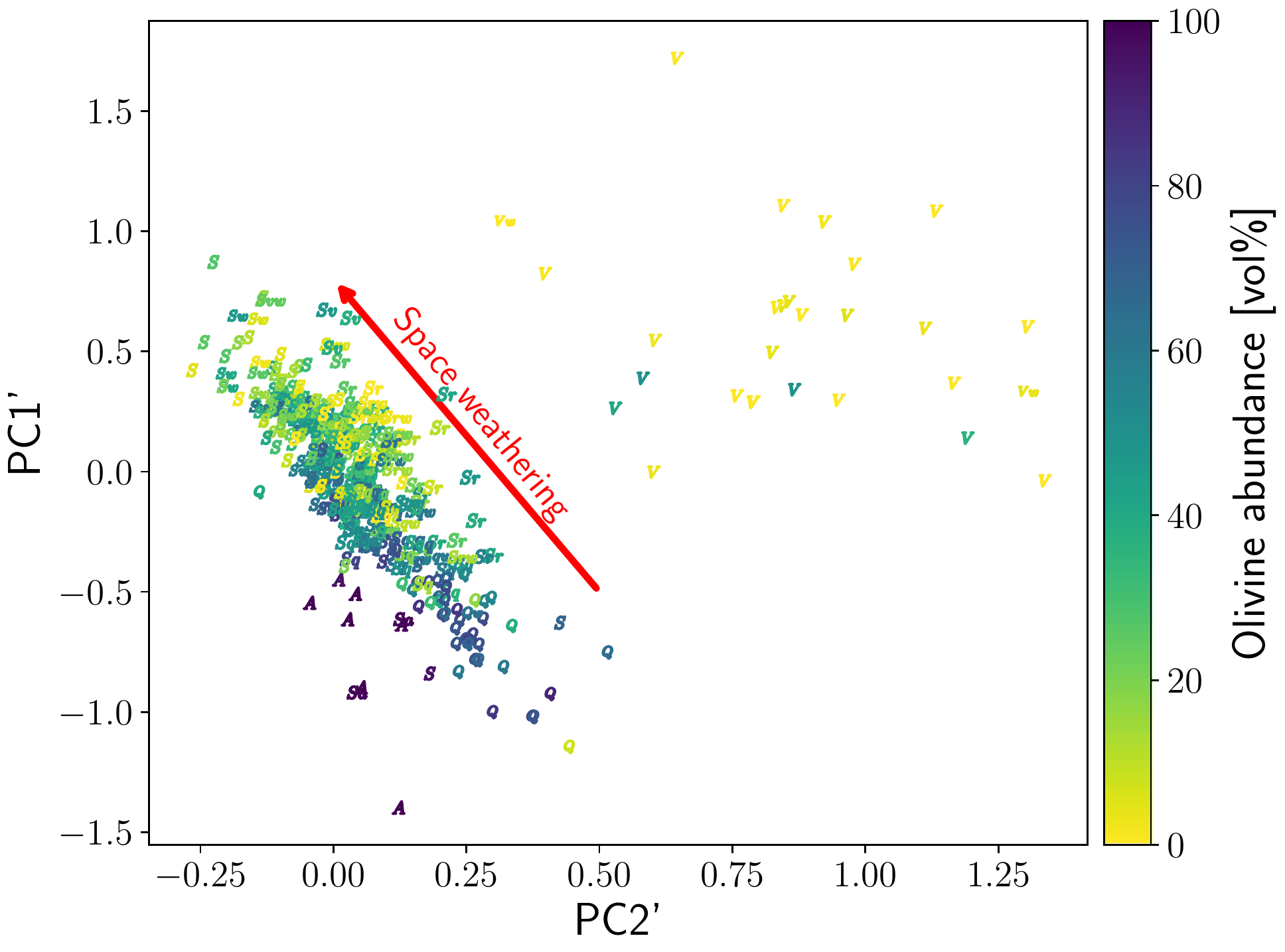}
{Principal components of S$^*$-complex asteroids. The predicted olivine fraction decreases in the direction of increasing space weathering, which is indicated by the red arrow.}
{fig:PCA}

\ocfigure{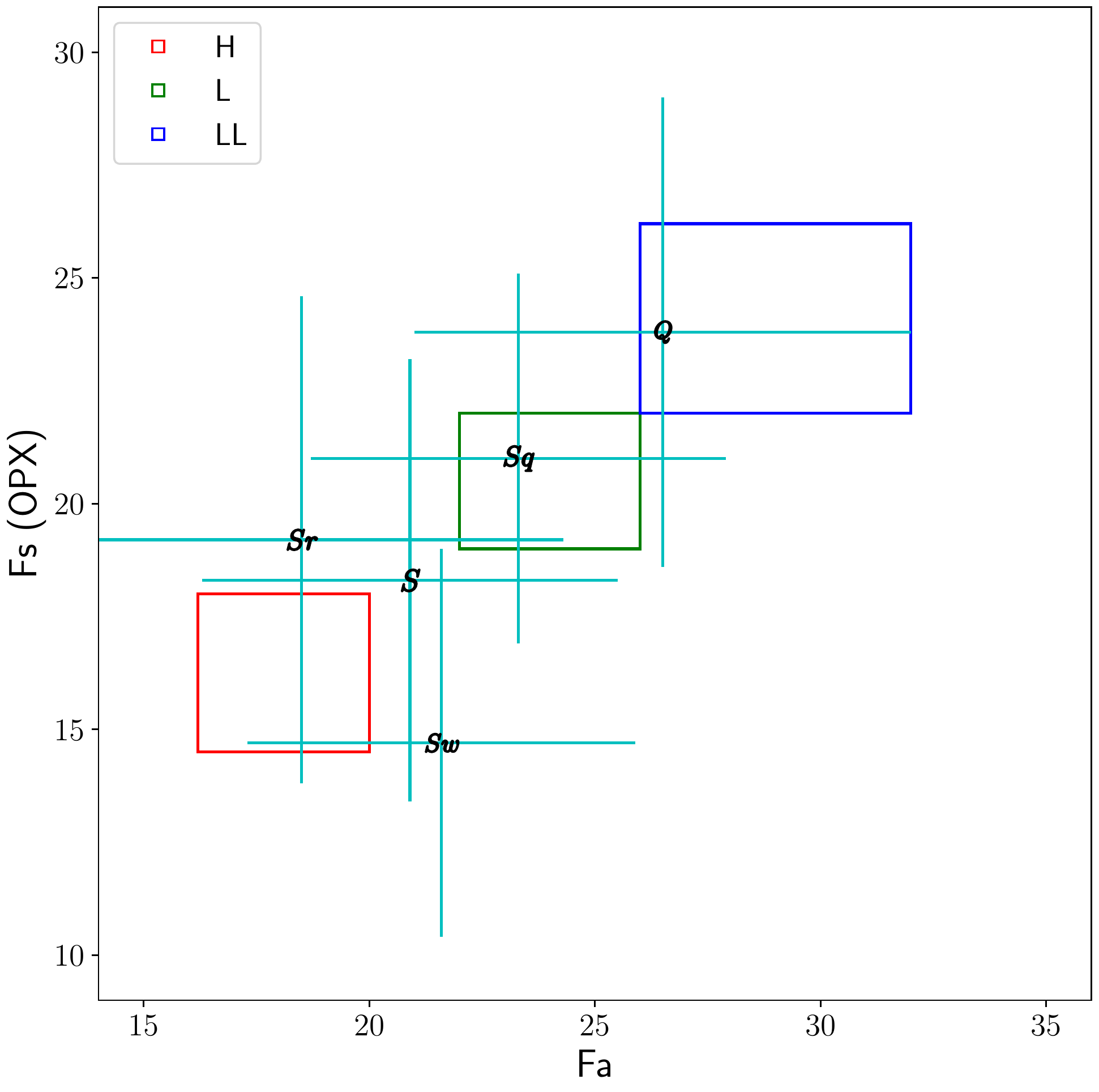}
{Iron contents in olivine and orthopyroxene in the mean S- and Q-type asteroids. The boxes roughly delimit the types of ordinary chondrites. Red,  green, and blue correspond to H, L, and LL types, respectively.}
{fig:Fa_vs_Fs}

%%%%%%%%%%%%%%%%%%%%%%%%%

\subsection{Comparison to legacy methods}

Legacy methods for a spectrum deconvolution that are optimised for an olivine-pyroxene-rich sample typically use the 1\textmu{}m and 2\textmu{}m bands, namely the positions of their central wavelengths (designated Band I Centre, BIC, and Band II Centre, BIIC) and the ratio of the 2\textmu{}m band area and 1\textmu{}m band area (band area ratio, BAR). BIC, BIIC, and BAR are used as an input for empirically derived relations where outputs are the olivine-to-pyroxene ratio and the chemistry of olivine and pyroxene. The compositional interval in which these methods are valid is typically rather restricted, and different relations exist for various mineralogies \citep[see e.g.][]{Gaffey_2002}.

We applied the equations from \citet{Cloutis_1986}, \citet{Gaffey_2002}, and \citet{Reddy_2015} to our test data, which consisted of selected olivine-orthopyroxene-dominated laboratory mixtures and meteorites (CPX lower than 10~vol\%). The empirical relation of \citet[BAR]{Cloutis_1986} was optimised for olivine-to-pyroxene ratios of S- and Q-type asteroids. The pyroxene chemistry equations from \citet[BIIC]{Gaffey_2002} must be solved iteratively, but the validity interval is much broader than \b{that based on} the equations in \citet[BIC]{Reddy_2015}. The comparison between our neural-network model and these legacy models is shown in Fig.~\ref{fig:NN_BAR_BC}. The BAR- and BIC-based methods agree well with our neural-network-based model and with the actual analytical compositional values. The BIIC-based method shows \b{an} overall lower agreement with analytical results and with our model. The $\RMSE$ \b{error estimate} of the BAR method is 10.5\b{~pp} for the olivine fraction, and the $\RMSE$s of the BIC and BIIC methods for orthopyroxene Fs are \num{6.8 and 13.7~pp}, respectively. We note that the application of these BAR- and BIC-based methods was optimised to rather narrow compositional intervals. The good agreement in this interval is therefore expected. In contrast, our neural-network-based method shows a \b{better} performance and is applicable to the whole range of compositions.

We also applied the BAR method on the asteroid spectra and plot the olivine fractions to the PCA in Fig.~\ref{fig:PCA_BAR}. The BAR method is less sensitive to space weathering than our neural-network model. This may be due to the fact the BAR method uses the combined 1\textmu{}m olivine-pyroxene band area, and the relatively low attenuation of the olivine bands does not influence the band area significantly. The neural-network method evaluates the whole spectrum shape and thus may be more sensitive to small (space-weathering-related) changes in the olivine bands. This was partially confirmed when we applied the method blindly to non-S-type asteroids. In the case of V-type asteroids, the BAR method results in up to \num{50}~vol\% of olivine (mean fraction of \num{27}~vol\%), which is unrealistic given the pyroxene-dominated composition of V-type asteroids. It therefore seems that the BAR method always divides the 1\textmu{}m band between olivine and pyroxene even \b{when} one of the constituents may be absent. For this reason, the BAR method is not a robust indicator for a broad range of mineral modal abundances.

\ocfigure{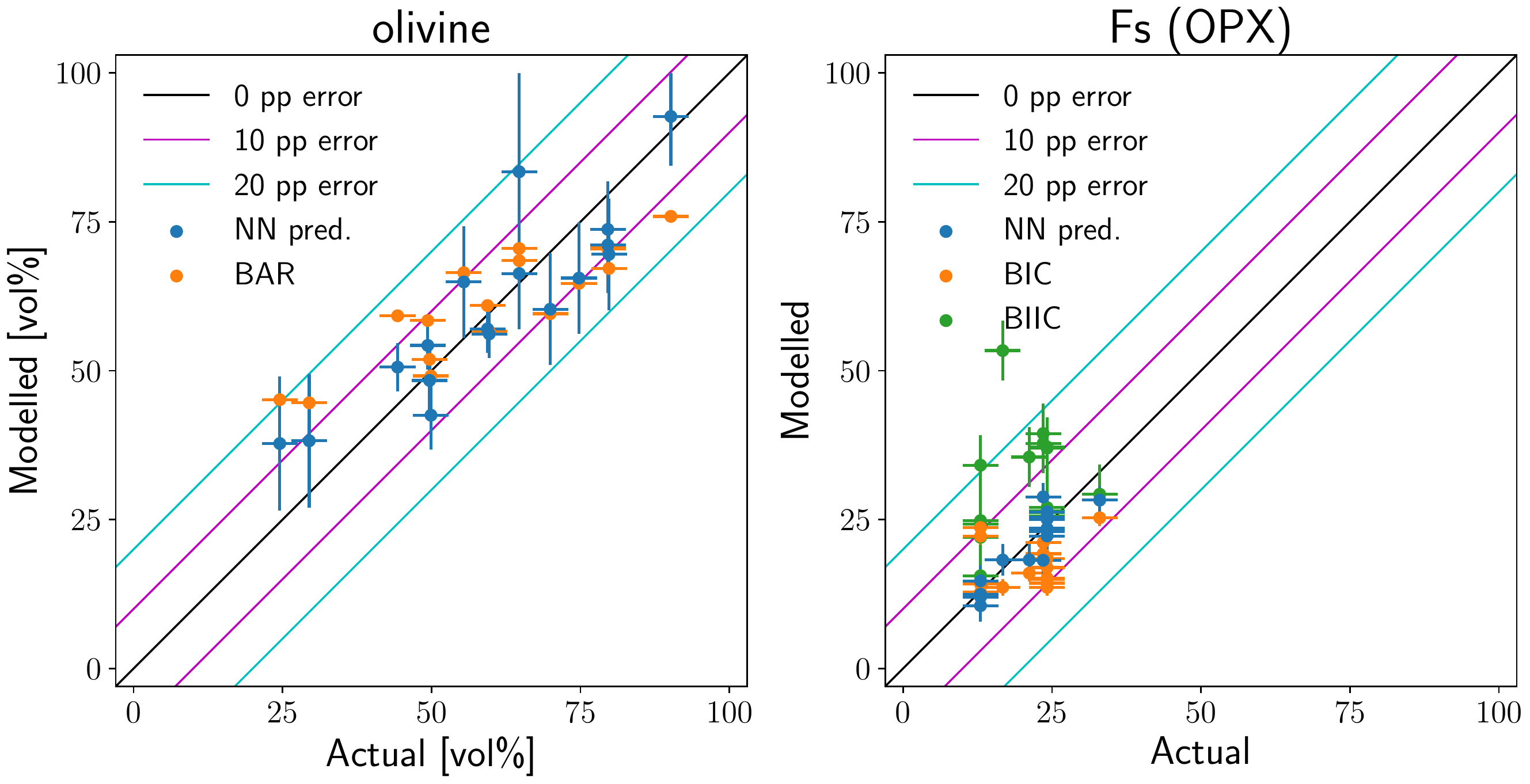}
{Scatter plots of the true and modelled olivine fraction and ferrosilite number of orthopyroxene in the olivine-pyroxene laboratory mixtures and meteorites in the test data. The diagonal lines delimit the accuracy of the models.}
{fig:NN_BAR_BC}

\ocfigure{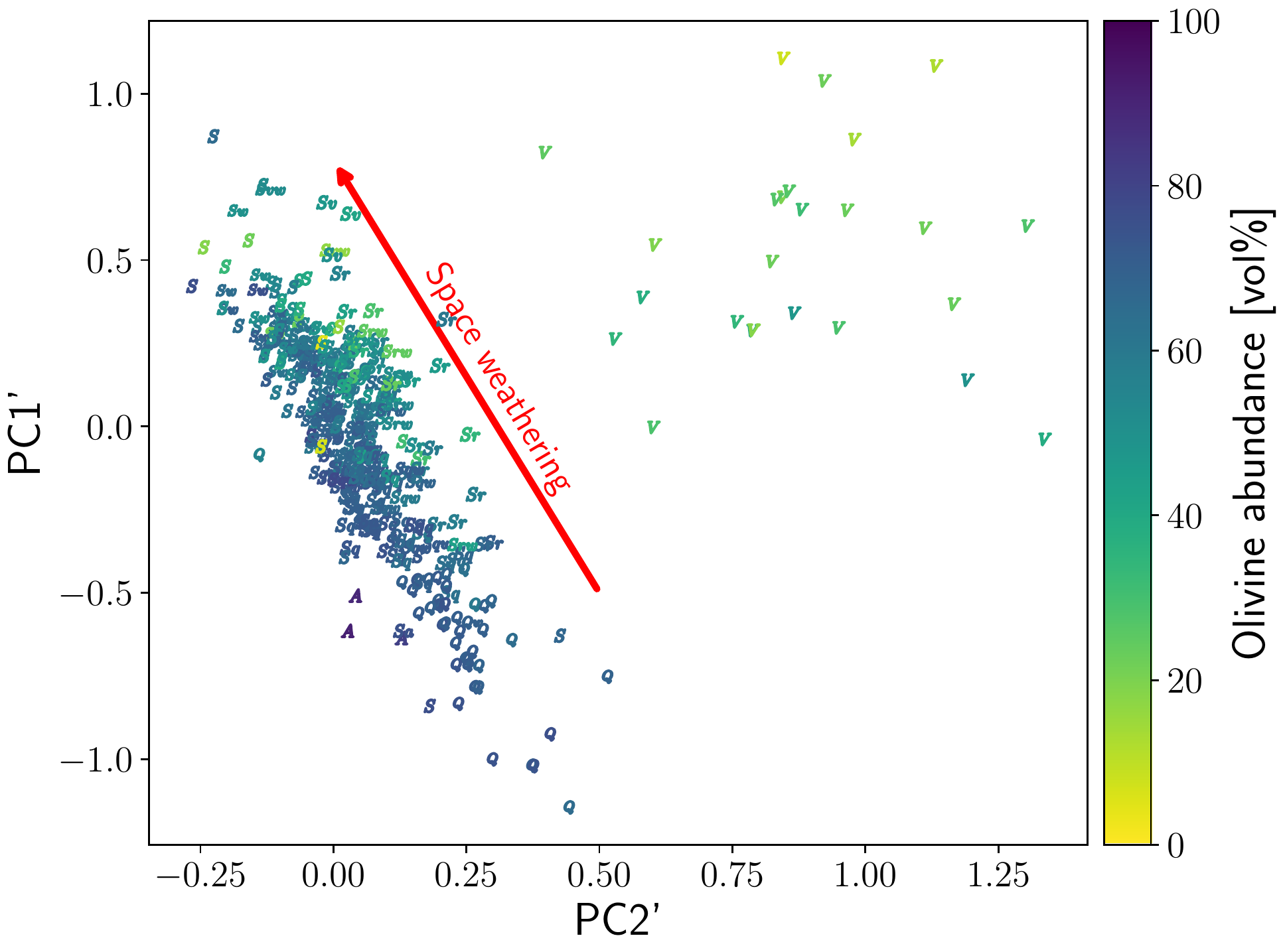}
{Principal components of S$^*$-complex asteroids. The direction of increasing space weathering is indicated by the red arrow. The model is based on BAR.}
{fig:PCA_BAR}

%%%%%%%%%%%%%%%%%%%%%%%%%%%%%%%%%%%%%%%%%%%%%%%%%%

\section{Conclusions}

We introduced a deep-learning-based method for estimating the modal and chemical compositions of olivine-pyroxene-rich mixtures, meteorites, and asteroids directly from their visible to near-infrared reflectance spectra. We used a feed-forward convolutional neural network with two hidden layers, which consisted of 24 and 8 different convolutional kernels.

We trained the model with real samples of silicates and their mixtures. The model predicted modal abundances and the mineral chemical compositions of olivine, orthopyroxene, and clinopyroxene. The predictions for most of the test-data samples had a precision better than 10~percentage points. Subsequently, we applied the method to derive the composition of S-complex asteroids with known taxonomy types. As expected, the V-type asteroids are dominated by orthopyroxene, and the A-type asteroids are made of almost pure olivine. The S-type and the Q-type asteroids have similar mineral chemical compositions among themselves, and the compositions are also similar to those of ordinary chondrites. The differences in predicted olivine modal abundances between the S-type and the Q-type asteroids are consistent with a lower resistance of olivine to space weathering. Standard deviations show that mineral chemical compositions are more homogeneous within the types, but modal abundances are potentially very different.

The main advantages of our neural-network model over legacy spectrum deconvolution methods are (1)~the overall low computational and memory requirements during classification runs, (2)~the versatility, demonstrated by \b{the scores of the used reliability metrics} over a wide range of olivine and pyroxene modal and chemical compositions, (3)~the ability to process complete spectra without the need of prior derivation of spectral parameters (i.e. fitting and quantification of absorption band properties), and (4)~the independence of our results of grain sizes, spectrally neutral phases, or shapes of continua. In the following years, space missions will generate extensive spectral datasets from asteroids or planets. Our code may be useful for these cases when a large number of spectra has to be evaluated in real time, for example, during hyperspectral remote sensing.

%%%%%%%%%%%%%%%%%%%%%%%%%%%%%%%%%%%%%%%%%%%%%%%%%%

\begin{acknowledgements}
This research is supported by Academy of Finland projects 335595 and 325805, NASA SSERVI Center for Asteroid and Lunar Surface Science (CLASS), and within institutional support RVO~67985831 of the Institute of Geology of the Czech Academy of Sciences. We would like to thank Francesca DeMeo and Richard Binzel for providing us the dataset on the asteroid spectra. We utilised data stored in the RELAB Spectral Database operated by Brown University and C-Tape database operated by The University of Winnipeg. This research has made use of NASA's Astrophysics Data System Bibliographic Services. The authors would like to thank the anonymous referee for the valuable comments and suggestions, which greatly improved the quality of the paper.
\end{acknowledgements}

\bibliographystyle{aa}
\bibliography{BIBL}

%%%%%%%%%%%%%%%%%%%%%%%%%%%%%%%%%%%%%%%%%%%%%%%%%%

\appendix

\section{On-line supplement with scripts and datasets}
\label{sect:supplement}

The data and Python scripts used in our work are \b{available online\footnote{\url{https://github.com/Sirrah91/Asteroid-spectra}} in version \mbox{v1.0}}. The datasets, modules, and models are in the corresponding sub-folders. For this work, we present the results of six models, which are distinguished in REAMDE.md file in the Models/\b{compositional} sub-folder.

\subsection{Datasets}

All datasets were interpolated to an equidistant grid from 450~nm to 2450~nm with a step of 5~nm. Data files in the Datasets folder are distinguished by the suffix. \b{That is the suffix '-denoised' means that the data were denoised with the Gaussian convolution kernel and the suffix `-norm' means that the data were normalised at 550~nm.} The description of the data, their labels, the wavelengths we used, and the known metadata are stored together in the data files (see README.md in the Datasets sub-folder for loading the data and metadata). Except for metadata, they also can be found in the README.md and are summarised in Table~\ref{tab:data_supplement}. The known information about the data and \b{metadata} is stored in Sample\_Catalogue.xlsx. This catalogue contains four sheets. Three sheets (RELAB, C-Tape, and TK) contain known metadata, and the fourth sheet \b{(Chemical\_analyses)} summarises the chemical analyses. The metadata include details about the samples (e.g. grain sizes, sample types and subtypes, sample textures, sample source, and origin). From column R to the right, we list the labels, the conversion of the modal abundances from wt\% into vol\%, and the references to the labels.

\begin{table*}[!ht]
    \small
    \caption{Datafile description}
    \label{tab:data_supplement}
    \centering
    \begin{tabular}{l c c c l l}
        \hline\hline
        name & wvl. range [nm] & denoised & normalised & labels & notes\\
        \hline
        Chelyabinsk-denoised-norm.npz & 450:5:2450 & \cmark & \cmark & composition & \citet{Kohout_2020}\\
        Kachr\_ol\_opx-denoised-norm.npz & 450:5:2450 & \cmark & \cmark & composition & \citet{Chrbolkova_2021}\\
        asteroid\_spectra-denoised-norm.npz & 450:5:2450 & \cmark & \cmark & taxonomy & \citet{DeMeo_2009} and \citet{Binzel_2019}\\
        combined-denoised-norm.npz & 450:5:2450 & \cmark & \cmark & composition & training, validation, and test data\\
        \hline
    \end{tabular}

    \vspace{1ex}
    {\raggedright Wavelength range in the format from:step:to.\par}
\end{table*}

\subsubsection*{Unit conversion}

To describe the conversion of modal abundances from wt\% into vol\%, or the mineral chemical composition from wt\% to the end-members, we used the molar masses and end-member densities from Table~\ref{tab:constants}.

\begin{table}[!ht]
    \caption{Densities and molar masses used for unit conversions.}
    \label{tab:constants}
    \centering
    \begin{tabular}{l l l l l l}
        \hline\hline
        \multicolumn{5}{c}{density [g\,cm$^{-3}$]}\\
        \hline
        Fa & Fo & Fs & En & Wo\\
        4.39 & 3.27 & 3.95 & 3.20 & 2.90\\
        \hline
        \multicolumn{5}{c}{molar mass [g\,mol$^{-1}$]}\\
        \hline
        FeO & MgO & CaO & &\\
        71.8440 & 41.3040 & 56.0774 & &\\
        \hline
        
    \end{tabular}
\end{table}

For an example of the conversion, we consider the ordinary chondrite with ID OC-SXS-022-D we find in the RELAB sheet in row 10 of the Sample\_Catalogue.xlsx table. In Table~2 of the reference article (in the Notes column AH), we found in the case of olivine chemical composition, for instance, FeO 28.02 wt\% and MgO 34.85 wt\%. The fayalite and forsterite numbers (columns V and W) were computed from these two numbers and the molar masses of FeO and MgO as
\begin{align}
    \mathrm{Fa} &= 100 \times \frac{\frac{28.02}{71.8440}}{\frac{28.02}{71.8440} + \frac{34.85}{41.3040}} = 31.08,\\
    \mathrm{Fo} &= 100 \times \frac{\frac{34.85}{41.3040}}{\frac{28.02}{71.8440} + \frac{34.85}{41.3040}} = 100 - \mathrm{Fa} = 68.92.
\end{align}
To convert modal abundances in wt\% into vol\%, we used the values in \b{Table~1 of the reference article}. First, we computed the approximate densities of olivine, orthopyroxene, clinopyroxene, and plagioclase, for which we know the individual chemical composition. The density estimates correspond to the weight averages of the mineral chemical composition and end-member densities. For example, the density of orthopyroxene for the OC-SXS-022-D sample with Fs24.7, En72.3, and Wo3.0 (columns X, Y, and Z) is
\begin{equation}
    \rho_{\mathrm{OPX}} = \frac{24.7 \times 3.95 + 72.3 \times 3.20 + 3.0 \times 2.90}{24.7 + 72.3 + 3.0} = 3.38\ \mathrm{g\, cm^{-3}},
\end{equation}
which can be found in column AX. With the knowledge of densities, the non-normalised volume fractions of individual minerals were computed as the modal abundance in wt\% divided by the density (columns BB to BE). The fractions were normalised to sum up to 100 to obtain the modal abundances in volume percent (columns R to U). 

\subsection{Modules}

All Python scripts we used are well documented via comment lines and can be found in the modules sub-folder. The neural-network pipeline (main.py script) is visualised in the block diagram in Fig.~\ref{fig:block_diagram}. The behaviour of the whole code is driven by the NN\_config.py, where the user selects labels, hyperparameters, and selection criteria for the data, and may change the wavelength grid of the data and their normalisation.

\ocfigure[!ht]{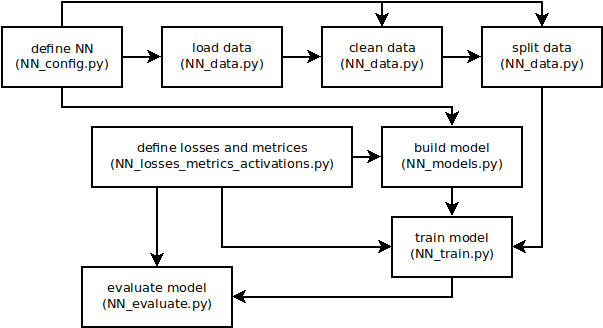}
{Schematic visualisation of the neural-network pipeline.}
{fig:block_diagram}

%%%%%%%%%%%%%%%%%%%%%%%%%%%%%%%%%%%%%%%%%%%%%%%%%%

\section{Test models}
\label{sect:app_models}

In addition to the baseline OL-OPX-CPX model, we tested the behaviour of the neural network in five additional tasks: evaluation of the chemical composition of (1)~pure olivine, (2)~pure orthopyroxene, (3)~pure and binary mixtures of olivine and orthopyroxene, (4)~a model that also contains the modal abundance of plagioclase, and (5)~a model that additionally contains a plagioclase chemical composition and the orthopyroxene wollastonite number. We note that for pure olivine, pure orthopyroxene, and pure and binary olivine-orthopyroxene mixtures, we used models with fewer convolutional kernels and increased the regularisation parameters to overcome any overfitting of these models. The hyperparameters that differ from those in Table~\ref{tab:hp} are listed in Table~\ref{tab:hp_app}. 

\begin{table}[!ht]
    \caption{Updated hyperparameters for less complex models.}
    \label{tab:hp_app}
    \centering
    \begin{tabular}{l l l l}
        \hline\hline
        \multicolumn{1}{c}{} & 
        \multicolumn{1}{c}{OL} &
        \multicolumn{1}{c}{OPX} &
        \multicolumn{1}{c}{OL + OPX}\\
        \hline
        Nodes/filters in hid. l. & 4 and 4 & 4 and 2 & 8 and 4\\
        $L_1$ trade-off parameter & 0.1 & 0.1 & 0.05\\
        \hline
    \end{tabular}
\end{table}

\subsection{Models with pure olivine and orthopyroxene}

For the very simple models for pure olivine and pure orthopyroxene, we expected to obtain more accurate predictions than in case of the final model because pure samples have a stronger effect on the loss function. Therefore, better precision is naturally expected. This was confirmed in both cases. Furthermore, for these models we did not need to omit samples with low iron content, which gave us six more enstatite-rich orthopyroxenes ($97.5 \leq \mathrm{En} \leq 100.0$). For the pure olivine model, we \num{obtained} $\RMSE = 4.4$\b{~pp}, and the worst prediction was only \num{8.4}~pp away from the actual value (i.e. all the predictions were within the 10~pp error interval). For the pure orthopyroxene model, we also obtained a better model with similar properties as for the pure \num{olivine model}: $\RMSE = 4.5$\b{~pp}, and the worst prediction was \num{9.6}~pp away from the actual value. The results are plotted in Figs.~\ref{fig:OL_pure}--\ref{fig:quantile_pures}. We note that the results in Fig.~\ref{fig:OL_pure} can be directly comparable with the pure samples from Fig.~\ref{fig:OL}. This does not hold for the pure orthopyroxene model because of the added enstatite-rich samples and the consequently different shuffling of data between training, validation, and test datasets.

\ocfigure[!t]{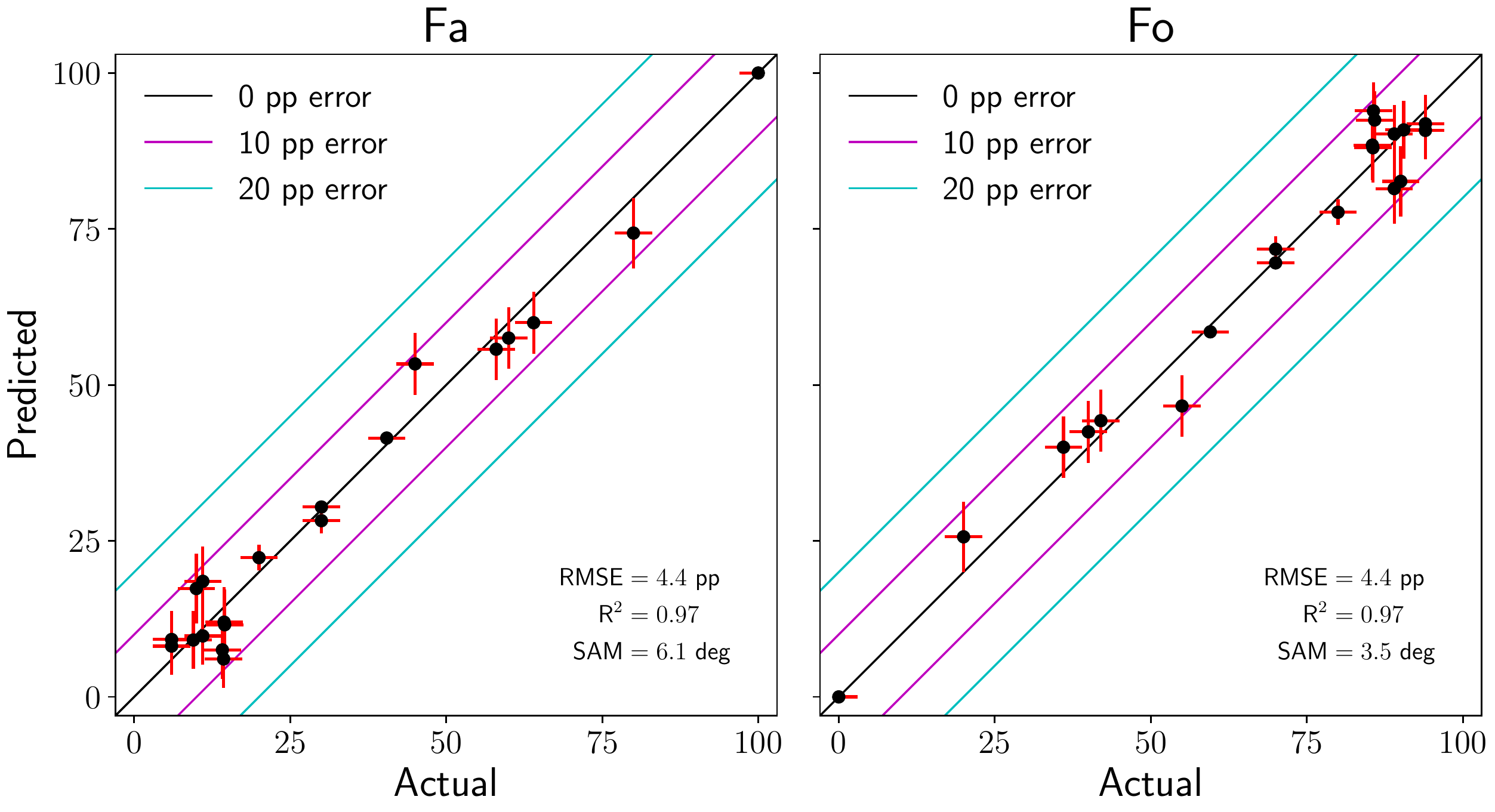}
{Scatter plots of the true and predicted chemical compositions of olivine in the test data. Left: Iron content. Right: Magnesium content. The colours of the points correspond to the actual modal abundance of olivine in the samples. \b{We show the model} with pure olivine.}
{fig:OL_pure}

\ocfigure[!ht]{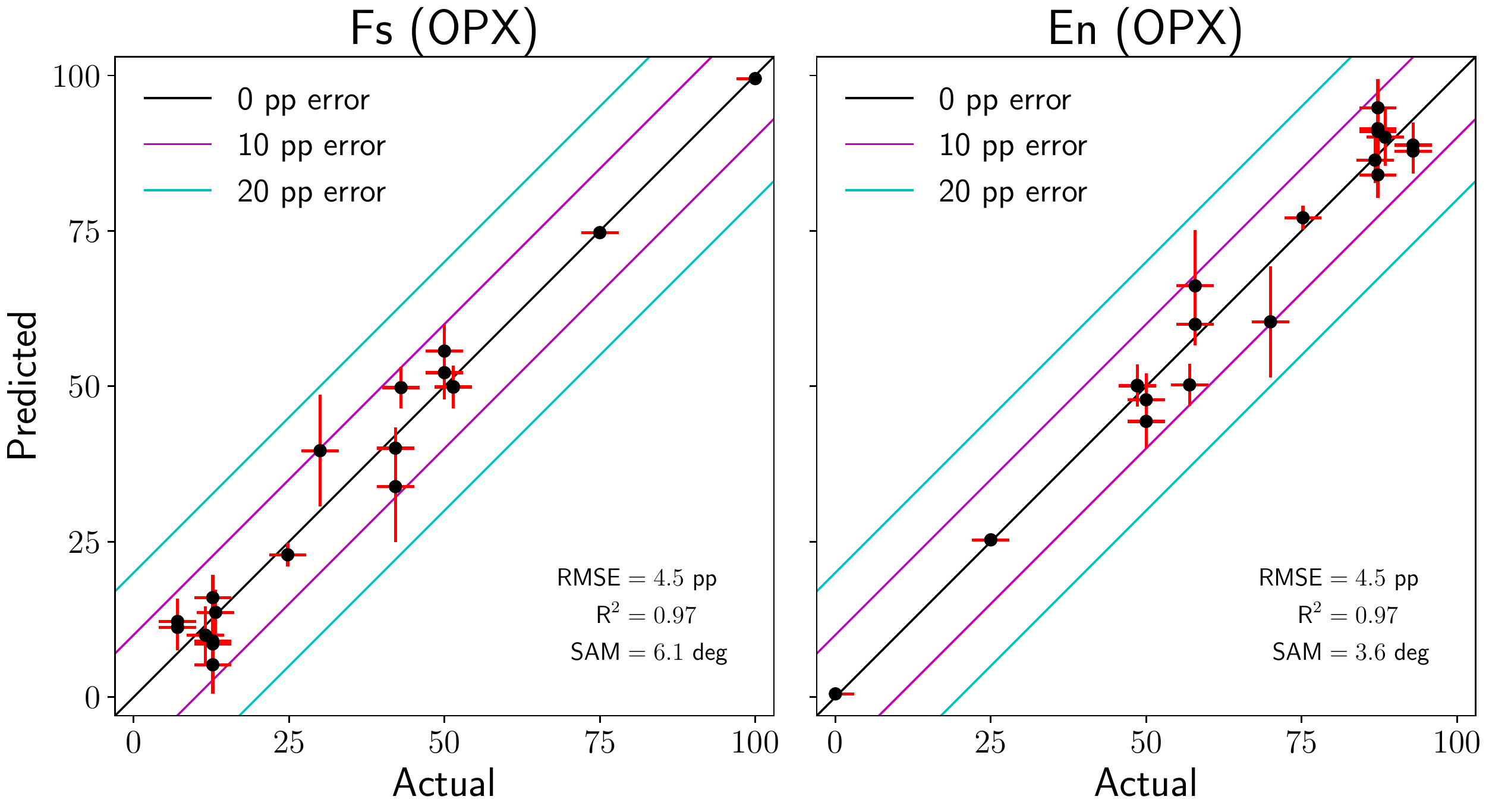}
{Scatter plots of the true and predicted chemical compositions of orthopyroxene in the test data. Left: Iron content. Right: Magnesium content. The colours of the points correspond to the actual modal abundance of orthopyroxene in the samples. \b{We show the model} with pure orthopyroxene.}
{fig:OPX_pure}

\ocfigure[!ht]{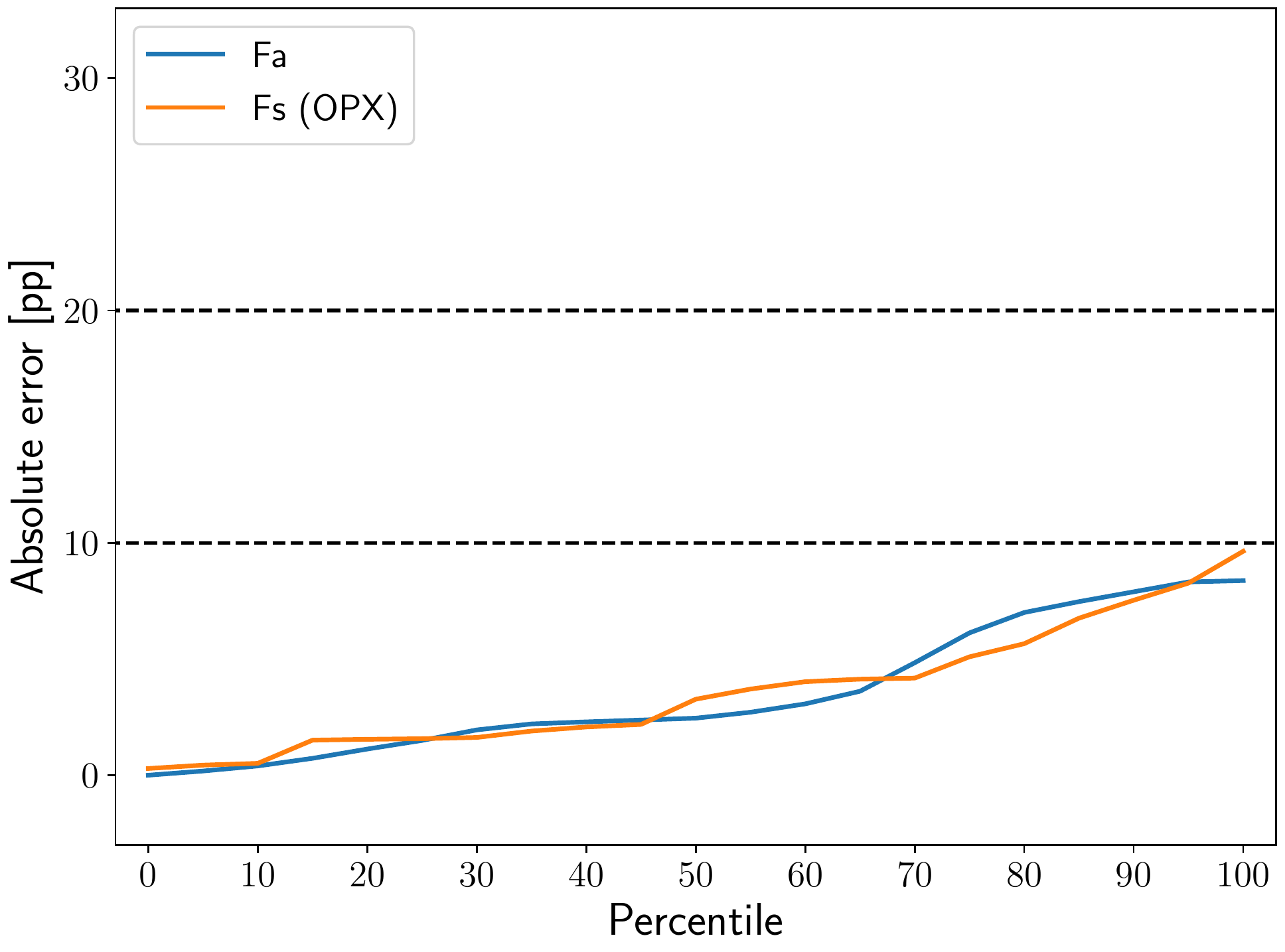}
{Quantiles of the absolute errors between the actual and predicted values. The dashed black lines indicate the 10~pp and 20~pp errors. \b{We show the models} with pure olivine and pure orthopyroxene together.}
{fig:quantile_pures}

%%%%%%%%%%%%%%%%%%%%%%%%%

\subsection{Model of olivine-orthopyroxene mixtures}

For a slightly more complex model of pure olivine, pure orthopyroxene, and their binary mixtures, we omitted low-iron olivines and kept low-iron orthopyroxenes (the six additional enstatite rich samples). In this case, the mineral chemical compositions were of the same precision as in the final model: $\RMSE$s were \num{6.4 and 5.2}\b{~pp} for olivine and orthopyroxene compositions, respectively. The advantage of this model is that there were no mismatches between orthopyroxene and clinopyroxene. This led to a high precision in mineral modal abundance, where only four predictions (\num{6}\%) were not within the 10~pp error interval, and over \num{90}\% of mineral chemical compositions were within the 10~pp error interval. These results are comparable to the final model when we ignore the orthopyroxene-clinopyroxene mismatches in modal abundances. The overall $\RMSE$ \b{for modal composition} was \num{4.7~pp and $R^2 = 0.99$}. The results of this model and the quantiles with absolute errors are plotted in Figs.~\ref{fig:modal_mix}--\ref{fig:quantile_mix}.

In the zeroth approximation, we can estimate that S-, Q-, V-, and A-type asteroids are olivine-orthopyroxene mixtures. We applied this model on the asteroid samples. For the mineral chemical compositions, the predictions are the same within one standard deviation as for the final model. As expected, the mineral modal abundances of the mean V-type and A-type asteroids are made of almost pyre orthopyroxene (\num{96.5} vol\%) and pure olivine (\num{100.0} vol\%), respectively. Similarly as in the final model, the mean Q type is dominated by olivine (\num{82.0} vol\%) and the mean S types are dominated by orthopyroxene (\num{76.0} vol\%). This difference is due to the space-weathered surface of S-type asteroids and to our non-space-weathered dataset. The standard deviation in modal abundances of S-type asteroids, which is equal to \num{21~pp}, indicates a different level of space weathering in this class. This is also shown in Fig.~\ref{fig:PCA_app}.

\ocfigure[!t]{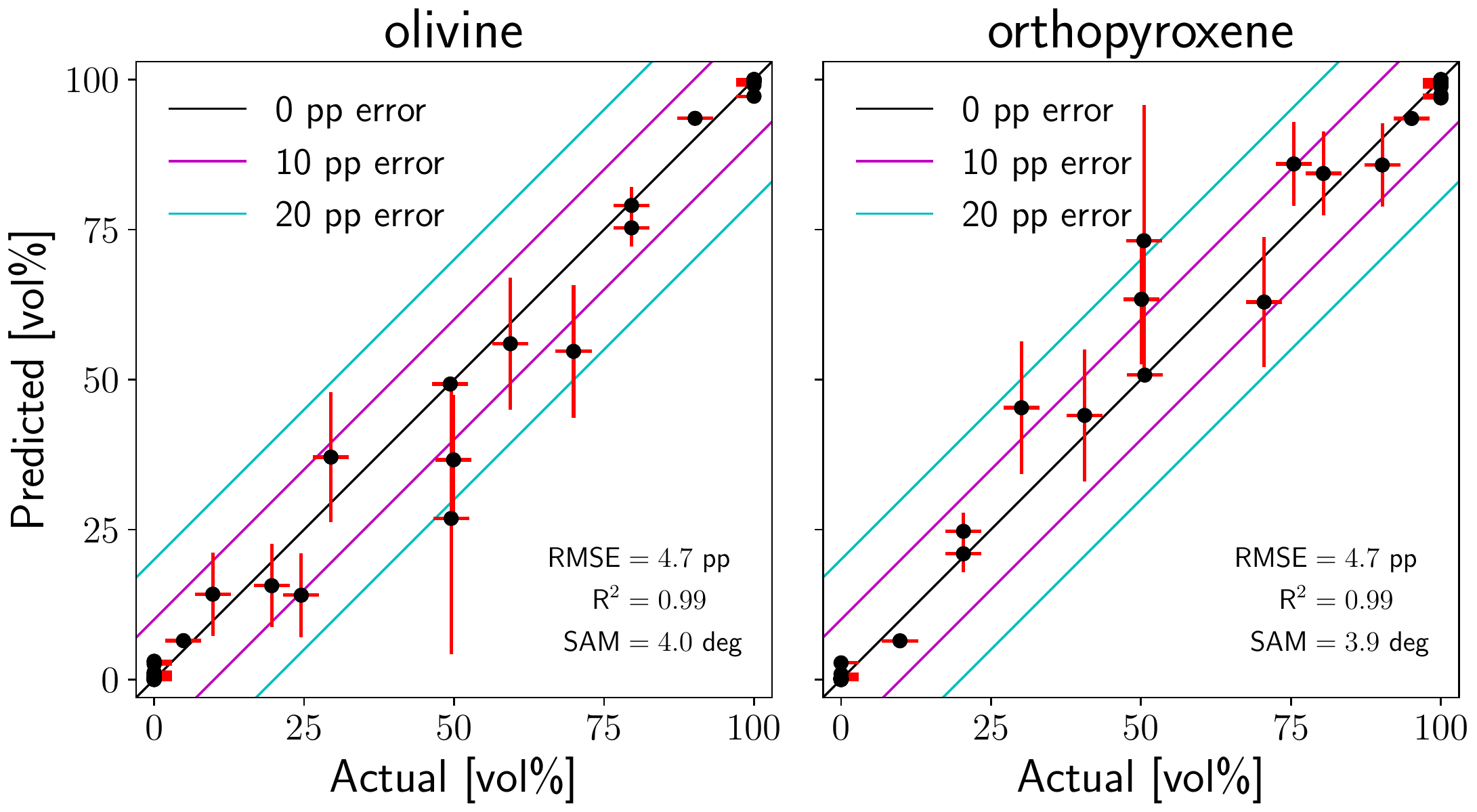}
{Scatter plots of the true and predicted modal compositions of the test data. The diagonal lines delimit the accuracy of the predictions. We show the model with pure olivine, pure orthopyroxene, and their binary mixtures.}
{fig:modal_mix}

\ocfigure[!t]{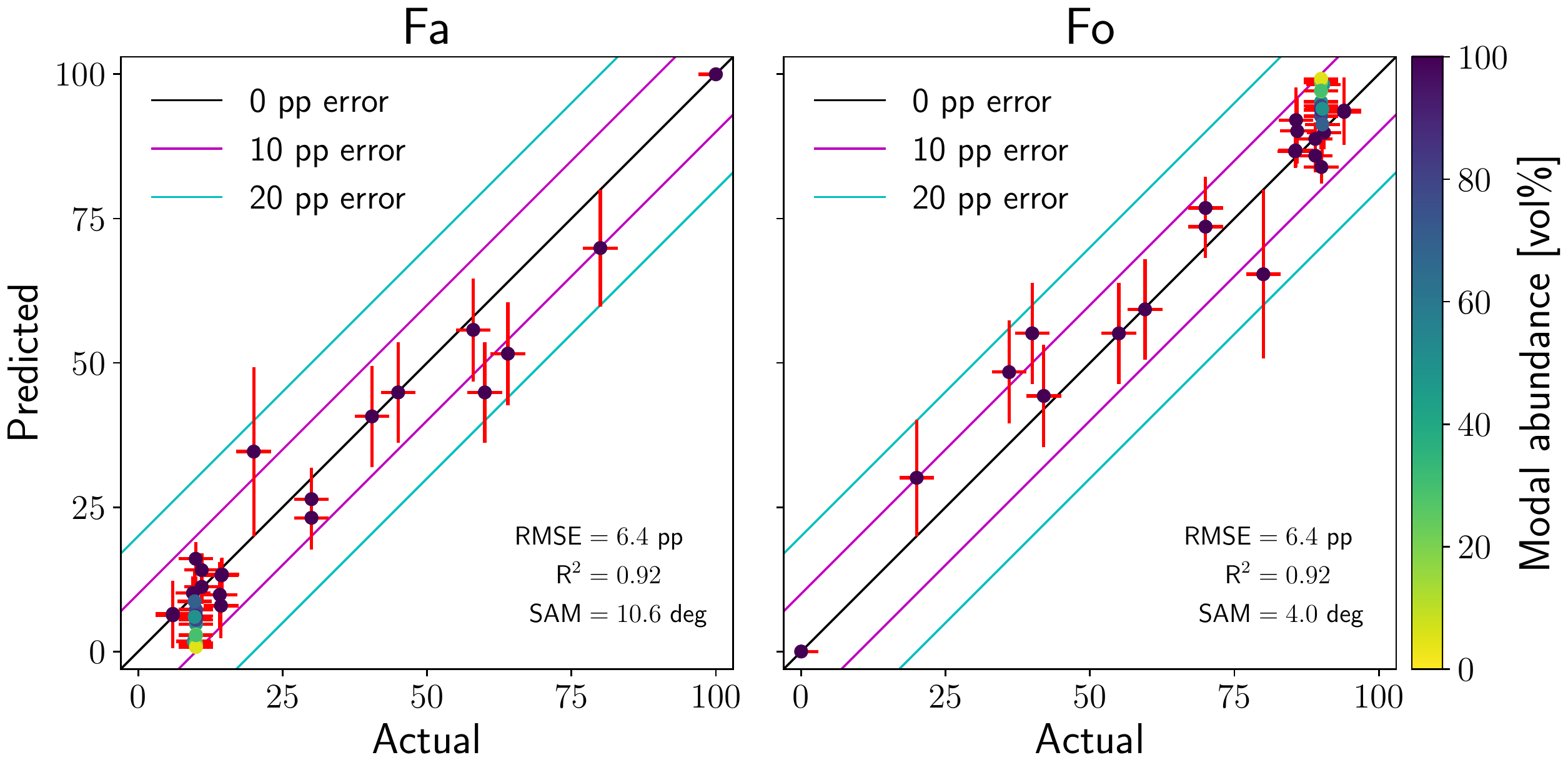}
{Scatter plots of the true and predicted chemical compositions of olivine in the test data. Left: Iron content. Right: Magnesium content. The colours of the points correspond to the actual modal abundance of olivine in the samples. We show the model with pure olivine, pure orthopyroxene, and their binary mixtures.}
{fig:OL_mix}

\ocfigure[!ht]{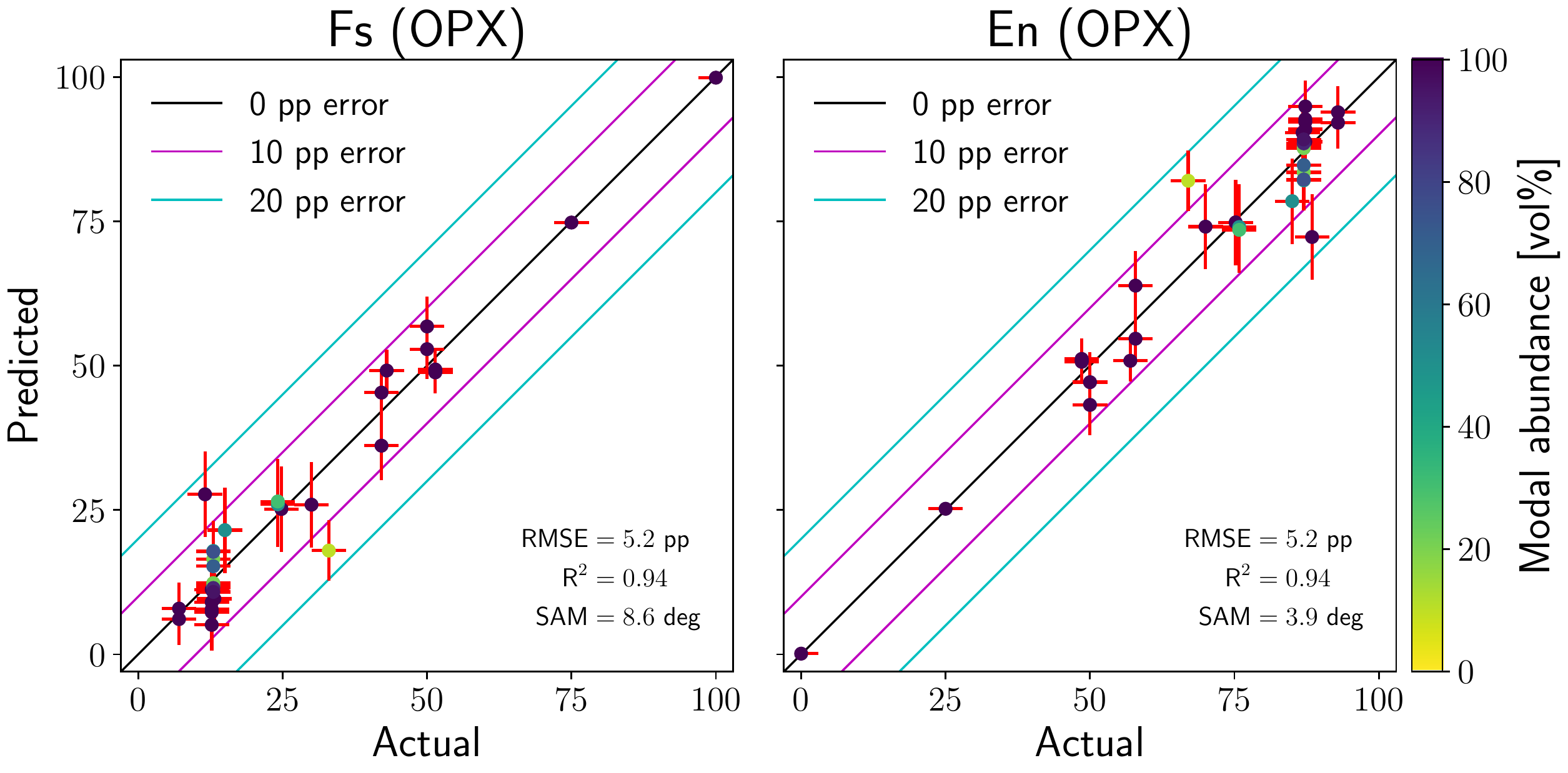}
{Scatter plots of the true and predicted chemical compositions of orthopyroxene in the test data. Left: Iron content. Right: Magnesium content. The colours of the points correspond to the actual modal abundance of orthopyroxene in the samples. We show the model with pure olivine, pure orthopyroxene, and their binary mixtures.}
{fig:OPX_mix}

\ocfigure[!ht]{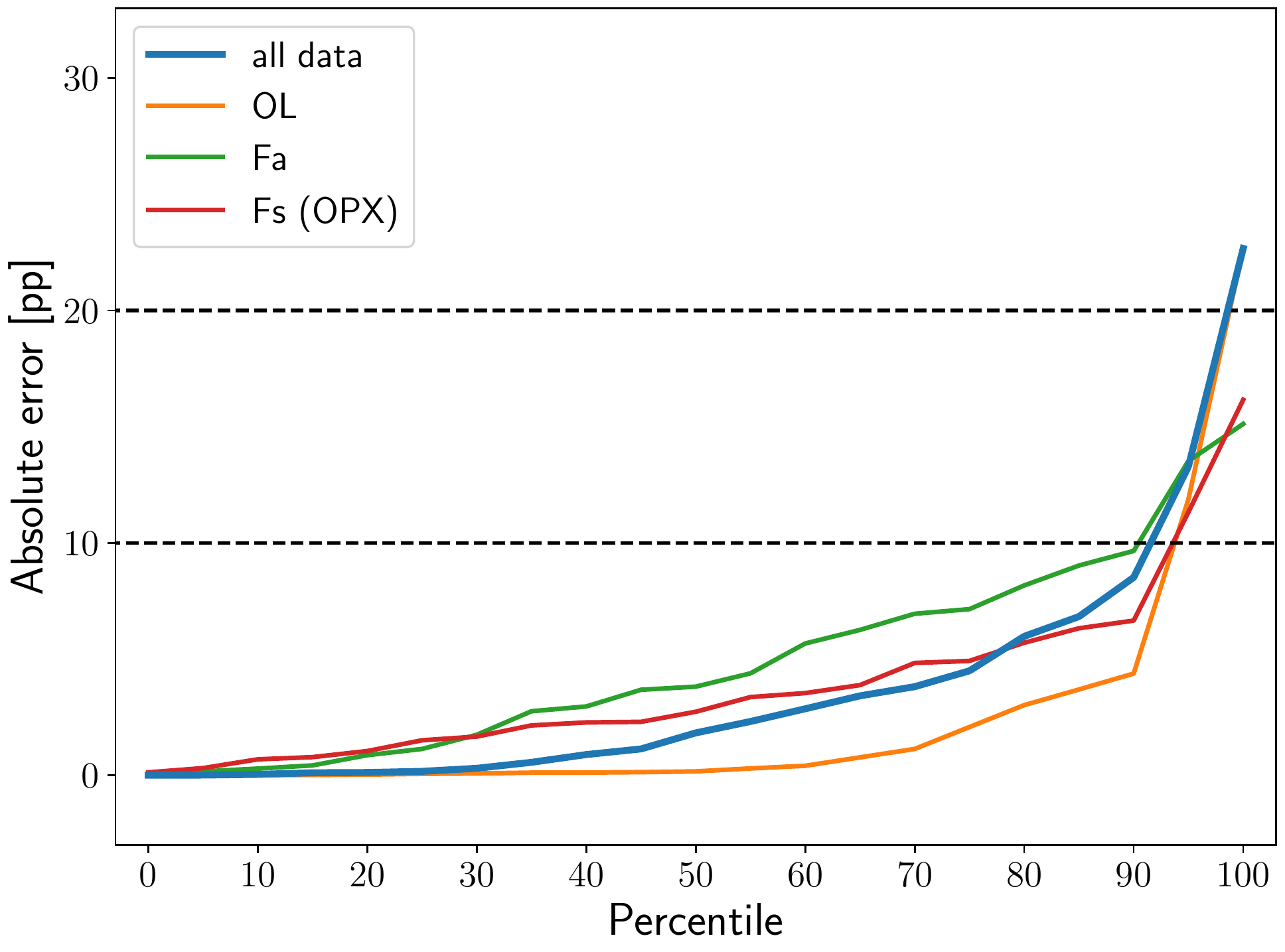}
{Quantiles of the absolute errors between the actual and predicted values. The dashed black lines indicate the 10~pp and 20~pp errors. We show the model with pure olivine, pure orthopyroxene, and their binary mixtures.}
{fig:quantile_mix}

\ocfigure[!ht]{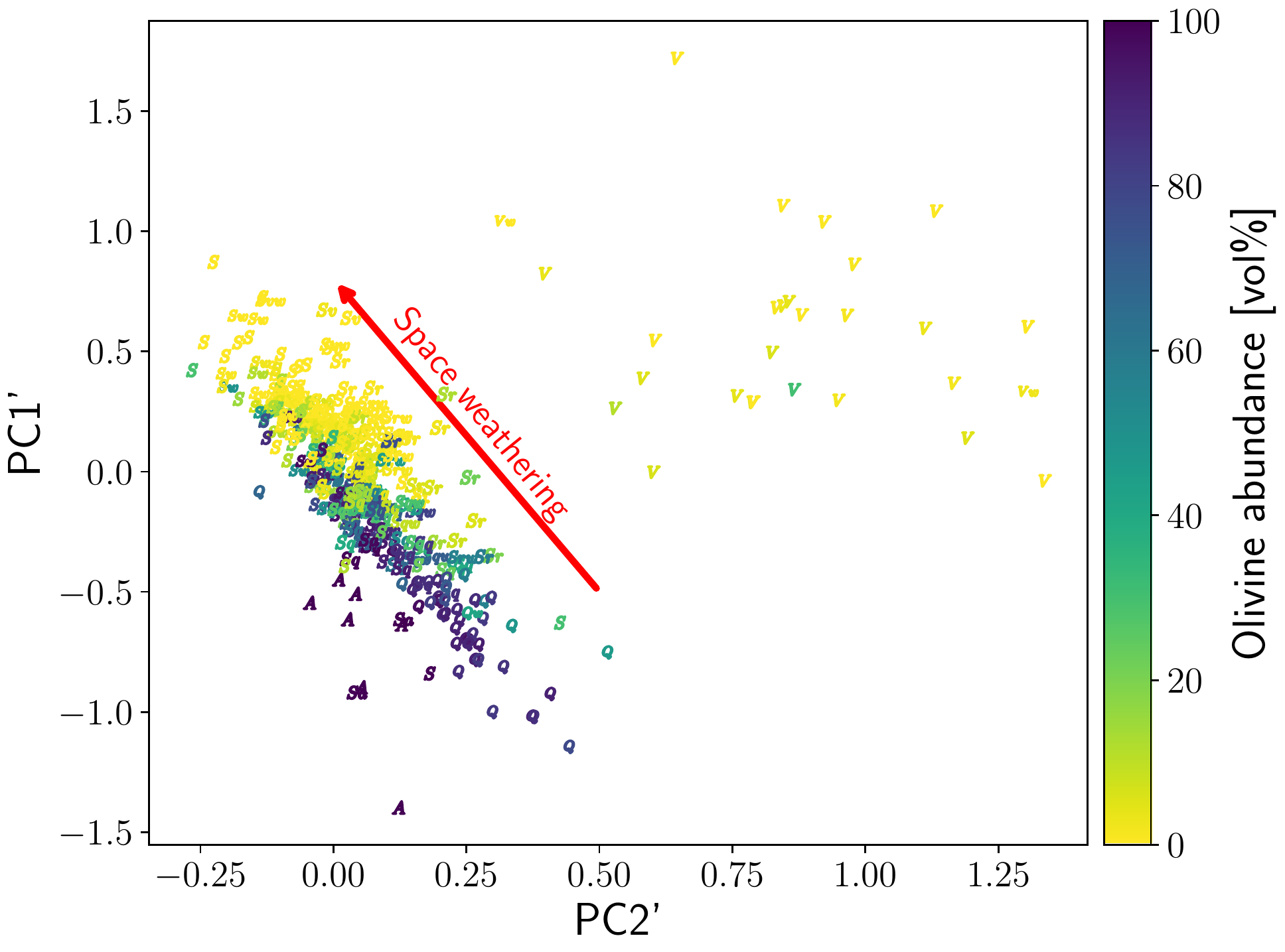}
{Principal components of S$^*$-complex asteroids. The predicted olivine fraction decreases in the direction of increasing space weathering, which is indicated by the red arrow. We show the model with pure olivine, pure orthopyroxene, and their binary mixtures.}
{fig:PCA_app}

%%%%%%%%%%%%%%%%%%%%%%%%%

\subsection{Model with plagioclase modal abundances}

\tcfigure[!ht]{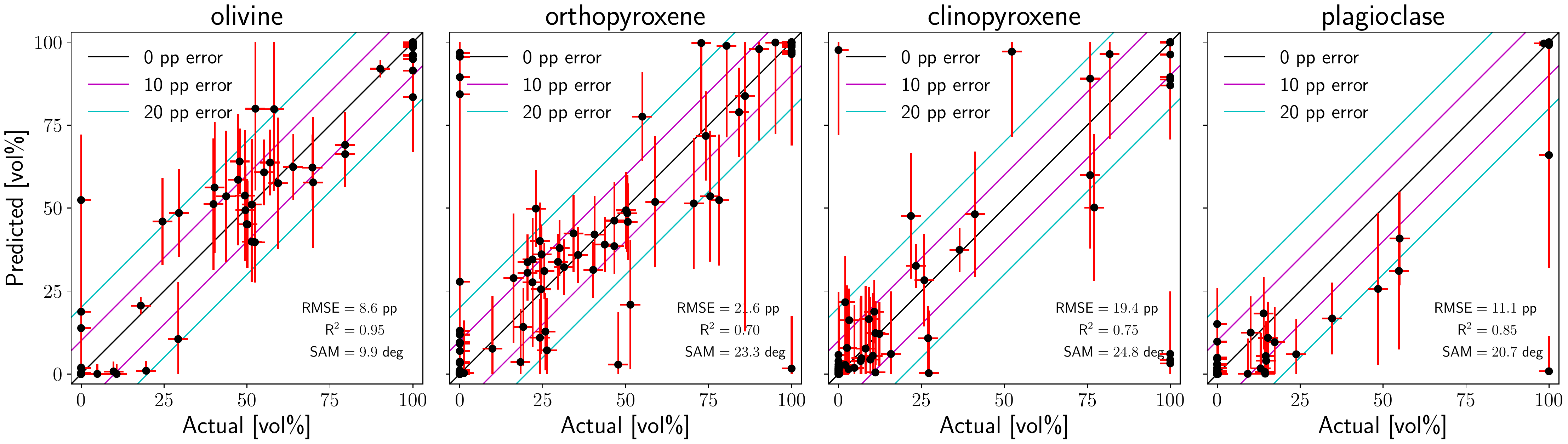}
{Scatter plots of the true and predicted modal compositions of the test data. We show the model with plagioclase modal abundance information.}
{fig:modal_all1}

\ocfigure[!ht]{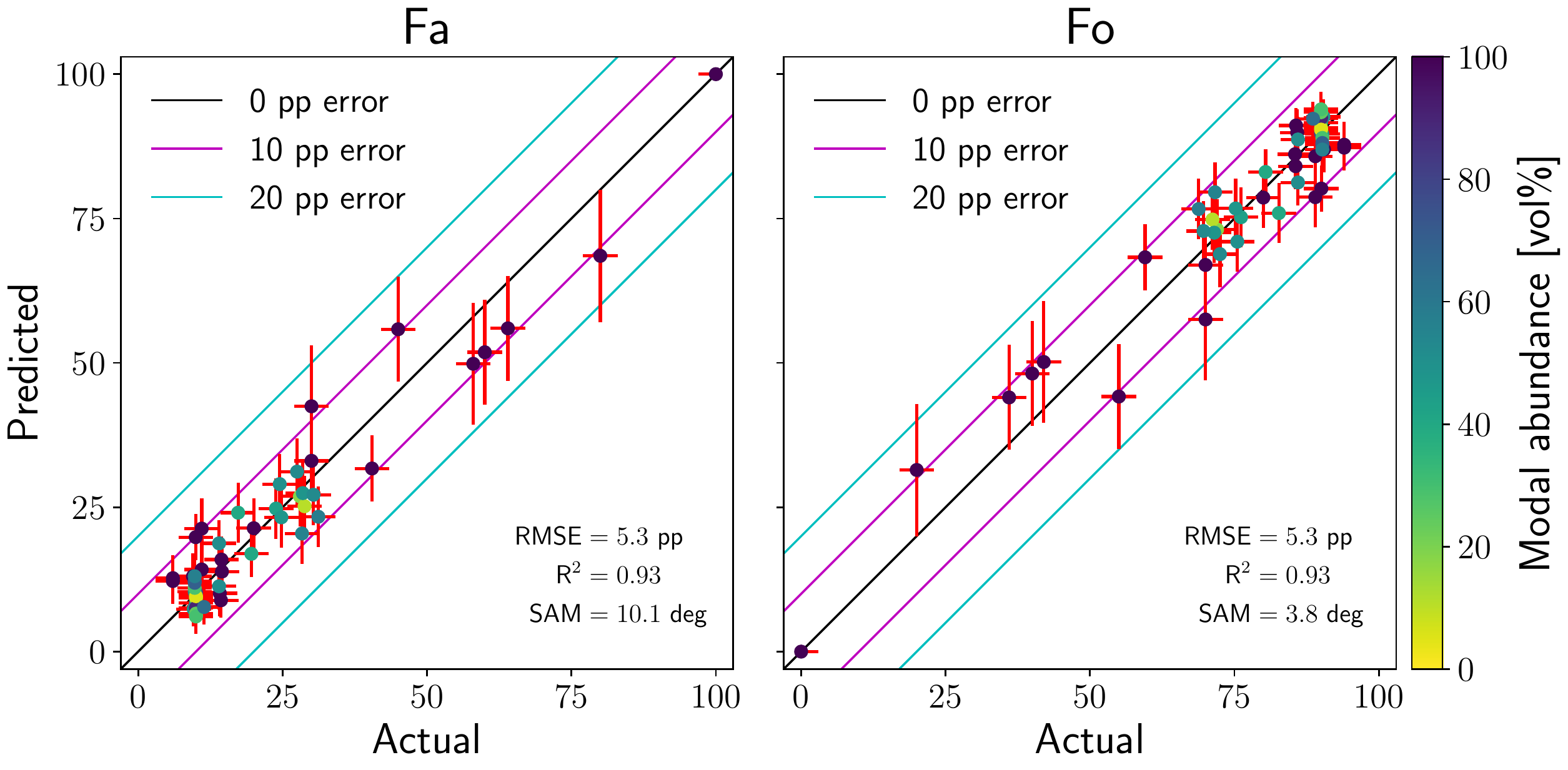}
{Scatter plots of the true and predicted chemical compositions of olivine in the test data. Left: Iron content. Right: Magnesium content. The colours of the points correspond to the actual modal abundance of olivine in the samples. We show the model with plagioclase modal abundance information.}
{fig:OL_all1}

\ocfigure[!ht]{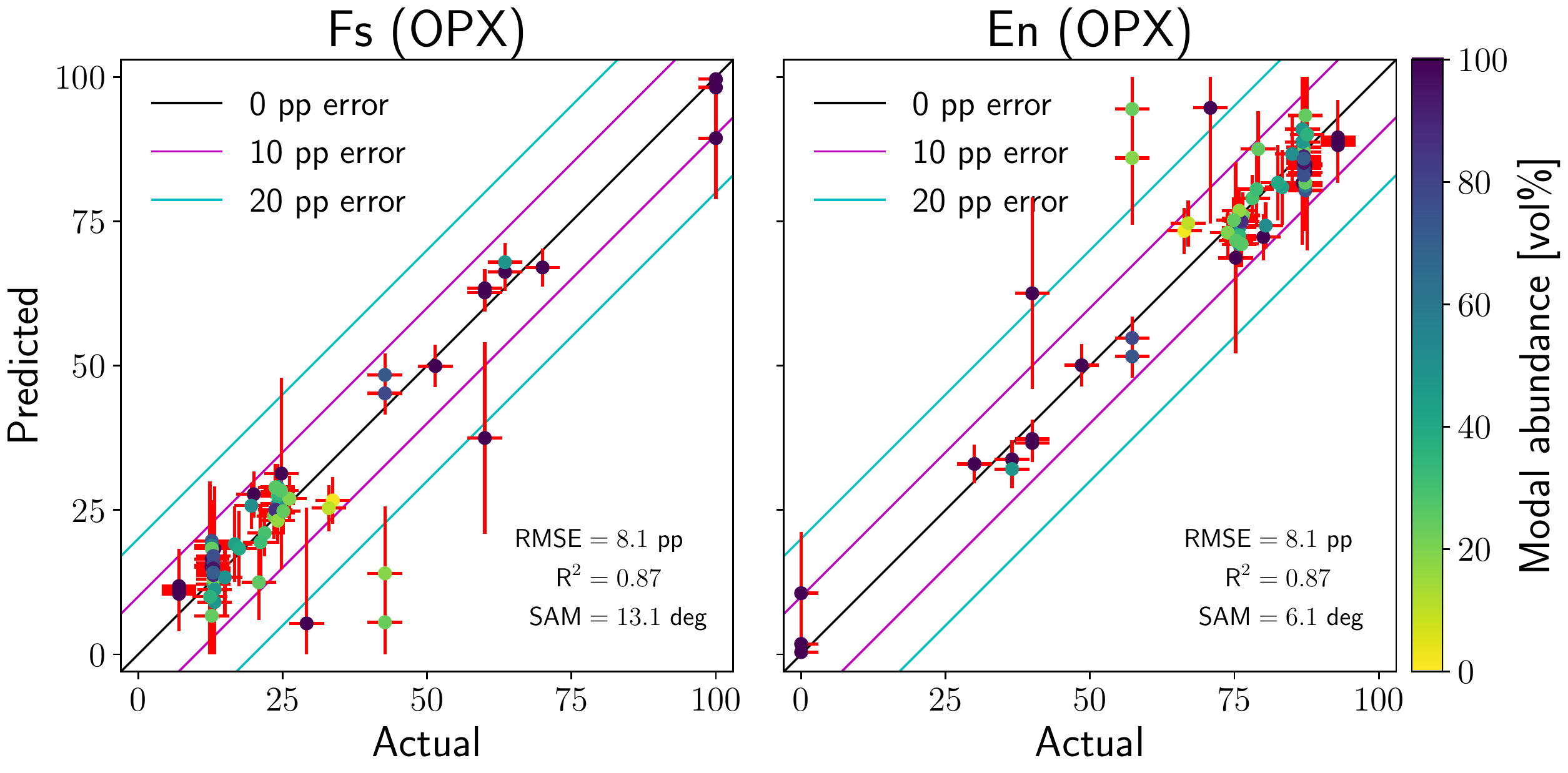}
{Scatter plots of the true and predicted chemical compositions of orthopyroxene in the test data. Left: Iron content. Right: Magnesium content. The colours of the points correspond to the actual modal abundance of orthopyroxene in the samples. We show the model with plagioclase modal abundance information.}
{fig:OPX_all1}

\ocfigure[!ht]{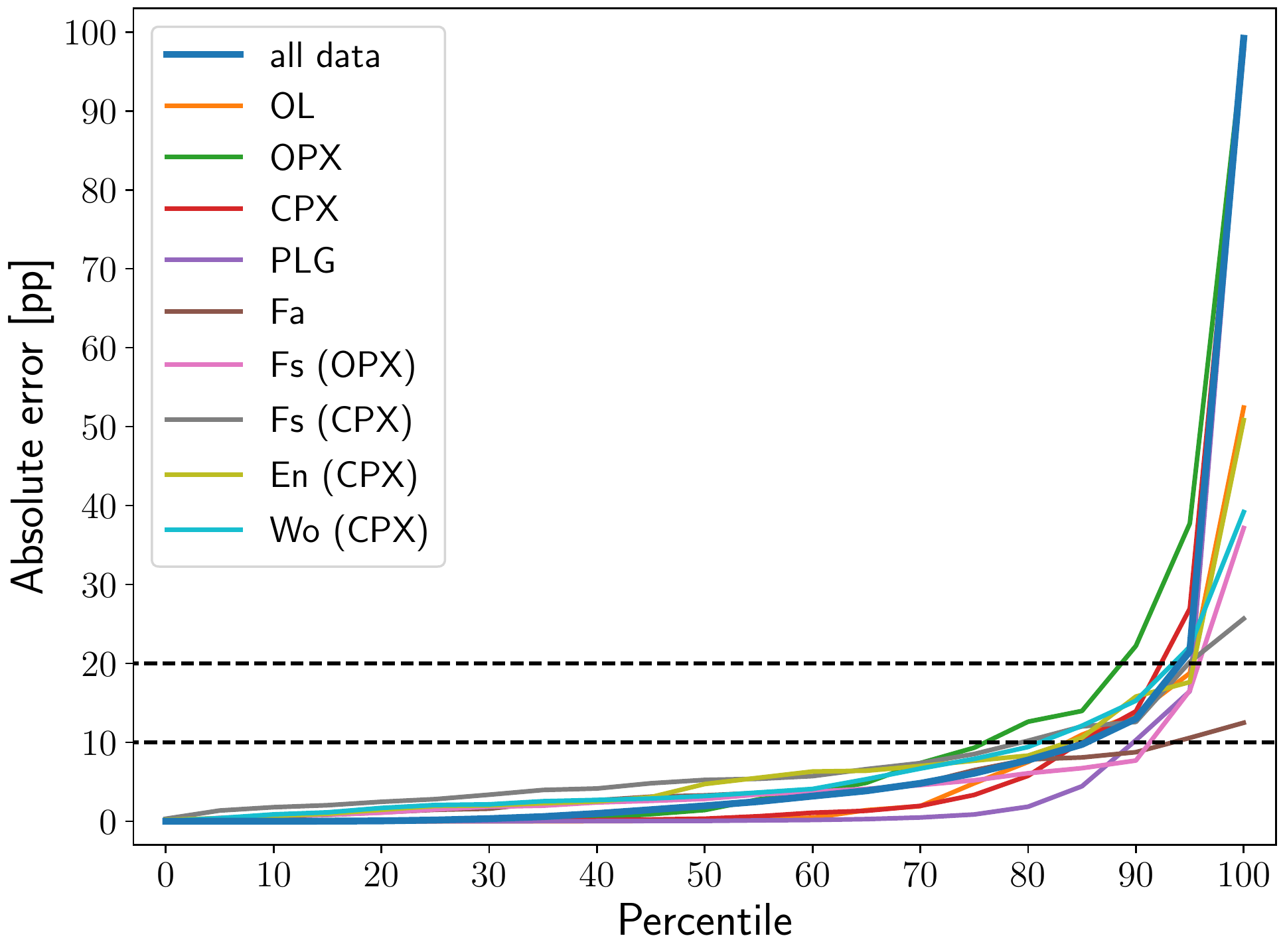}
{Quantiles of the absolute errors between the actual and predicted values. The  dashed black lines indicate the 10~pp and 20~pp errors. We show the model with plagioclase modal abundance information.}
{fig:quantile_all1}

\wfigure[!ht]{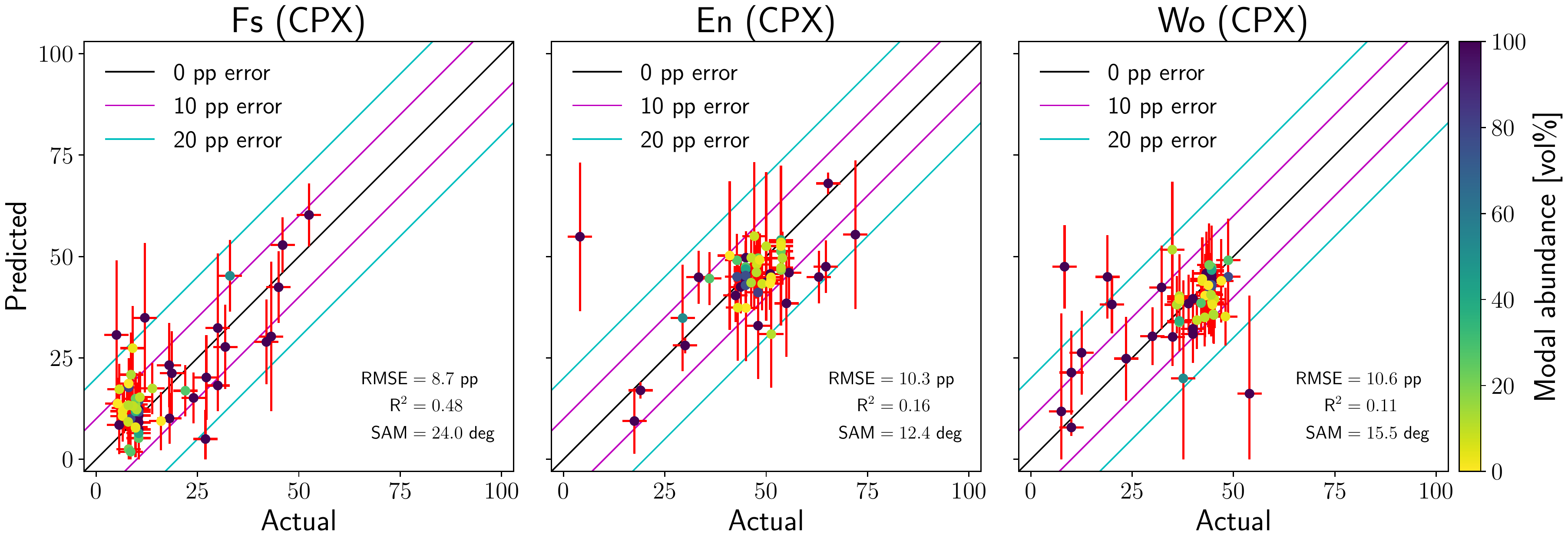}
{Scatter plots of the true and predicted chemical compositions of clinopyroxene in the test data. Left: Iron content. Middle: Magnesium content. Right: Calcium content. The colours of the points correspond to the actual modal abundance of clinopyroxene in the samples. We show the model with plagioclase modal abundance information.}
{fig:CPX_all1}

In the following two models, we also incorporated modal and complete (modal and chemical) information about plagioclase (PLG). This includes anorthosite (An), albite (Ab), and orthoclase (Or). The latter model also included orthopyroxene wollastonite information. This slightly changed the numbers and classification of our samples into different mixture categories in Table~\ref{tab:data}. In total, we had 44 pure plagioclase spectra and 15 additional mixtures containing a significant portion of plagioclase (e.g. lunar soils). On the other hand, we did not know the specific plagioclase chemical composition for all the non-plagioclase mixtures. In the model in Appendix~\ref{sect:complete_model}, we therefore removed 36 non-plagioclase mixtures.

The additional information about the modal fraction of plagioclase in the model had significant effect on the predictions of the mineral modal compositions. They were significantly worse (see Fig.~\ref{fig:modal_all1}). In addition to the orthopyroxene-clinopyroxene mismatches, there are additional outliers in the olivine modal composition, and the accuracy of predictions was worse in general. In the 10~pp error interval, there are only \num{82\%, 77\%, 84\%, and 89\%} predicted modal compositions of olivine, orthopyroxene, clinopyroxene, and plagioclase, respectively. This can be caused either by a limited dataset with an intermediate plagioclase modal composition or by the fact that the plagioclase 1.3~\textmu{}m band is due to Fe$^{2+}$ cations that naturally substitute Ca$^{2+}$ cations in the structure of plagioclase \citep{Burns_1989} rather than due to the main Ca$^{2+}$, Na$^+$, or K$^+$ cations. Therefore, the shape of the plagioclase spectrum may vary significantly, while the abundance of the main cations is almost the same, and the 1.3~\textmu{}m band of plagioclase can be misinterpreted, for example, as the contribution of the redmost band of olivine or of the weak band of orthopyroxene. We note that the predictions of the mineral chemical compositions were not affected by the additional information (see Figs.~\ref{fig:OL_all1}--\ref{fig:CPX_all1}).

%%%%%%%%%%%%%%%%%%%%%%%%%

\subsection{Model with complete information}
\label{sect:complete_model}

\tcfigure[!ht]{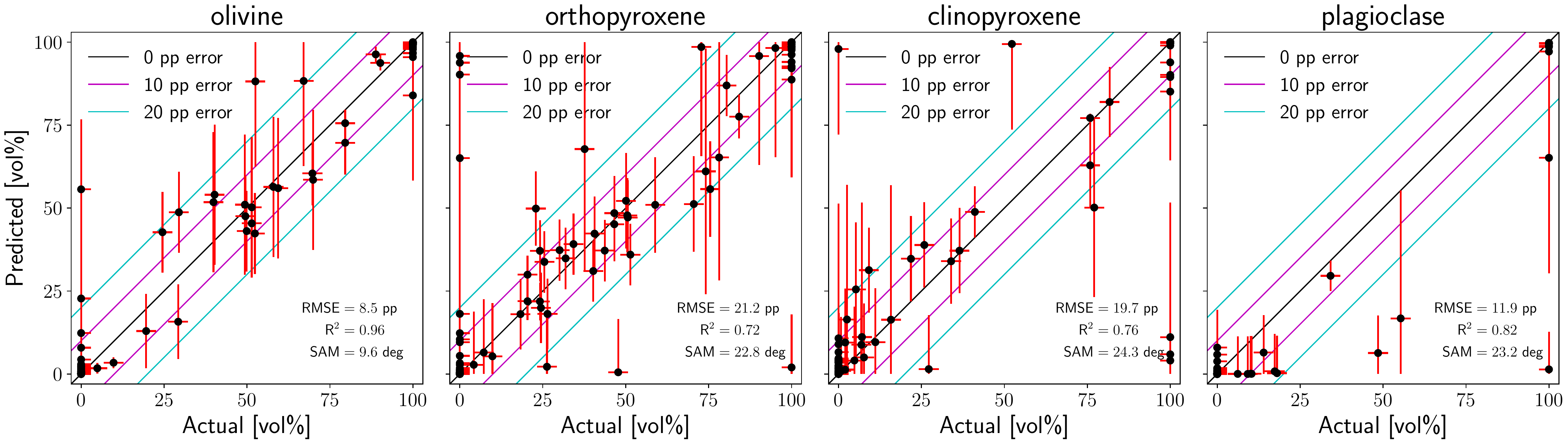}
{Scatter plots of the true and predicted modal compositions of the test data. We show the model with complete orthopyroxene and plagioclase information.}
{fig:modal_all2}

The most complex model also contains the orthopyroxene wollastonite number and the chemical composition of plagioclase. For this model, we applied additional penalisation terms to avoid predictions in regions of non-existent mineral solid solution (i.e. the region where $\mathrm{An} > 15$ and $\mathrm{Ab} < 50$ and $\mathrm{Or} > 15$ for plagioclase and $\mathrm{Wo} > 10$ for orthopyroxene). As expected from the previous test models, the mineral modal abundances and mineral chemical compositions of olivine, orthopyroxene, and clinopyroxene were not affected by this additional information (see Fig.~\ref{fig:modal_all2}--\ref{fig:CPX_all2}); the maximum predicted wollastonite number of orthopyroxene was lower than one. This model totally failed in predictions of the plagioclase chemical compositions (see Fig.~\ref{fig:PLG_all2}) because the shape of plagioclase spectrum is determined by the presence or absence of Fe$^{2+}$ cations and not by the main cations. Considering the 10~pp error interval, \num{88\%, 81\%, 85\%, and 93\%} on olivine, orthopyroxene, clinopyroxene, and plagioclase modal composition are in it. For the mineral chemical compositions, \num{93\%} of olivine Fa or Fo are within the 10~pp error interval; \num{91\%, 92\%, and 100\%} of orthopyroxene Fs, En, and Wo; \num{70\%, 77\%, and 81\%} of clinopyroxene Fs, En, and Wo; but only \num{55\%, 44\%, and 88\%} of plagioclase An, Ab, and Or are within the interval (see Fig.~\ref{fig:quantile_all2}).

\ocfigure[t]{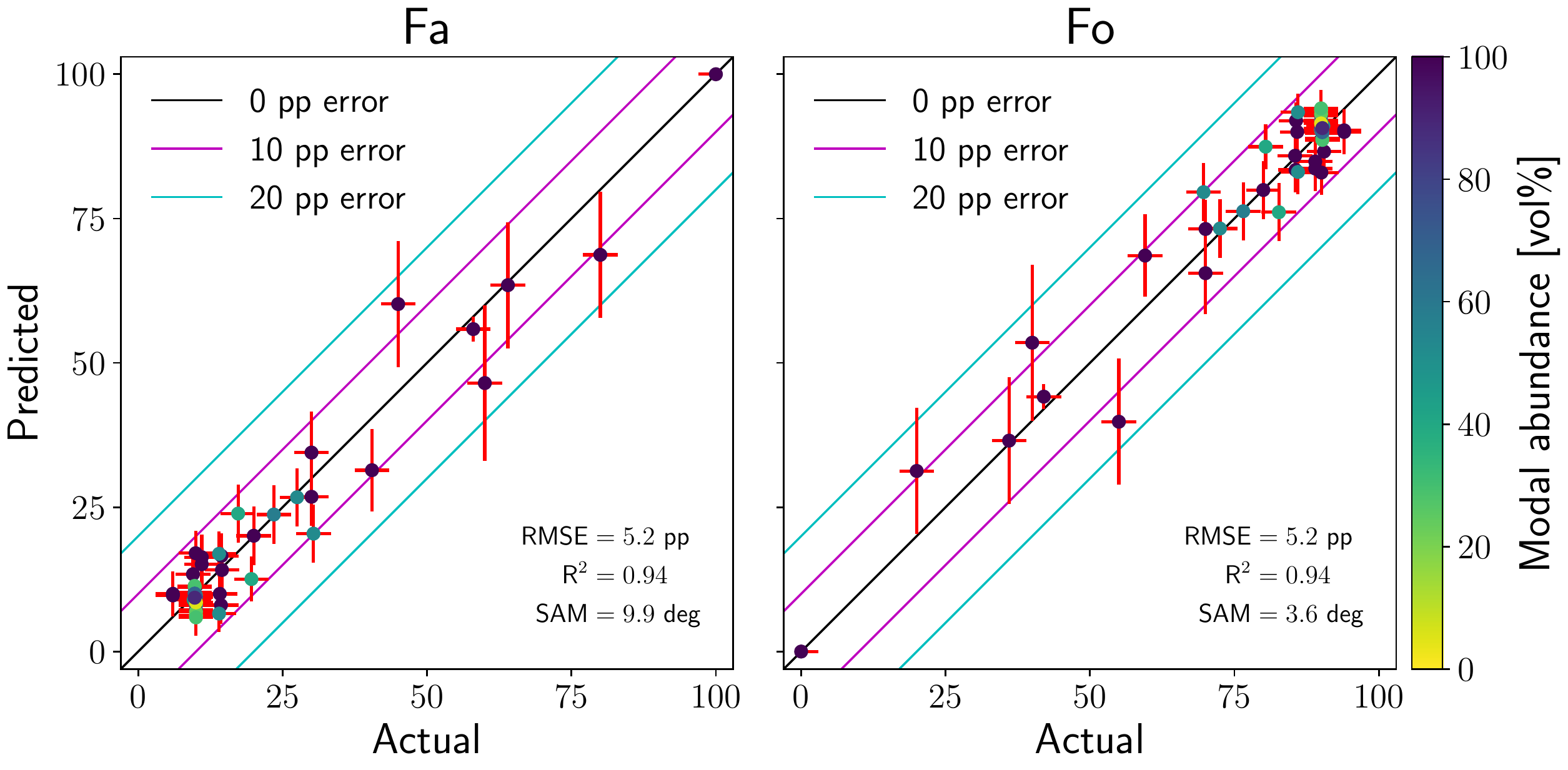}
{Scatter plots of the true and predicted chemical compositions of olivine in the test data. Left: Iron content. Right: Magnesium content. The colours of the points correspond to the actual modal abundance of olivine in the samples. We show the model with complete orthopyroxene and plagioclase information.}
{fig:OL_all2}

\wfigure[!ht]{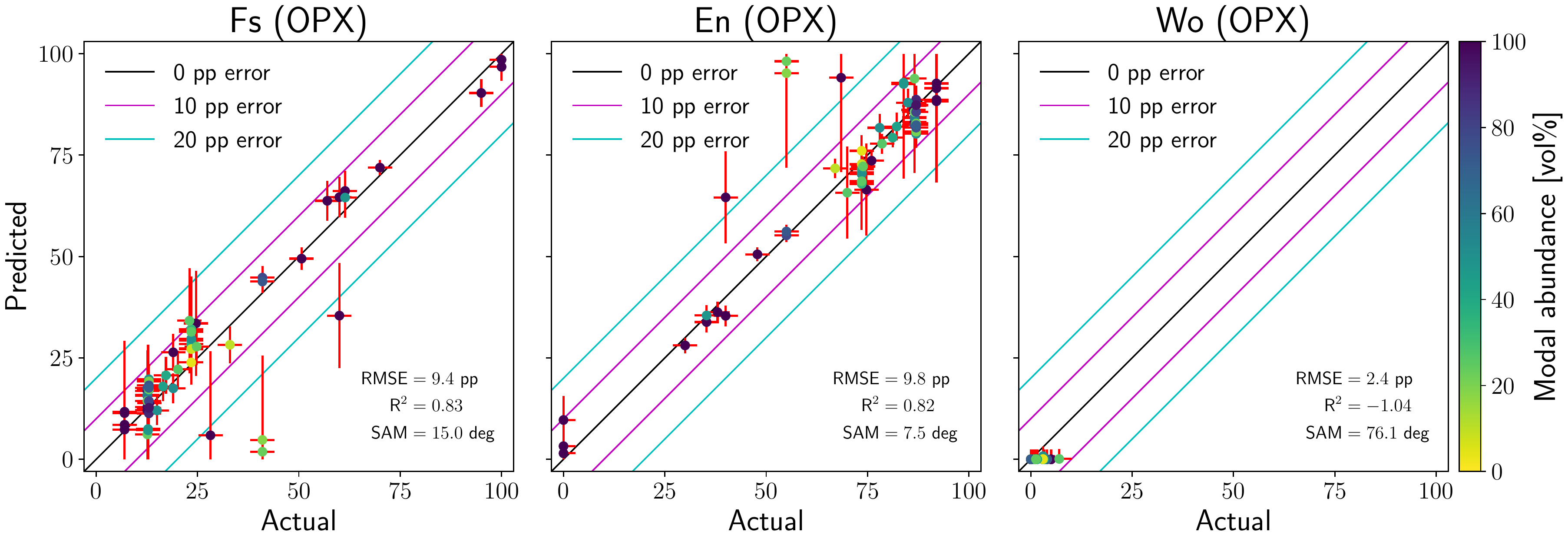}
{Scatter plots of the true and predicted chemical compositions of orthopyroxene in the test data. Left: Iron content. Right: Magnesium content. The colours of the points correspond to the actual modal abundance of orthopyroxene in the samples. We show the model with complete orthopyroxene and plagioclase information.}
{fig:OPX_all2}

\wfigure[!ht]{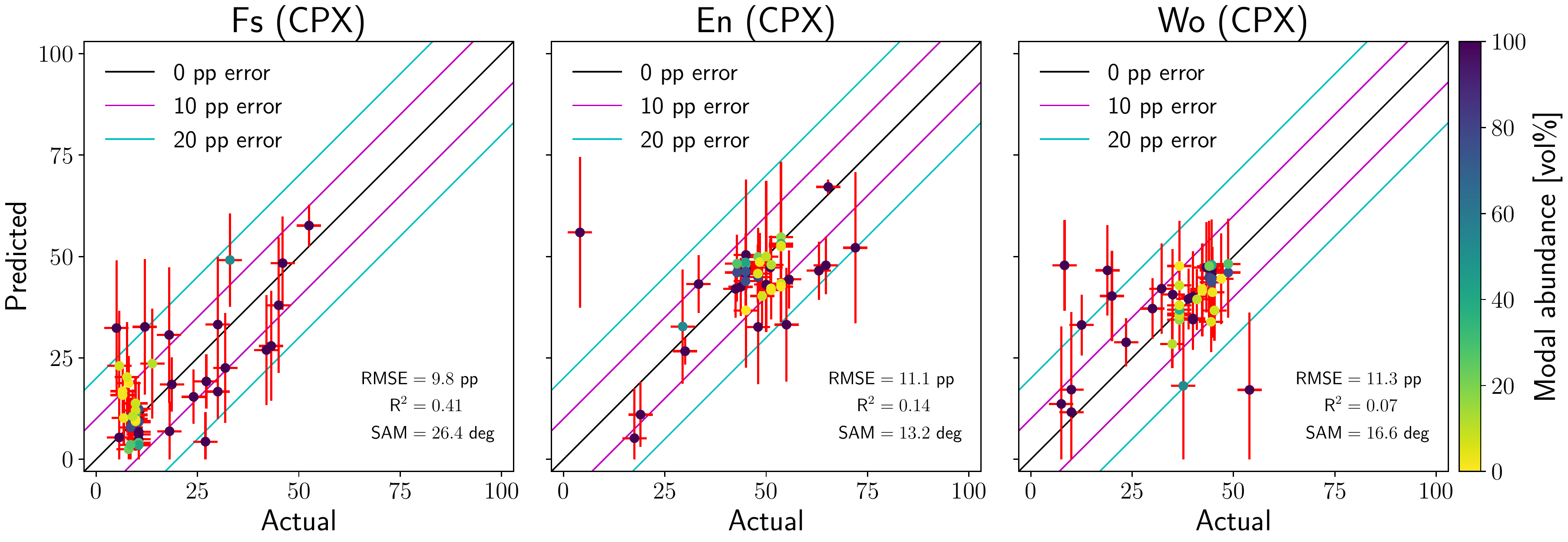}
{Scatter plots of the true and predicted chemical compositions of clinopyroxene in the test data. Left: Iron content. Middle: Magnesium content. Right: Calcium content. The colours of the points correspond to the actual modal abundance of clinopyroxene in the samples. We show the model with complete orthopyroxene and plagioclase information.}
{fig:CPX_all2}

\wfigure[!ht]{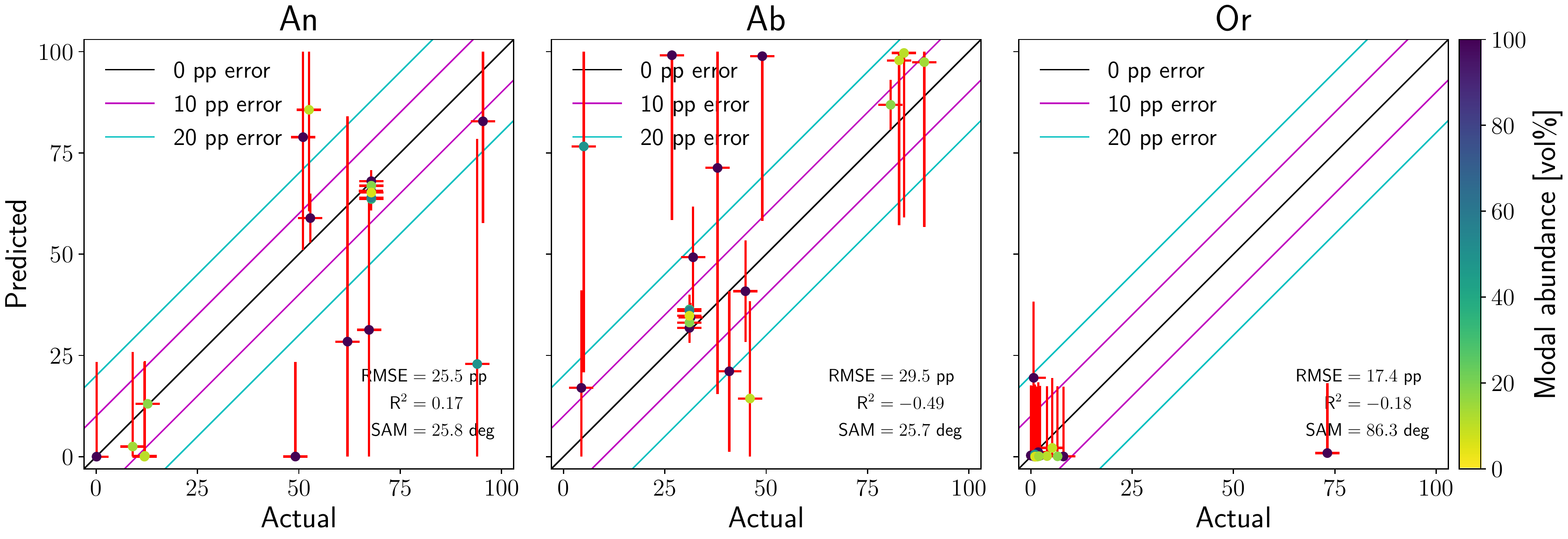}
{Scatter plots of the true and predicted chemical compositions of plagioclase in the test data. Left: Calcium content. Middle: Sodium content. Right: Potassium content. The colours of the points correspond to the actual modal abundance of plagioclase in the samples. We show the model with complete orthopyroxene and plagioclase information.}
{fig:PLG_all2}

\ocfigure[!ht]{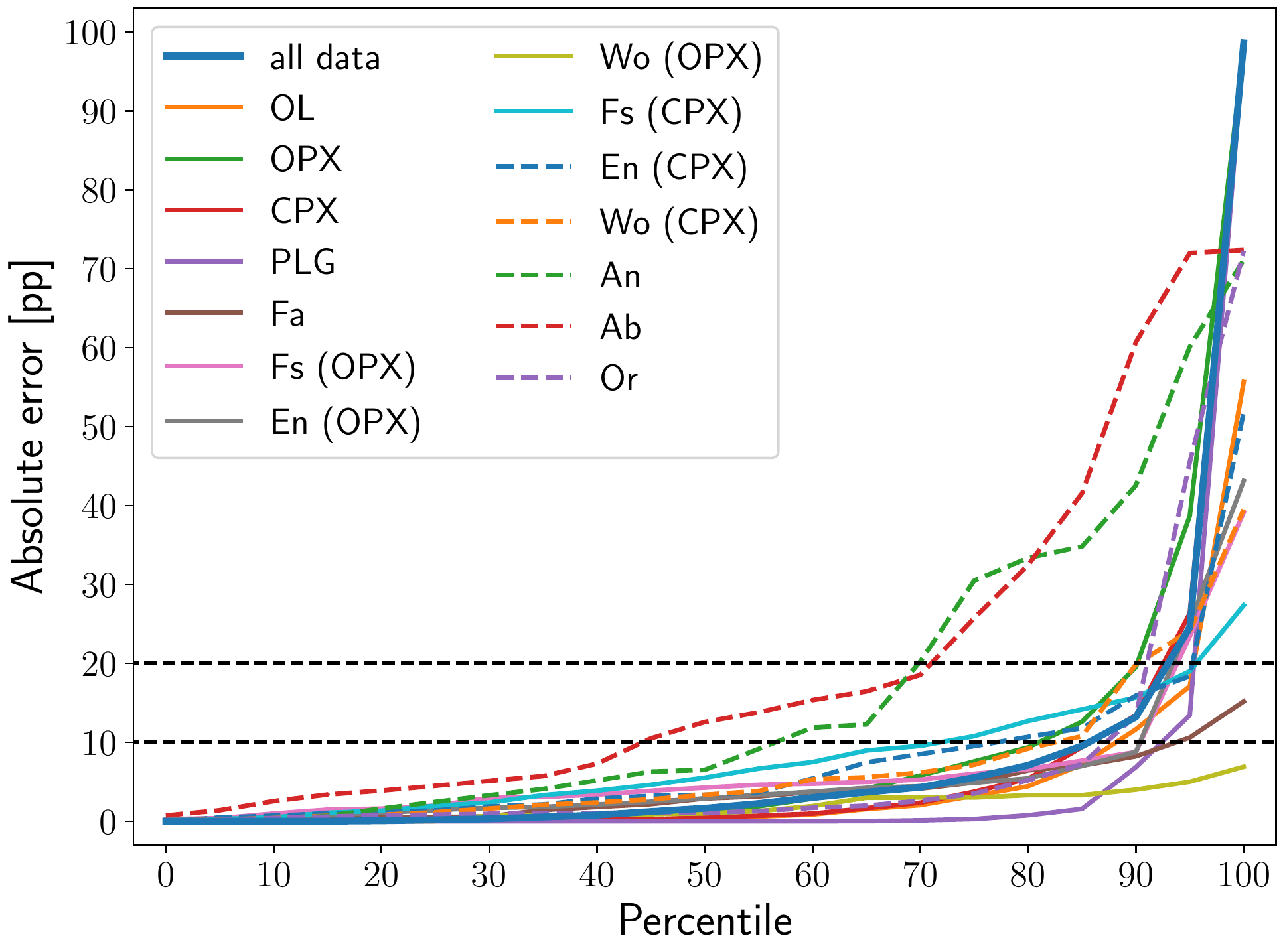}
{Quantiles of the absolute errors between the actual and predicted values. The dashed black lines indicate the 10~pp and 20~pp errors. We show the model with complete orthopyroxene and plagioclase information.}
{fig:quantile_all2}

%%%%%%%%%%%%%%%%%%%%%%%%%

\subsection{Summary of the test models}

Based on the models we tested, it seems that when we added another chemical property to our samples, it did not worsen the predictions of the previous properties. The test shows that the chemical properties of olivine and orthopyroxene can be predicted with very high accuracy, and the chemical properties of clinopyroxene can be predicted with acceptable accuracy. Moreover, the wollastonite number in case of orthopyroxene can be neglected. Furthermore, the models show that the chemical properties of plagioclase cannot be reliably obtained from the reflectance spectra.

On the other hand, additional information about mineral modal abundances caused misinterpretations of the model. The misinterpretations were mostly between orthopyroxene and clinopyroxene, which are natural and cannot be fully removed. Other misinterpretations were between the 1.3~\textmu{}m absorption band of plagioclase, which overlaps with the weak 1.25~\textmu{}m band of orthopyroxene, and the redmost absorption band of olivine.
For these reasons, we decided to omit plagioclase modal and chemical compositions and the wollastonite number of orthopyroxene from the final model.

%%%%%%%%%%%%%%%%%%%%%%%%%%%%%%%%%%%%%%%%%%%%%%%%%%

\section{Evaluation of the Chelyabinsk meteorite}
\label{sect:chelyabinsk}

The model predictions for impact-melted and shock-darkened mixtures of the Chelyabinsk meteorite are listed in Tables~\ref{tab:IM} and \ref{tab:SD}. The designation follows \citet{Kohout_2020}. In the top part of the tables, we repeat the XRD-derived values of \citet{Reddy_2014}.

\begin{table}[!ht]
    \caption{Modal abundances and mineral chemical composition of the Chelyabinsk meteorite mixed with the impact-melted phase.}
    \label{tab:IM}
    \centering
    \begin{tabular}{l c c c c c c c}
        \hline\hline
        \multicolumn{1}{c}{} &
        \multicolumn{3}{c}{modal [vol\%]} & 
        \multicolumn{1}{c}{} &
        \multicolumn{1}{c}{OPX} &
        \multicolumn{2}{c}{CPX}\\
        & OL & OPX & CPX & Fa & Fs & Fs & Wo\\
        \hline
        XRD & 66.2 & 33.8 & 0.0 & 28.6 & 23.9 & N/A & N/A\\
        \hdashline
        IM 0\% & 64.3 & 27.8 & \phantom{0}7.9 & 22.3 & 23.2 & 13.5 & 40.0 \\
        IM 10\% & 64.0 & 29.0 & \phantom{0}7.0 & 20.5 & 24.1 & 13.7 & 40.1 \\
        IM 20\% & 62.8 & 29.2 & \phantom{0}8.0 & 20.2 & 25.1 & 13.5 & 40.0 \\
        IM 30\% & 64.2 & 27.3 & \phantom{0}8.5 & 19.0 & 25.6 & 14.2 & 38.7 \\
        IM 40\% & 62.9 & 27.7 & \phantom{0}9.4 & 17.8 & 27.1 & 15.1 & 38.4 \\
        IM 50\% & 64.0 & 27.6 & \phantom{0}8.4 & 17.3 & 27.3 & 15.9 & 38.3 \\
        IM 60\% & 68.2 & 25.6 & \phantom{0}6.2 & 17.2 & 28.0 & 15.6 & 38.7 \\
        IM 70\% & 65.7 & 25.6 & \phantom{0}8.6 & 17.4 & 28.1 & 16.3 & 38.7 \\
        IM 80\% & 66.0 & 27.5 & \phantom{0}6.5 & 17.3 & 27.5 & 17.4 & 39.0 \\
        IM 90\% & 64.3 & 28.0 & \phantom{0}7.7 & 17.3 & 28.8 & 17.9 & 37.6 \\
        IM 95\% & 68.0 & 24.3 & \phantom{0}7.7 & 18.3 & 28.0 & 17.6 & 38.4 \\
        IM 100\% & 63.3 & 28.7 & \phantom{0}8.0 & 16.4 & 29.0 & 19.2 & 38.4 \\
        \hline
    \end{tabular}
        
    \vspace{1ex}
    {\raggedright XRD: \citet{Reddy_2014}. Bottom part: Model predictions. Impact-melt designation from \citet{Kohout_2020}. Error bars as derived from the model testing are \num{7.0, 6.5, 5.6, and 5.7}\b{~pp} for OL, OPX, Fa, and Fs (OPX), respectively.\par}
\end{table}

\begin{table}
    \caption{Modal abundances and mineral chemical composition of the Chelyabinsk meteorite mixed with the shock-darkened phase.}
    \label{tab:SD}
    \centering
    \begin{tabular}{l c c c c c c c}
        \hline\hline
        \multicolumn{1}{c}{} &
        \multicolumn{3}{c}{modal [vol\%]} & 
        \multicolumn{1}{c}{} &
        \multicolumn{1}{c}{OPX} &
        \multicolumn{2}{c}{CPX}\\
        & OL & OPX & CPX & Fa & Fs & Fs & Wo\\
        \hline
        XRD & 66.2 & 33.8 & 0.0 & 28.6 & 23.9 & N/A & N/A\\
        \hdashline
        SD 0\% & 58.6 & 32.9 & \phantom{0}8.6 & 20.5 & 22.6 & 11.5 & 40.5 \\
        SD 5\% & 58.3 & 33.5 & \phantom{0}8.1 & 19.8 & 22.2 & 11.9 & 40.7 \\
        SD 10\% & 58.6 & 34.1 & \phantom{0}7.3 & 20.0 & 22.8 & 11.9 & 41.1 \\
        SD 20\% & 59.7 & 33.9 & \phantom{0}6.4 & 19.5 & 23.6 & 13.0 & 40.1 \\
        SD 30\% & 59.6 & 34.0 & \phantom{0}6.4 & 18.4 & 24.6 & 12.3 & 40.5 \\
        SD 40\% & 58.0 & 34.2 & \phantom{0}7.8 & 17.2 & 24.5 & 13.6 & 39.7 \\
        SD 50\% & 60.9 & 32.8 & \phantom{0}6.4 & 16.2 & 24.7 & 15.3 & 39.1 \\
        SD 60\% & 64.1 & 30.7 & \phantom{0}5.2 & 15.0 & 26.8 & 15.1 & 39.2 \\
        SD 70\% & 67.9 & 25.8 & \phantom{0}6.3 & 14.6 & 26.8 & 15.8 & 38.4 \\
        SD 80\% & 65.4 & 27.0 & \phantom{0}7.6 & 14.3 & 25.2 & 17.3 & 37.7 \\
        SD 90\% & 64.6 & 27.3 & \phantom{0}8.1 & 13.8 & 25.4 & 17.3 & 38.0 \\
        SD 95\% & 63.9 & 28.7 & \phantom{0}7.4 & 14.5 & 25.8 & 17.9 & 37.5 \\
        SD 100\% & 63.8 & 28.9 & \phantom{0}7.3 & 13.8 & 26.8 & 18.0 & 38.0 \\
        \hline
    \end{tabular}
        
    \vspace{1ex}
    {\raggedright XRD: \citet{Reddy_2014}. Bottom part: Model predictions. Shock-darkening designation from \citet{Kohout_2020}. Error bars as derived from the model testing are \num{7.0, 6.5, 5.6, and 5.7}\b{~pp} for OL, OPX, Fa, and Fs (OPX), respectively.\par}
\end{table}

%%%%%%%%%%%%%%%%%%%%%%%%%%%%%%%%%%%%%%%%%%%%%%%%%%

\section{Evaluation of ion-irradiated samples}
\label{sect:kachr}

The model predictions for the ion-irradiated samples of pure olivine and pure pyroxene are summarised in Table~\ref{tab:kachr_app}. The designation follows \citet{Chrbolkova_2021}.

\begin{multicols}{1}

\begin{table*}[!ht]
    % \small
    \caption{Modal abundances and mineral chemical composition of the ion-irradiated pure olivine and pyroxene samples.}
    \label{tab:kachr_app}
    \centering
    \begin{tabular}{l c c c c | c c c c}
        \hline\hline
        \multicolumn{1}{c}{} &
        \multicolumn{4}{c|}{olivine} & 
        \multicolumn{4}{c}{orthopyroxene}\\
        & OL & OPX & CPX & Fa & OL & OPX & CPX & Fs\\
        \hline
        actual & 100.0 & 0.0 & 0.0 & \phantom{0}9.9 & \phantom{0}0.0 & 94.4 & \phantom{0}5.6 & 32.9\\
        \hdashline
        fresh & \phantom{0}98.8 & \phantom{0}\phantom{0}1.1 & \phantom{0}\phantom{0}0.0 & \phantom{0}\phantom{0}9.8 & \phantom{0}\phantom{0}0.1 & \phantom{0}92.3 & \phantom{0}\phantom{0}7.6 & \phantom{0}16.5 \\
        1e14 H$^+$\,cm$^{-2}$ & \phantom{0}98.8 & \phantom{0}\phantom{0}1.1 & \phantom{0}\phantom{0}0.0 & \phantom{0}\phantom{0}9.6 & - & - & - & - \\
        1e15 H$^+$\,cm$^{-2}$ & \phantom{0}98.8 & \phantom{0}\phantom{0}1.1 & \phantom{0}\phantom{0}0.0 & \phantom{0}\phantom{0}9.9 & - & - & - & - \\
        1e16 H$^+$\,cm$^{-2}$ & \phantom{0}98.7 & \phantom{0}\phantom{0}1.2 & \phantom{0}\phantom{0}0.0 & \phantom{0}10.7 & \phantom{0}\phantom{0}0.1 & \phantom{0}88.0 & \phantom{0}11.9 & \phantom{0}15.3 \\
        1e17 H$^+$\,cm$^{-2}$ & \phantom{0}98.8 & \phantom{0}\phantom{0}1.1 & \phantom{0}\phantom{0}0.0 & \phantom{0}11.4 & \phantom{0}\phantom{0}0.0 & \phantom{0}79.8 & \phantom{0}20.2 & \phantom{0}10.1 \\
        2e17 H$^+$\,cm$^{-2}$ & \phantom{0}99.5 & \phantom{0}\phantom{0}0.3 & \phantom{0}\phantom{0}0.2 & \phantom{0}18.2 & \phantom{0}\phantom{0}0.0 & \phantom{0}71.3 & \phantom{0}28.7 & \phantom{0}\phantom{0}7.3 \\
        5e17 H$^+$\,cm$^{-2}$ & \phantom{0}99.2 & \phantom{0}\phantom{0}0.5 & \phantom{0}\phantom{0}0.3 & \phantom{0}12.3 & \phantom{0}\phantom{0}0.0 & \phantom{0}71.8 & \phantom{0}28.2 & \phantom{0}\phantom{0}6.9 \\
        1e18 H$^+$\,cm$^{-2}$ & \phantom{0}99.2 & \phantom{0}\phantom{0}0.4 & \phantom{0}\phantom{0}0.4 & \phantom{0}14.2 & \phantom{0}\phantom{0}0.0 & \phantom{0}62.6 & \phantom{0}37.4 & \phantom{0}\phantom{0}5.1 \\
        \hdashline
        fresh & \phantom{0}99.2 & \phantom{0}\phantom{0}0.8 & \phantom{0}\phantom{0}0.0 & \phantom{0}\phantom{0}7.7 & \phantom{0}\phantom{0}0.0 & \phantom{0}75.0 & \phantom{0}25.0 & \phantom{0}16.2 \\
        1e16 He$^+$\,cm$^{-2}$ & \phantom{0}99.0 & \phantom{0}\phantom{0}1.0 & \phantom{0}\phantom{0}0.0 & \phantom{0}12.2 & \phantom{0}\phantom{0}0.0 & \phantom{0}74.3 & \phantom{0}25.7 & \phantom{0}16.9 \\
        3e16 He$^+$\,cm$^{-2}$ & \phantom{0}98.5 & \phantom{0}\phantom{0}1.5 & \phantom{0}\phantom{0}0.0 & \phantom{0}16.0 & \phantom{0}\phantom{0}0.0 & \phantom{0}75.9 & \phantom{0}24.1 & \phantom{0}16.5 \\
        6e16 He$^+$\,cm$^{-2}$ & \phantom{0}99.3 & \phantom{0}\phantom{0}0.7 & \phantom{0}\phantom{0}0.0 & \phantom{0}16.5 & \phantom{0}\phantom{0}0.0 & \phantom{0}80.1 & \phantom{0}19.8 & \phantom{0}15.5 \\
        1e17 He$^+$\,cm$^{-2}$ & \phantom{0}99.7 & \phantom{0}\phantom{0}0.3 & \phantom{0}\phantom{0}0.0 & \phantom{0}19.7 & \phantom{0}\phantom{0}0.0 & \phantom{0}79.0 & \phantom{0}21.0 & \phantom{0}15.0 \\
        \hdashline
        fresh & \phantom{0}99.3 & \phantom{0}\phantom{0}0.7 & \phantom{0}\phantom{0}0.1 & \phantom{0}10.4 & \phantom{0}\phantom{0}0.2 & \phantom{0}90.0 & \phantom{0}\phantom{0}9.8 & \phantom{0}20.7 \\
        1e15 Ar$^+$\,cm$^{-2}$ & \phantom{0}99.3 & \phantom{0}\phantom{0}0.7 & \phantom{0}\phantom{0}0.1 & \phantom{0}11.4 & \phantom{0}\phantom{0}0.3 & \phantom{0}92.8 & \phantom{0}\phantom{0}6.9 & \phantom{0}20.9 \\
        3e15 Ar$^+$\,cm$^{-2}$ & \phantom{0}99.3 & \phantom{0}\phantom{0}0.7 & \phantom{0}\phantom{0}0.0 & \phantom{0}14.3 & \phantom{0}\phantom{0}0.3 & \phantom{0}93.8 & \phantom{0}\phantom{0}5.9 & \phantom{0}19.6 \\
        6e15 Ar$^+$\,cm$^{-2}$ & \phantom{0}99.4 & \phantom{0}\phantom{0}0.6 & \phantom{0}\phantom{0}0.0 & \phantom{0}14.3 & \phantom{0}\phantom{0}0.4 & \phantom{0}95.2 & \phantom{0}\phantom{0}4.4 & \phantom{0}19.6 \\
        1e16 Ar$^+$\,cm$^{-2}$ & \phantom{0}99.4 & \phantom{0}\phantom{0}0.6 & \phantom{0}\phantom{0}0.0 & \phantom{0}15.3 & \phantom{0}\phantom{0}0.2 & \phantom{0}93.7 & \phantom{0}\phantom{0}6.2 & \phantom{0}20.6 \\
        2e16 Ar$^+$\,cm$^{-2}$ & \phantom{0}99.4 & \phantom{0}\phantom{0}0.5 & \phantom{0}\phantom{0}0.0 & \phantom{0}17.9 & \phantom{0}\phantom{0}0.4 & \phantom{0}94.7 & \phantom{0}\phantom{0}4.9 & \phantom{0}19.8 \\
        6e16 Ar$^+$\,cm$^{-2}$ & \phantom{0}99.6 & \phantom{0}\phantom{0}0.4 & \phantom{0}\phantom{0}0.0 & \phantom{0}19.2 & \phantom{0}\phantom{0}0.1 & \phantom{0}90.7 & \phantom{0}\phantom{0}9.2 & \phantom{0}18.4 \\
        1e17 Ar$^+$\,cm$^{-2}$ & \phantom{0}99.6 & \phantom{0}\phantom{0}0.4 & \phantom{0}\phantom{0}0.0 & \phantom{0}21.3 & \phantom{0}\phantom{0}0.0 & \phantom{0}82.1 & \phantom{0}17.8 & \phantom{0}15.1 \\
        \hline
    \end{tabular}
        
    \vspace{1ex}
    {\raggedright Space-weathering designation from \citet{Chrbolkova_2021}. Actual modal abundances of pyroxenes are in wt\%.\par}
\end{table*}
\end{multicols}

\end{document}